\documentclass[a4paper,openright, twoside, 12pt]{book}

\usepackage{latexsym,amsmath,amssymb,amscd,epsfig,fancyhdr,graphicx,amsfonts, epigraph, mathrsfs, tabularx, booktabs, empheq}

\begin{document}

%%%%%%%%%%%%%%%%%%%%%%
%%%    Nuovi comandi di Jarah            %%%
%%%%%%%%%%%%%%%%%%%%%%

\newcommand{\dd}{\textrm{d}}
\newcommand{\beq}{\begin{equation}}
\newcommand{\eeq}{\end{equation}}
\newcommand{\bea}{\begin{eqnarray}}
\newcommand{\eea}{\end{eqnarray}}
\newcommand{\beas}{\begin{eqnarray*}}
\newcommand{\eeas}{\end{eqnarray*}}
\newcommand{\defi}{\stackrel{\rm def}{=}}
\newcommand{\non}{\nonumber}
\newcommand{\bquo}{\begin{quote}}
\newcommand{\enqu}{\end{quote}}
\renewcommand{\(}{\begin{equation}}
\renewcommand{\)}{\end{equation}}
\def\de{\partial}
\def\Om{\ensuremath{\Omega}}
\def\Tr{ \hbox{\rm Tr}}
\def\rc{ \hbox{$r_{\rm c}$}}
\def\H{ \hbox{\rm H}}
\def\HE{ \hbox{$\rm H^{even}$}}
\def\HO{ \hbox{$\rm H^{odd}$}}
\def\HEO{ \hbox{$\rm H^{even/odd}$}}
\def\HOE{ \hbox{$\rm H^{odd/even}$}}
\def\HHO{ \hbox{$\rm H_H^{odd}$}}
\def\HHEO{ \hbox{$\rm H_H^{even/odd}$}}
\def\HHOE{ \hbox{$\rm H_H^{odd/even}$}}
\def\K{ \hbox{\rm K}}
\def\Im{ \hbox{\rm Im}}
\def\Ker{ \hbox{\rm Ker}}
\def\const{\hbox {\rm const.}}
\def\o{\over}
\def\im{\hbox{\rm Im}}
\def\re{\hbox{\rm Re}}
\def\bra{\langle}\def\ket{\rangle}
\def\Arg{\hbox {\rm Arg}}
\def\exo{\hbox {\rm exp}}
\def\diag{\hbox{\rm diag}}
\def\longvert{{\rule[-2mm]{0.1mm}{7mm}}\,}
\def\a{\alpha}
\def\b{\beta}
\def\e{\epsilon}
\def\l{\lambda}
\def\ol{{\overline{\lambda}}}
\def\ochi{{\overline{\chi}}}
\def\th{\theta}
\def\s{\sigma}
\def\oth{\overline{\theta}}
\def\ad{{\dot{\alpha}}}
\def\bd{{\dot{\beta}}}
\def\oD{\overline{D}}
\def\opsi{\overline{\psi}}
\def\dag{{}^{\dagger}}
\def\tq{{\widetilde q}}
\def\L{{\mathcal{L}}}
\def\p{{}^{\prime}}
\def\W{W}
\def\N{{\cal N}}
\def\hsp{,\hspace{.7cm}}
\def\bo{\ensuremath{\hat{b}_1}}
\def\bfo{\ensuremath{\hat{b}_4}}
\def\co{\ensuremath{\hat{c}_1}}
\def\cfo{\ensuremath{\hat{c}_4}}
\newcommand{\C}{\ensuremath{\mathbb C}}
\newcommand{\Z}{\ensuremath{\mathbb Z}}
\newcommand{\R}{\ensuremath{\mathbb R}}
\newcommand{\rp}{\ensuremath{\mathbb {RP}}}
\newcommand{\cp}{\ensuremath{\mathbb {CP}}}
\newcommand{\vac}{\ensuremath{|0\rangle}}
\newcommand{\vact}{\ensuremath{|00\rangle}                    }
\newcommand{\oc}{\ensuremath{\overline{c}}}

\newcommand{\Vol}{\textrm{Vol}}

\newcommand{\half}{\frac{1}{2}}

\def\changed#1{{\bf #1}}

%%%%%%%%%%%%%%%%%%%%%%%%%%%%%%%%
%%%                                                                                               %%%
%%%%%%%%%%%%%%%%%%%%%%%%%%%%%%%%

%Front Page

\frontmatter

\pagestyle{empty}

\newpage

\oddsidemargin +1.5cm \evensidemargin +0.5cm

\begin{titlepage}
\begin{center}

  \begin{figure}[t]
    \centering
    \includegraphics[width=0.18\textwidth]{logo_unipi.epsi}\hspace{1.2cm} 
    \includegraphics[width=0.15\textwidth]{logo_galilei.epsi}\hspace{1.2cm} 
    \includegraphics[width=0.19\textwidth]{logo_dott.epsi} 
\vspace{1cm} 
  \end{figure}

  {\LARGE   Graduate Course in Physics} \vspace{0.5cm}

  {\LARGE   University of Pisa} \vspace{0.5cm}

  {\large Department of Physics ``Enrico Fermi''} \vspace{0.5cm}

  {\large   School of Graduate Studies in Basic Sciences ``Galileo Galilei''}
  \vspace{0.5cm} 

  {\large  XXI Cycle --- 2006--2008 \\ \vspace{\stretch{3}}}

  {\Large   Ph.D. Thesis \\ \vspace{\stretch{1}}}

 \parbox[c]{14cm}{\LARGE \bf \center Surface Layers in the Gravity/Hydrodynamics Correspondence \vspace{0.3cm}}

 \vspace{\stretch{3}}

 {\large
   \begin{tabular}[h]{lcccr}
     \parbox[t]{6cm}{{\it Candidate}\vspace{0.3cm} \\ Giovanni Ricco}& & & &
     \parbox[t]{6cm}{{\it Supervisor}\vspace{0.3cm} \\ Dr. Jarah Evslin}
   \end{tabular}
 }

\end{center}
\end{titlepage}

% Acknoledgments

\chapter*{Acknowledgments}

This PhD Thesis has been completed at the Department of Physics of University of Pisa. This work includes the most recent results of my research that has been conducted under the supervision of Dr Jarah Evslin. For the sake of homogeneity of the subject, some results about non-tachyonic non-supersymmetric vacua in Superstring Theory, that are part of my research work during the PhD, are not reported in this Thesis.\\

The research presented here has greatly benefited from Prof. Juan Maldacena's inspiring inputs, for which I am deeply grateful.\\ 

I am in debt for the organization of materials of the first chapter to the wonderful lectures held by Prof. Damour during the LXXXVII Session of the \'Ecole d'\'Et\'e de Physique des Houches,  I attended in Les Houches in 2006.\\

I would like to thank the Graduate Council of Physics of University of Pisa for giving me the opportunity to complete my research. I am particularly grateful to the Head of the Graduate Council in Physics, Prof. Kenichi Konishi, for his patience and for having encouraged me to reflect on my PhD experience.\\

I would like to thank Prof. Emilian Dudas and the whole CPhT of the \'Ecole Polytechnique of Paris for their kind hospitality. Also, I would like to thank the Theoretical Group of the Department of Physics of the University of Turin for hosting me during the third year of my PhD.\\		 

Without the guidance and support of  Dr Jarah Evslin, I could not have completed this Thesis. In particular, I would like to thank him for his time and for his invaluable insights and comments during our work together. I am also grateful to him for correcting thousands of English mistakes, flipping hundreds of pluses into minuses,  sharing dozens of Chinese dinners, dragging me into  a couple of days of physics and sailing in the Adriatic sea, and for roasting a (two!) whole pig(s) during a memorable Croatian night this summer.\\
 
I would like to acknowledge the encouragement and support of Prof. Mihail Mintchev throughout my years in Pisa.\\

I would like to acknowledge the support of Fernando D'Aniello and Francesco Mauriello, with whom I have shared a long experience within the ADI - Associazione Dottorandi Italiani -,  and Marco Broccati of the FLC Cgil - Federazione Lavoratori della Conoscenza of the Confederazione Generale Italiana del Lavoro -, who has taught me a lot during our years of collaboration. \\

I am also very grateful to all the PhD fellows at the Department of Physics for helpful intellectual discussions, their moral support and for making me laugh. In particular, I am in debt to Matteo Giordano, Walter Vinci, Bjarke Gud{\hspace{-.05cm}}${\bar{}}$nason and Guglielmo Paoletti. Finally, I would like to thank Giuliana Manta, Ilaria D'Angelo and Leone Cavicchia, my flatmates in Pisa, for their great help and for sharing with me many wonderful sunny Sundays on the roof of our house.\\

%Table of Contents
\tableofcontents

% Style of pages

\pagestyle{fancy}\fancyhf{}
\renewcommand{\chaptermark}[1]{\markboth{#1}{}}
\renewcommand{\sectionmark}[1]{\markright{\thesection.\ #1}{}}
\fancyhead[LE,RO]{\bfseries\thepage}
\fancypagestyle{plain}{\fancyhead{} \renewcommand{\headrulewidth}{0pt}}
\oddsidemargin +1.5cm \evensidemargin +0.5cm
\fancyhead[RE]{\bfseries\leftmark}
\fancyhead[LO]{\slshape\rightmark}
\setlength{\headwidth}{14cm}

\mainmatter

% Introduction

\chapter*{Introduction}
\addcontentsline{toc}{chapter}{Introduction}

It has long been known that the dynamics of a $p$--dimensional gravitational theory is captured by quantities on $(p-1)$--dimensional hypersurfaces \cite{EIH}.  It was argued by Damour \cite{D1979}, based on an analogy by Hartle and Hawking \cite{HH1972}, that in the case of certain black hole solutions these surface quantities describe the flow of a viscous $(p-1)$--dimensional fluid.  Perhaps the most concrete realization of this idea is the one to one map between large wavelength features of asymptotically AdS$_p$ black brane solutions and $(p-1)$-dimensional conformal fluid flows presented recently in Refs.~\cite{Bhattacharyya:2008jc} and \cite{Bhattacharyya:2008mz}.  

Indeed, within the framework of string theory's AdS/CFT correspondence,  an initially surprising relationship between the vacuum equations of Einstein gravity in an asymptotically locally AdS$_{d+1}$ space and the equations of hydrodynamics in $d$ dimensions has been unearthed. In particular a one to one correspondence has been found between, on the one hand, a class of regular, long wavelength locally asymptotically AdS$_{d+1}$ black brane solutions to the vacuum Einstein equations with a negative cosmological constant and, on the other, all long-wavelength solutions of the $d$ dimensional hydrodynamical equations $$\nabla_\mu T^{\mu \nu} = 0$$ of conformal fluid flows. The AdS/hydrodynamics correspondence provides an explicit black brane solution for every history of a particular conformal fluid so long as the fluid variables are constant over distances which are large compared with the inverse temperature. 

The nonrelativistic scaling limit - long distances, long times, low speeds and low amplitudes - of the relativistic equations of hydrodynamics, connects the relativistic conservation equation to the incompressible non-relativistic Navier-Stokes equations 
\[
\label{NS} 
\frac{\partial {\vec v}}{\partial t}+ {\vec v}\cdot \nabla  {\vec v} = -{\vec \nabla} p
+ \nu \nabla^2 {\vec v} + {\vec f}  \ ,
\]
that have been investigated for nearly two centuries but whose  extremely rich dynamics still remainsl to be completely clarified. A particularly interesting phenomenon is turbulence. In fact, most fluid flows become turbulent under a wide range of conditions. Even though turbulent flows are complicated  phenomena of a statistical nature, it has been proposed that they could be actually governed by a new and simple universal mathematical structure analogous to a fixed point of the renormalization group flow equations ({\it e.g.} \cite{Polyakov:1992yw}).

Given the fluid-gravity correspondence, it is of great interest to investigate the gravity dual solutions of turbulent flows, and the conditions under which turbulence may be expected. These turbulent solutions in the gravity dual may cast light on the possible turbulent decay of gravitational solutions near spacelike singularities - where indeed chaotic evolution is expected \cite{BKL}. \\ 

It is well known, for example from the Richardson's cascade model \cite{Rich}, that to realize steady state turbulence, one has to inject energy into a system. 

A first possible way to do this is to apply an external perturbation - a forcing function - deforming the fluid.  Another way is to consider boundary conditions. In the context of the fluid-gravity correspondence, this first approach has been investigated in Ref.~\cite{Bhattacharyya:2008ji} where it was argued that a laminar fluid flow and the dual gravity solution could decay to turbulent configurations. The problem with this method is that forcing functions are  necessarily inhomogeneous and bring rather complicated solutions.

The second approach is closer to the intuition deriving from classical hydrodynamics. Indeed many solutions of the Navier-Stokes equations describing fluids subject to hard wall type boundary conditions are known, even though it is still not clear how to generate gravitational duals of these boundary conditions.\\ 

In this Thesis we study boundary conditions in the AdS/hydrodynamics correspondence as a preliminary investigation to the implementation of this second approach. In particular we study the gravity dual of a boundary layer separating two nonrelativistic, incompressible, fluid solutions with different properties: a stationary fluid on the left side and a moving fluid on the right. This simple setting is sufficient to produce turbulent motion in the fluid. For this configuration we derive gravitational duals of the boundary, following the prescription of Ref.~\cite{Bhattacharyya:2008jc}. While the fluid-gravity correspondence maps fluids on the left and on the right of the layer into two vacuum gravity solutions, the non trivial point is how to glue these two regions together.

We consider two distinct methods, that have different limits of validity, with respect to the ``thickness of the layer''. The first one consists in gluing the two gravitational solutions by applying the Israel matching conditions \cite{Israel:1966rt} to determine the stress tensor on the surface layer that separates the two sides. One can understand this approach as letting the gravitational solution continuously interpolate between the two solutions over a finite distance $d$ and then taking the limit as this distance tends to zero while keeping the extrinsic curvature of the interpolating layer finite.

The second possibility is to consider the gravity dual of a velocity function interpolating between the two fluid solutions.  Clearly this function will not be a  a solution of the Navier-Stokes equation in this region and therefore the dual will not be a solution of the vacuum Einstein equations in this region. Nevertheless one can follow this idea to see where it leads. In this case, one cannot take the interpolation distance $d$ to zero, because the dual is not defined when derivatives are large with respect to the inverse of the temperature $T$. 

A first interesting result that we will describe in detail in the last chapter is that both of these methods yield stress tensors that do not depend on the ``thickness of the layer'' $d$. Moreover  the two methods yield bulk stress tensors which differ by a finite amount.  In particular, we will see that the disagreement between the two calculations of the stress tensor arises entirely from the higher derivative terms.  Of course the fluid map is not defined at small $d$, as it yields a divergent series, and so no divergences appear within the range of validity of either approach. 

The solutions that we find have to be handled with care. Indeed, as we shall point out, the stress tensors derived using the two different methods do not satisfy null energy condition and thus, as we will discuss, it is not clear whether such a boundary layer may exist. \\

This Thesis is organised as follows. In Chapter 1 we shall give a short review of ``classic'' work on black holes of the seventies  which led to the {\it membrane paradigm}, that is a picture of black holes as analogues to dissipative branes endowed with finite electrical resistivity, and finite surface viscosity.  We shall review the derivation of the classical surface viscosity of black holes, that has been re-derived in the quantum context of the AdS/CFT duality. We shall also provide an introduction to black hole thermodynamics and a (sketchy) derivation of the phenomenon of Hawking radiation, that of its crucial importance in fixing the coefficient between the area of the horizon and black hole ``entropy''.

Chapter 2 contains a presentation of properties of relativistic fluids and a discussion of the special case of conformally invariant fluids that will be relevant in the construction of gravity dual solutions. We shall review the extremely useful Weyl invariant formalism that we will apply in the perturbative study of dissipative fluid solutions up to the second order in the derivative expansion. Finally, we will introduce the non-relativistic scaling limit that reduces relativistic conservation equations of a conformal fluid to the celebrated Navier-Stokes formula and we will study the residual conformal symmetry of this equation.

In Chapter 3 we shall introduce the basic scheme for constructing gravitational solutions dual to fluid flows. We shall present - in various forms - the AdS/hydrodynamic map up to the second order in the derivative expansion, and we shall discuss its conformal invariance. Then we shall turn to some of the physical properties of these solutions. We shall conclude the chapter with the discussion of the non-relativistic limit of the AdS/hydrodynamic map under the non-relativistic scaling limit introduced in Chapter 2.

In Chapter 4 we shall present some techniques and issues that are relevant in the study of black holes and singular hypersurfaces in gravity. In particular we shall start by reviewing various possible {\it energy conditions} that are relevant for black holes physics. Then we discuss the issue of the violation of the loosest among these conditions, the null energy condition, and its physical consequences. A relevant part of the chapter will be devoted to the introduction of some geometric notions useful for the description of hypersurfaces and in particular the {\it extrinsic curvature}. These concepts will be used in applying the seminal work of Israel on singular hypersurfaces in General Relativity to our solutions. 

Chapter 5 contains a derivation and discussion of the main results reported in Ref. \cite{Evslin:2010ss}. In particular we consider boundaries between nonrelativistic flows, applying the usual boundary conditions for viscous fluids.  We find that a na\"ive application of the correspondence to these boundaries yields a surface layer in the gravity theory whose stress tensor is not equal to that given by the Israel matching conditions.  In particular, while neither stress tensor satisfies the null energy condition and both have nonvanishing momentum, only Israel's tensor has stress.  The disagreement arises entirely from corrections to the metric due to multiple derivatives of the flow velocity, which violate Israel's finiteness assumption in the thin wall limit.  

Finally we summarise the findings and we discuss some open issues. An Appendix is devoted to the notation and conventions that we have adopted.

% First chapter

\chapter{Black holes as dissipative branes}

Solutions of general relativity describing spherically symmetric objects were obtained soon after the formulation of Einstein's theory. The singular behaviour of these solutions - named black holes - was interpreted, following Subrahmanyan Chandrasekhar's proposal, as the description of a potential well due to a gravitational collapse of very massive astrophysical objects of mass $M$ when their radius shrinks below the limit $2 G M/ c^2$, known as Schwarzschild radius. Early works assumed black holes to be  {\it passive objects}, {\it i.e.} as given geometrical backgrounds. This viewpoint changed in the 1970's, when black holes started being considered as {\it dynamical} objects, able to exchange mass, angular momentum and charge with the external world. The study of the global dynamics of black holes was pioneered by Penrose \cite{P1969}, Christodoulou and Ruffini \cite{C1970,CR1971}, Hawking \cite{H1971}, and Bardeen, Carter and Hawking \cite{BCH1973}.  Later on this approach evolved with the study of the {\it local dynamics} of black hole's horizons in what is called the ``membrane paradigm'' \cite{Thorne:1986iy}) thanks to the works of Hartle and Hawking \cite{HH1972}, Hanni and Ruffini \cite{HR1973},  Damour \cite{D1979,D1978,D1982}, and Znajek \cite{Z1978}. According to this point of view the horizon of a black hole is interpreted as a brane with dissipative properties described for instance by an electrical resistivity and a surface viscosity.

\section{Black hole solutions}

A few months after the publishing of the General Theory of Relativity, Karl Schwarzschild proposed a solution of Einstein's equations in the case of a spherically symmetric object with mass $M$, that can be regarded as the general relativistic analog of the gravitational field of a mass point. In $3+1$ dimensions,  the metric of the Schwarzschild solution can be written as
\begin{equation}
\label{metric}
\mathrm{d}s^2=-A(r)c^2 \mathrm{d} t^2+B(r)\mathrm{d}r^2+r^2
\left(\mathrm{d}\theta^2+\mathrm{sin}^2\theta \mathrm{d}\varphi^2 \right)
\end{equation}
where $t$ denotes the time coordinate measured by a stationary clock at infinity, and where the coefficients $A \left( r \right)$ and $B \left( r \right)$ have the form
\begin{align}
A(r) &=  1-\frac{2GM}{c^2 r}\ ,\\
B(r) &=  \frac{1}{A(r)} \ .
\end{align}
This solution has  an apparently singular behaviour at the so called {\it Schwarzschild radius} 
 \begin{equation}
 r_S  = \frac{2GM}{c^2} \ .
\end{equation}
For example, the sphere located at $r = r_S$ has the observable characteristic of being an infinite-redshift surface. Indeed, the redshift  of a clock at rest in the metric, whose ticks are �read� from infinity, via electromagnetic signals,
\begin{equation}
\frac{\mathrm{d}s_{at\, r=\infty}}{\mathrm{d}s_{at\, r}} = \frac{\sqrt{- g_{00}(r= \infty)}}{\sqrt{-g_{00}(r)}}= \frac{1}{\sqrt{1-\frac{2G M}{c^2 r}}}
\end{equation}
goes to infinity for $r \to r_S$. Similarly the force needed to keep a particle at rest at a radius $r > r_S$ goes to infinity as $r \to r_S$.

The 2-dimensional surface at $r = r_S$, can be regarded as a 3-dimensional  hypersurface $\mathcal{H}$ in spacetime. It is possible to see that the hypersurface $\mathcal{H}$ is a fully regular submanifold of a locally regular spacetime via a change of coordinates near $r = r_S$. Introducing the ingoing Eddington-Finkelstein coordinates $(v, r, \theta, \varphi)$, where $v = t + r_*$, with $r_*$, the so-called {\it tortoise} coordinate, defined by
\begin{equation}
\label{tortoise}
r_* = \int \frac{\mathrm{d}r}{A\left( r \right)} = \int \frac{dr}{1 - \frac{2 G M}{c^2 r}} =  r + \frac{2 G M}{c^2}\log{\left(\frac{c^2 r}{2 G M}-1\right)} \ ,
\end{equation}
the line element takes the form
\begin{equation}
\label{EF}
\mathrm{d}s^2 = - \left(1-\frac{2GM}{c^2 r}\right) \mathrm{d} v^2 + 2 \mathrm{d}v \mathrm{d}r + r^2(\mathrm{d}\theta^2 + \sin^2 \theta \mathrm{d}\varphi^2) \ .
\end{equation}
In the new coordinates, the metric presents a manifestly regular geometry at $r= r_S$. An important property of $\mathcal{H}$ is that it is a null hypersurface, {\it i.e.} a co-dimension-1 surface locally tangent to the light cone. 

The vector normal to the hypersurface, $\ell_\mu$,  such that $\ell_\mu dx^\mu = 0$ for all infinitesimal displacements $dx_\mu$ on the hypersurface, is a null vector {\it i.e.} $\ell_\mu \ell^\mu=0$. Since $\ell^{\mu}$ is a vector tangent to the hypersurface, we can consider its integral lines $\ell^{\mu} = d x^{\mu}/dt$, which lie within the horizon. These integral curves are called the {\it generators} of the horizon. They are null geodesics curves, lying entirely within the horizon. The physical consequence of this fact is that, classically, $\mathcal{H}$ is a boundary between two distinct regions of spacetime: an external one from which light can escape to infinity and an internal one where light is trapped and out of which cannot escape. The internal region of $\mathcal H$ is an infinite spacetime volume which ends, in its future, in a spacelike singularity at r = 0, where the curvature blows up as $r^{-3}$.

This kind of matter configuration is called {\it black hole} - since from the classical point of view it absorbs all the light that hits it, just like a perfect black body in thermodynamics -, while the hypersurface $\mathcal H$  is called the (future) {\it horizon}. Black holes are conjectured to be the final, stationary state reached by any type of matter configuration undergoing a gravitational collapse.  

The Schwarzschild solution was generalized in independent works by Reissner, and by Nordstr\"om, considering electrically charged spherically symmetric objects. In this case, the coefficients $A \left( r \right)$ and $B \left( r \right)$ in (\ref{metric}) are given by  (setting for simplicity $G = c = 1$ here and in what follows)
\begin{align}
A(r) &=  1-\frac{2M}{r}+\frac{Q^2}{r^2} \ ,\\
B(r) &=  \frac{1}{A(r)} \ .
\end{align}
It is easy to realise that now there exist two different horizons: an outer and an inner one, defined by 
\begin{equation}
r_{\pm}=M \pm \sqrt{M^2-Q^2} \ ,
\end{equation} 
which are the two roots of $A(r) =0$.  

A more general black hole solution, due to Kerr and Newman, is obtained considering  in addition to gravity $(g_{\mu\nu})$ also electromagnetic interactions mediated by long range fields $(A_{\mu})$. The final, stationary configuration of such a black hole is described by three parameters, its total mass $M$, its total angular momentum $J$, and its total electric charge $Q$, and is given by the following Kerr-Newman solution
\begin{align}
\mathrm{d}s^2 &=  - \frac{\Delta}{\Sigma} \, \omega_t^2 + \frac{\Sigma}{\Delta} \, 
\mathrm{d}r^2 + \Sigma \mathrm{d}\theta^2 + \frac{\sin^2 \theta}{\Sigma} \, \omega_{\varphi}^2 \, , \\
A_{\mu}\, \mathrm{d}x^{\mu} &= - \frac{Q r}{\Sigma} \omega_t \, , \nonumber \\
\frac{1}{2} \, F_{\mu\nu} \, \mathrm{d}x^{\mu} \wedge \mathrm{d}x^{\nu} &= \frac{Q}{\Sigma^2} 
\, (r^2 - a^2 \cos^2 \theta) \, \mathrm{d}r \wedge \omega_t + \frac{2Q}{\Sigma^2} \, a r \cos 
\theta \sin \theta \, \mathrm{d} \theta \wedge \omega_{\varphi} \, , \nonumber
\end{align}
where
\begin{equation}
a = \frac{J}{M} \, , \quad \Delta = r^2 - 2Mr + a^2 + Q^2 \, , \quad \Sigma = 
r^2 + a^2 \cos^2 \theta \nonumber \, ,
\end{equation}
\begin{equation}
\omega_t = dt - a \sin^2 \theta \, d\varphi \, , \quad \omega_{\varphi} = (r^2 + a^2) 
\, d\varphi - a \, dt \, .
\end{equation}

The Kerr-Newman solution describes a black hole with regular horizon if and only if the three parameters $M$, $J$, $Q$ satisfy the 
inequality
\begin{equation}
a^2 + Q^2 \leq M^2 \, .
\end{equation}
A special class of black holes, called {\it extremal}, are those for which the inequality is saturated. A Schwarzschild black  hole ($a = Q = 0$) can never be extremal, while a Reissner-Nordstr\"om  black hole ($a=0$) is extremal when $\vert Q \vert = M$, and a Kerr black hole ($Q=0$) is extremal when $J = M^2$.

\section{Global dynamics of black holes}

Until the end of the 60's black holes were seen purely as geometric backgrounds. A famous {\it gedanken} experiment was proposed by Penrose in 1969 \cite{P1969} showing that in principle it would be possible to extract energy from a black hole. The idea of Penrose was to consider a charged test particle - with energy $E_1$, angular momentum $p_{\varphi_1}$, 
and electric charge $e_1$ - coming from infinity and falling into a Kerr black hole, moving on generic nonradial orbits. By Noether's theorem, the time-translation, axial and $U(1)$ gauge  symmetries of the background guarantee the conservation of $E$, $p_\varphi$ and $e$ during the ``fall'' of the test particle. Now it is possible to imagine a process in which the test particle splits near the horizon of the black hole into two particles characterised  respectively by energy, angular momentum and charge $E_2$, $p_{\varphi_2}$, $e_2$, and $E_3$,  $p_{\varphi_3}$, $e_3$. 

\begin{figure}
\begin{center}
\includegraphics[width=10cm]{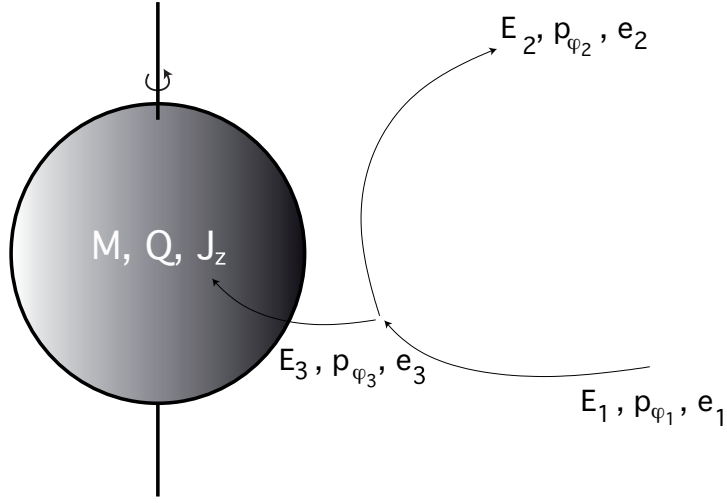}
\caption{Penrose process: a particle ``$1$'' splits in two particles ``$2$'' and ``$3$'' while particle $2$ escapes at $\infty$, particle 3 is absorbed by the black hole.}
\label{Penrose-process}
\end{center}
\end{figure}

The black hole is described as a kind of {\it gravitational soliton}, that is as a physical object, localized ``within the horizon ${\mathcal H}$'', possessing a total mass  $M$, a total angular momentum $J$ and a total electric charge $Q$. Upon dropping a massive, charged test particles one expects a change in the values of $M$, $J$ and $Q$. Therefore in the Penrose experiment the black hole should evolve, due to absorption of particle 3, from an initial Kerr-Newman black hole state $(M,J,Q)$ to a final one  ($M+\delta M$, $J + \delta J$, $Q + \delta Q$) where
\begin{align}
\label{particlefall}
\delta M  &=  E_3 = E_1-E_2,\\
\delta J &=  J_3  =  J_1-J_2,\\
\delta Q &=  e_3 =  e_1-e_2.
\end{align}
Penrose found that , under certain conditions, one finds that particle 3 can be absorbed by the BH, and that particle 2 may come out at infinity with more energy than the incoming particle 1, indeed the process can lead to a decrease of the total mass of the black hole: $\delta M < 0$. \\

Christodoulou and Ruffini \cite{C1970,CR1971} proposed a detailed analysis of the Penrose work, casting light on the existence of a fundamental irreversibility in black holes dynamics. Let us consider for the sake of simplicity a Reissner-Nordstr\"om black hole, and a process of the kind described in (\ref{particlefall}). Considering an on-shell particle 
of mass $\mu$, the Hamilton-Jacobi equation reads
\begin{equation}
g^{\mu \nu}\left(p_{\mu}-eA_{\mu}\right)
\left(p_{\nu}-eA_{\nu}\right)=- \mu^2,
\end{equation}
where a $\left(-+++ \right)$ signature is adopted. The momentum $p_\mu$ is equal to
\begin{equation}
p_{\mu}=\partial{S}/ \partial{x^{\mu}} \ ,
\end{equation}
where $S$ is the action that in an axisymmetric\footnote{A black hole is said to be axisymmetric if there exists a one parameter
group of isometries which correspond to rotations at infinity.} and time-independent background can be taken as a linear function of $t$ and $\varphi$
\begin{equation}
\label{act}
S=-Et+p_{\varphi}\varphi+S \left(r,\theta \right).
\end{equation}
$E=-p_t=-p_0$ is the conserved energy, $p_{\varphi}$ is the conserved $\varphi$-component of angular momentum while the last 
term contains contributions depending on the radial distance $r$ and angular coordinate $\theta$.

The Hamilton-Jacobi equation using eq. (\ref{act}) and the inverse metric for a Reissner-Nordstr\"om black hole reads
\begin{equation}
-\frac{1}{A(r)}\left(p_0-eA_0\right)^2+A(r)p_r^2+\frac{1}{r^2}\left(p_{\theta}^2+\frac{1}{\sin^2
\theta} p_{\varphi}^2\right)=-\mu^2 \ ,
\end{equation}
that can be recast as
\begin{equation}
\left(p_0-e A_0\right)^2=A(r)^2p_r^2+A(r)\left( \mu^2+\frac{L^2}{r^2}\right) .
\end{equation}
Solving for $E= - p_0$, and inserting the expression for the electric potential 
\begin{equation}
A_0 = - \frac{Q}{r} \ ,
\end{equation}
one finds that the energy (where the sign discriminates between particles and antiparticles) is
\begin{equation}
E=\frac{eQ}{r}\pm\sqrt{A(r)^2p_r^2+A(r)\left(\mu^2+\frac{L^2}{r^2}\right)} \ .
\end{equation}
This expression generalises the formula for flat spacetime $E=\pm \sqrt{\mu^2 + {\bf p}^2 }$ to a black hole background.

It is possible to pin down the conserved energy of particle 3 when it crosses the horizon $r_+$ of the Reissner-Nordstr\"om black hole (assuming that the condition $Q<M$ for a regular horizon is fulfilled), {\it i.e.} the limit $r \to r_+$ of  $E_3$. The expression for $E_3$ is
\begin{equation}
E_3=\frac{e_3 \, Q}{r_+}+|p^r| \ ,
\end{equation}
where $p^r=g^{rr}p_r=A(r)p_r$ has  a finite limit on the horizon and the absolute value of $p^r$ comes from the limit of a positive square-root. Now remembering that $\delta M = E_3$ and $\delta Q = e_3$, the expression above can be re-written as
\begin{equation}
\delta M=\frac{Q \delta Q}{r_+(M,Q)}+|p^r| ,
\end{equation}
and from the positivity of $|p^r|$ follows the inequality
\begin{equation}
\delta M \ge \frac{Q \delta Q}{r_+(M,Q)}.
\label{Irreversibility}
\end{equation}
Inequality (\ref{Irreversibility}) spells out the irreversibility  of black holes energetics. Indeed, there can exist two types of process: reversible ones in which a particle of charge $+e$ with $|p^r|=0$ is absorbed for which the inequality (\ref{Irreversibility}) is saturated; and irreversible ones for which it is a strict inequality. Reversible processes are clearly quite peculiar and difficult to obtain, therefore one expects that irreversibility will occur in most black holes processes. The problems of reversibility of process in black holes physics replicate in some sense the relation between reversible and irreversible processes in thermodynamics.\\

The calculation performed for a Reissner-Nordstr\"om black hole can be replicated in the more complex case of a Kerr-Newman black hole, yielding  
\begin{equation}
\delta M - \frac{a\delta J+r_+Q\delta Q}{r_+^2+a^2}=\frac{r_+^2+a^2 \cos^2 \theta}{r_+^2+a^2}|p^r|
\end{equation}
where as before $a=J/M$, and the bound $Q^2+(J/M)^2 \le M^2$ must be fulfilled. Once again we can write the variation in mass as an inequality
\begin{equation}
\delta M  \ge \frac{a\delta J+r_+Q\delta Q}{r_+^2+a^2} \ .
\label{Irreversibility2}
\end{equation}
Processes for which this bound is saturated are called ``reversible'' because, after having produced a change $\delta M$, $\delta J$ and $\delta Q$, it is possible to perform a new process such that $\delta' J = - \delta J$,  $\delta' Q = -\delta Q$ (and the corresponding, saturated $\delta' M = -\delta M$)return the system to its initial state. On the contrary, any elementary process for which equation (\ref{Irreversibility2}) holds as a strict inequality cannot be reversed.

Christodoulou and Ruffini  considered a sequence of infinitesimal reversible changes in the state of the black hole, obtained by dropping in the black hole particles for which $p^r \rightarrow 0$, and studied states which are reversibly connected to some initial state with defined mass $M$, angular momentum $J$ and charge $Q$. Equation (\ref{Irreversibility2}) simplifies to the  partial differential equation for $\delta M$,
\begin{equation}
\delta M=\frac{a \delta J+r_+Q \delta Q}{r_+^2+a^2},
\end{equation}
which is integrable and that has solution in the Christodoulou-Ruffini mass formula
\begin{equation}
M^2=\left(M_{\rm irr}+\frac{Q^2}{4 M_{\rm irr}}\right)^2+\frac{J^2}{4
M_{\rm irr}^2} \ .
\label{massformula}
\end{equation}

The mass formula is composed of two terms: the square of the sum of the irreducible mass and of the Coulomb energy ($\propto  \, Q$), and the the rotational energy ($\propto  \, J$). Irreducible mass, defined as 
\begin{equation}
M_{\rm irr}=\frac{1}{2}\sqrt{r_+^2+a^2}
\end{equation} 
enters as an integration constant. To understand its role, it is useful to differentiate the mass expression and insert it in (\ref{Irreversibility2}). One obtains 
\begin{equation}
\delta M_{\rm irr}\ge 0 \ ,
\end{equation}
where the equality holds only for reversible transformations, while the relation holds as strict inequality for all irreversible processes. 

The irreversible increase of the irreducible mass has a striking similarity to the second law of thermodynamics. Therefore  we can interpret the quantity $M-M_{\rm irr}$ as the free energy of the black hole, {\it i.e.} the maximum extractable energy by depleting (in a reversible) way $J$ and  $Q$ is $M - M_{\rm irr}$.  Indeed the free energy has both Coulomb and  rotational contributions (it vanishes in the case of a Schwarzschild black hole ($J = 0 = Q$)). As a consequence of this relation, black holes can no longer be considered to be passive object since they actually store energy - up to 29 \%  of their mass as rotational energy, and up to 50 \% as Coulomb energy - that can be extracted.\\

There exists a link between the irreducible mass and the area of the horizon of a Kerr-Newman black hole. Indeed, the metric of a Kerr-Newman black hole, when using $r = r_+$  ($\Delta =0$ and $\mathrm{d}r = 0$) for the inner geometry of the horizon, is 
\begin{align}
\mathrm{d} \sigma^2 &= \gamma_{AB} (x^C) \, \mathrm{d}x^A \, \mathrm{d}x^B \nonumber\\ 
&= (r_+^2 + a^2 \cos^2 \theta) \, \mathrm{d} \theta^2 + \frac{\sin^2 \theta (r_+^2 + a^2)^2}{(r_+^2 + a^2 \cos^2\theta)} \, \mathrm{d} \varphi^2  \ .
\end{align}
Therefore the area of a timeslice of the horizon is proportional to irreducible mass 
\begin{equation}
\label{AM}
A_{\rm KN} = \int \int (r_+^2 + a^2) \sin \theta \, d\theta \, d\varphi = 4\pi (r_+^2 +  a^2) = 16\pi M_{\rm irr}^2 \, .
\end{equation}
\\

More generally, Hawking proved \cite{H1971} a theorem stating that the area $A$ of successive time sections  of the horizon of a black hole cannot decrease
\begin{equation}
\delta A \geq 0 \, .
\end{equation}
Moreover the the sum of the area of a system of separated black holes also cannot decreases
\begin{equation}
\label{eq1.12}
\delta \left( \sum_a A_a \right) \geq 0 \, .
\end{equation}
These properties are consequences of Einstein's equations, when assuming the weak energy condition.  \\

Hawking's result suggested the possibility of a formulation of ``thermodynamic laws'' for black holes, in particular it suggests the possibility of defining an entropy for black holes proportional to the area of the horizon. 

While the area law is reminiscent of the second law of the thermodynamics, the analog of the first law can be obtained by varying the mass formula (\ref{massformula})
\begin{equation}
\mathrm{d}M\left(Q,J,A\right)=V\mathrm{d}Q+\Omega \mathrm{d}J+\frac{g}{8 \pi}\mathrm{d}A \ ,
\label{firstlaw}
\end{equation}
where 
\begin{align}
V &=  \frac{Qr_+}{r_+^2+a^2},\\
\Omega &= \frac{a}{r_+^2+a^2}, \\
g &=\frac{1}{2}\frac{r_+-r_-}{r_+^2+a^2}.
\end{align}
$V$ can be interpreted as the electric potential of the black hole, and $\Omega$ as its angular velocity. The coefficient $g$, called {\it surface gravity}, in the Kerr-Newman case, can be rewritten as
\begin{equation}
g=\frac{\sqrt{M^2-a^2-Q^2}}{r_+^2+a^2} \ ,
\end{equation}
and is zero for extremal black holes. The surface gravity of a Schwarzschild black hole reduces to the usual  formula for the surface gravity of a massive star, $g = GM / r_S^2 = 
M / (2M)^2 = 1/(4M)$.  It is also possible to prove that the horizon has constant surface gravity for a stationary black hole, this result is known as the ``zeroeth law of black holes thermodynamics''\footnote{In the ``membrane'' approach to black-hole physics - that will be discussed later on -, the uniformity of $g$ for stationary  black hole states can be viewed as a consequence of the Navier-Stokes equation since the ``surface pressure'' of a black hole happens to be equal to $p \equiv g / 8\pi$.}.\\

The expression (\ref{firstlaw}), once again, suggests the possibility of interpreting the area term as some kind of entropy. 

\section{Black hole thermodynamics}

In 1974,  Bekenstein tried to push the thermodynamical analogy for black hole physics further by taking into account quantum effects. In particular he proposed Carnot-cycle-type arguments: it is possible in principle to extract work from a black hole  by slowly lowering into it an infinitesimal a box of radiation, in this way all the energy of the box of radiation, $mc^2$, would be converted into work.  The efficiency of this Carnot cycle is defined as
\begin{equation}
\eta=1-\frac{T_{\mathrm{cold}}}{T_{\mathrm{hot}}}, 
\end{equation}
where $T_{\mathrm{cold}}$ and $T_{\mathrm{hot}}$ are respectively the two source temperatures between which the ``engine'' operates. Classically, $T_\mathrm{cold}$ is expected to be zero and therefore $\eta$ is expected to be 1. Bekenstein observed that quantum effects should arise modifying the classical result. Indeed the uncertainty principle poses restriction to the size of box of thermal radiation at temperature $T$: since the typical wavelength of radiation is $\lambda \sim 1/T$, the box will have a minimum finite size $\sim \lambda$. From this limit on this size of the box, Bekenstein  derived an upper bound a  bound on the efficiency $\eta$ and therefore concluded that there will exist a nonzero temperature of the black hole  $T_{\mathrm{BH}} \ne 0$.\\

A second argument proposed by Bekenstein consider a reversible process in which a particle with  $p^r=0$ hits the horizon of a black hole at $r=r_+$. The absorption of such a particle should not increase the surface area of the black hole. However, for this to be true {\it both} the (radial) position and the (radial)  momentum of the particle must be exactly fixed: namely, $r=r_+$ and $p^r=0$. This is in contradiction with limits posed by Heisenberg's  uncertainty principle. To state this argument formally, we need the covariant component $p_r$ of the radial momentum 
\begin{align}
p^r &= g^{rr} p_r = A(r) p_r \nonumber\\
& = \frac{\left(r-r_+\right)\left(r-r_-\right)}{r^2}p_r  \simeq  \displaystyle \delta r  \frac{\left(r_+-r_-\right)}{r_+^2}p_r\\
& \simeq \left(\frac{\partial A}{\partial r}\right)_{r_+}\delta r p_r.
\end{align}
It is useful to notice that the  surface gravity $g$ is proportional to  the partial derivative of $A$ with respect to $r$, 
\begin{equation}
\left(\frac{\partial A}{\partial r}\right)_{r_+}=2g.
\end{equation}
Therefore we can re-write the expression for $p^r$ as
\begin{equation}
p^r \simeq 2 g \delta r \delta p_r .
\end{equation}

From the uncertainty relation 
\begin{equation}
\delta r \delta p_r \geq \frac{1}{2} \hbar \ ,
\end{equation}
we get
\begin{equation}
p^r  \geq g \hbar \ ,
\end{equation}
Substituting the bound on $p^r$ in the expression for $\delta M$, we find the following inequality
\begin{equation}
\delta M-\frac{Q \delta Q}{r_+}=|p^r|\geq g \hbar,
\end{equation}
which using expression (\ref{AM}) for irreducible mass, can be written as
\begin{equation}
\delta A \geq 8 \pi \hbar.
\end{equation}

This result show that there is a strong limitation arising from the quantum level on the possibility of reversible processes. A particle falling into a black hole always produces an irreversible process that increases the area of the black hole by a quantity of order $\hbar$. Such a process represents a loss of information - the information about the particle - for the world outside the horizon. For this reason, Bekenstein suggested the introduction of an entropy for black holes \cite{Bekenstein:1973ur} of the form
\begin{equation}
S_{\mathrm{BH}}=\hat{\alpha}\frac{c^3}{\hbar G}A,
\end{equation}
whose coefficient - which Bekenstein was not able to fixed in a unique, and convincing, manner - had to be dimensionless and of order one, $\hat{\alpha}\approx \mathcal{O}\left( 1 \right)$ (Bekenstein proposed $\hat{\alpha}= \ln 2/ 8 \pi$).\\
  
An important consequence of the definition of an entropy for black holes is the natural definition of a temperature  
\begin{equation}
T_{BH}=\frac{1}{8 \pi \hat{\alpha}}\frac{\hbar}{c}g.
\end{equation}
The existence of a temperature implies the possibility that black holes may emit radiation, contrasting the assumption about their ``black'' behaviour. Indeed, soon after Bekenstein's original suggestions, Stephen Hawking - who was skeptical about the possibility that black holes could radiate - in 1974 discovered the universal phenomenon of quantum radiance of black holes \cite{Hawking:1974sw} and was capable to fix in an unambiguous way the coefficient  $\hat{\alpha}$ to be 
\begin{equation}
\hat{\alpha}=\frac{1}{4}  \ .
\end{equation}
We shall discuss the Hawking computation in detail in the next section.

\section{Black hole quantum radiation}

The surprising phenomenon of Hawking radiation, that is the thermal radiation with a black body spectrum predicted to be emitted by black holes due to quantum effects, was first proposed by Hawking in 1974 \cite{Hawking:1974sw}. In this section we will follow a simplified derivation due to Damour and Ruffini \cite{Damour:1976jd, Damour:2004kwa}, considering for simplicity a $3+1$-dimensional Schwarzschild black hole.\\

Let us consider a massless scalar field $\varphi (x)$ propagating in a Schwarzschild background. The Klein-Gordon equation coupled to gravity spells out
\begin{equation}
\label{KGeq}
\Box_g \, \varphi = \frac{1}{\sqrt g} \, \partial_{\mu} (\sqrt g \, g^{\mu\nu} \, \partial_{\nu} \, \varphi) \, = 0 .
\end{equation}
The solutions of this equation - given the symmetries of the background - can be decomposed  into mode functions, given by the product of a Fourier decomposition into frequencies, spherical harmonics and a radial dependent factor, namely
\begin{equation}
\label{sol}
\varphi_{\omega \ell m} (t,r,\theta,\varphi) = \frac{e^{-i\omega t}}{\sqrt{2 \pi \vert 
\omega \vert}} \, \frac{u_{\omega \ell m} (r)}{r} \, Y_{\ell m} (\theta , \varphi) \, .
\end{equation}
Introducing a ``tortoise radial'' coordinate $r_*$, substituting  the explicit form of the metric in the new coordinates and using the generic solution (\ref{sol}), the Klein-Gordon equation boils down to a radial equation for $u_{\omega \ell m} (r)$:
\begin{equation}
\frac{\partial^2 u}{\partial r_*^2} + (\omega^2 - V_{\ell} \, [r (r_*)]) \, u = 0 \, ,
\end{equation}
where the effective radial potential $V_{\ell}$ has the form
\begin{equation}
V_{\ell} (r) = \left( 1 - \frac{2M}{r} \right) \left( \frac{\ell (\ell + 1)}{r^2} + 
\frac{2M}{r^3} \right) \, .
\end{equation}
The effective potential $V_{\ell} (r)$ vanishes both at $r \to \infty$ (which corresponds to $r_* \to + \infty$) as the massless centrifugal potential $\ell(\ell+1)/r^2$, and at the horizon $r=2M$ ($r_* \to + \infty$). Therefore, the potential is negligible in these two regions and the solution of the wave equation is expected to behave essentially as in flat space. The general solutions $\varphi$ in the asymptotic region will be
\begin{equation}
\label{modes}
\varphi_{\omega \ell m} \sim \frac{e^{-i \omega \left( t \pm r_* \right)}}{\sqrt{2 \pi |\omega|}}\frac{1}{r}Y_{\ell m}\left( \theta,\varphi \right).
\end{equation}
Conversely, the potential generated by the gravitational coupling is effective only in the intermediate region where it represents a positive potential barrier combining the effect of curvature and centrifugal effects. Modes generated near the horizon must penetrate this barrier to escape to infinity, hence a {\it  grey body factor} which diminishes the amplitude of the quantum modes will appear in the solution of the Klein-Gordon equation.
\\

\begin{figure}
\includegraphics[width=12cm, height=6.5cm]{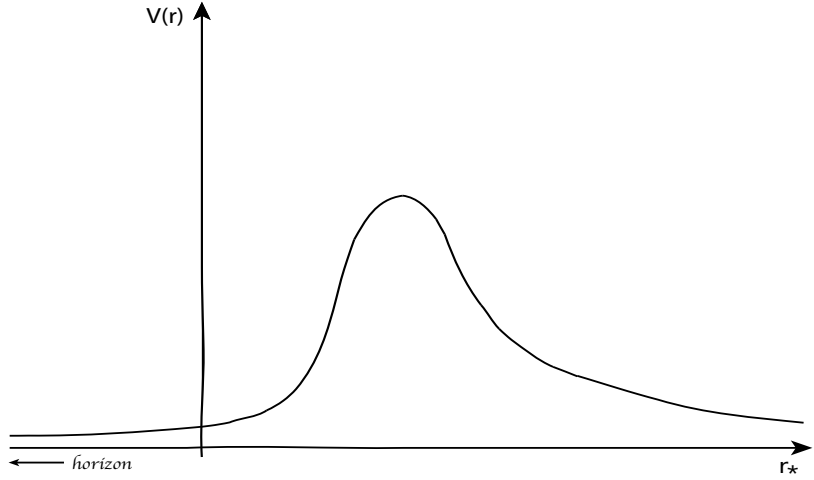}
\caption{Representation of the effective gravitational potential  $V_{\ell} (r)$ in the neighbourhood of a black hole. Spacetime is essentially flat both at infinity and near the horizon. In the central region the  tidal-centrifugal barrier produce a grey body factor.}
\label{Potential}
\end{figure}

To  quantize the scalar field $\varphi$ we will follow a rather standard procedure decomposing the field in the eigenfunctions of the Klein Gordon equation, with coefficients given by creation and annihilation operators. In the case of a black hole background the definition of  positive and negative frequencies must be handled with care.

Let us start reviewing the general formalism in the case of a quantum operator $\hat\varphi (x)$ describing real massless particles in a background spacetime which becomes
 stationary both in the infinite past, and in the infinite future. The operator $\hat\varphi (x)$ can be decomposed both with respect to some ``in'' basis of modes, describing free, incoming particles,
\begin{equation}
\hat\varphi (x) = \sum_i \hat a_i^{\rm in} \, p_i^{\rm in} (x) + (\hat a_i^{\rm in})^+ 
\, n_i^{\rm in} (x) \, ,
\end{equation}
and with respect to an ``out'' basis of modes, describing outgoing particles, 
\begin{equation}
\hat\varphi (x) = \sum_i \hat a_i^{\rm out} \, p_i^{\rm out} (x) + (\hat a_i^{\rm 
out})^+ \, n_i^{\rm out} (x) \, ,
\end{equation}
where $\{\hat a^{\rm in}, (\hat a^{\rm in})^+\}$ and $\{ \hat a^{\rm out}, (\hat a^{\rm  out})^+\}$ are the two sets of annihilation and creation operators, such that
\begin{equation}
[\hat a_i^{\rm in} , (\hat a_j^{\rm in})^+] = \delta_{ij} \ .
\end{equation}
The two sets of annihilation and creation operators correspond to a decomposition in modes which can physically be considered as incoming positive-frequency ones 
$(p_i^{\rm in} (x))$, incoming negative-frequency ones ($n_i^{\rm in} (x)$); which can be taken to be the complex conjugate of $p_i^{\rm in} (x)$ in our case, outgoing 
positive-frequency ones $(p_i^{\rm out} (x))$ and outgoing negative-frequency ones  $(n_i^{\rm out} (x))$.  Mode functions can be normalized as
\begin{equation}
(p_i^{\rm in} , p_j^{\rm in}) = \delta_{ij} \ , \qquad  (p_i^{\rm in} , n_j^{\rm in}) = 0\ , \qquad (n_i^{\rm in} , n_j^{\rm in}) = - \delta_{ij} \ ,
\end{equation}
where $(\,\,, \,)$ denotes the standard (conserved) Klein-Gordon scalar product
\begin{equation}
(\varphi_1,\varphi_2) \sim i \int d\sigma^{\mu} \left( \varphi_1^* \partial_{\mu} \varphi_2 - \partial_{\mu}\varphi_1^* \varphi_2 \right) \ .
\end{equation}
Vacuum states $\vert {\rm in} \rangle$, $\vert {\rm out} \rangle$ are defined as respectively the states annihilated by $a_i^{\rm in}$ and $a_i^{\rm out}$. The mean number of $i$-type ``out'' particles present in the vacuum $\vert {\rm in} \rangle$ is given by
\begin{equation}
\label{bog}
\langle N_i \rangle = \langle {\rm in} \vert \, (a_i^{\rm out})^+ \, a_i^{\rm out} \,  \vert {\rm in} \rangle = \sum_j \vert T_{ij} \vert^2
\end{equation}
where $T_{ij} \equiv (p_i^{\rm out} , n_j^{\rm in})$ is the transition amplitude (also named Bogoliubov coefficients), between the  incoming negative-frequency mode $n_j^{\rm in}$ and the outgoing positive-frequency one $p_i^{\rm out}$.
\\

In the case of a black hole background, the application of the general formalism, detailed above, can be problematic since the background is not asymptotically stationary in the infinite future - in the interior of the black hole the Killing vector $\partial/\partial t$ is spacelike -, while it can be considered asymptotically stationary in the infinite past only by ignoring the formation of the black hole. In his work, Hawking pointed out that it is possible overcome these problems by considering the high-frequency modes coming from the infinite past and  the outgoing modes, viewed in the asymptotically flat region and in the far future. Outgoing modes can be unambiguously decomposed into modes with positive and negative frequency, because, as explained above, their asymptotic behaviour is given by a sum of essentially flat-spacetime modes as seen in (\ref{modes}). 
\\

We want to calculate the average number of outgoing particles  seen in the $\vert {\rm in} \rangle$ vacuum, that is the transition amplitude $\sum_j |T_{ij}|^2$, defined in (\ref{bog}), from an initial negative frequency mode $n_j^{in}$ into a final outgoing positive frequency one $p_i^{out}$, observed at infinity. To compute the sum $\sum_j |T_{ij}|^2$, we have to find out what is an initial negative frequency mode $n_j^{in}$. As pointed out before, there is a physically infinite redshift between the surface of the horizon and asymptotically flat space at infinity. Therefore, particle with finite frequency at infinity will correspond to a wave packets with very high frequency near the horizon, hence very localized. A second important observation is that the near horizon geometry is regular and with a finite radius of curvature, hence it is sensible to approximate it locally by a flat spacetime. In conclusion, the technical issue that we need to discuss is how to describe a negative frequency mode $n_j^{in}$ in a small neighbourhood of the horizon, that we can assume to be a Minkowski flat vacuum.
\\

Let us introduce the technical criterion, that we will use later on, for characterizing positive and negative frequency modes in a locally flat space-time. Consider, in Minkowski space, a wave  packet
\begin{equation}
\label{wp}
\varphi_- (x) = \int_{{\mathcal C}^-} d^4 k \, \tilde\varphi (k) \, 
e^{ik_{\mu} x^{\mu}}
\end{equation}
made only of negative frequencies, {\it i.e.} such that the 4-momenta $k^{\mu}$  in the wave packet are all contained in the past light cone ${\mathcal 
C}^-$ of $k^{\mu}$. A convenient technical criterion for characterizing, in $x$-space,  such a negative-frequency wave packet is the well-known condition that $\varphi_- (x)$ 
be analytically continuable to complexified spacetime points $x^{\mu} + i \,  y^{\mu}$ with $y^{\mu}$ lying in the {\it future} light cone, $y^{\mu} \in {\mathcal 
C}^+$. If one performs a complex shift of the spacetime coordinate, $x_{\mu} \rightarrow  x^{\mu} + i y^{\mu}$, where $y^{\mu}$ is timelike-or-null and  future-directed ($y^{\mu} \in C^+$), then, the $e^{i k_{\mu}x^{\mu}}$  term will be suppressed by a $e^{-k_{\mu}y^{\mu}}$ term, where the scalar product $k_{\mu}y^{\mu}$ is positive because it involves two
 timelike vectors that point in opposite directions (we use the ``mostly plus'' signature). This ensures that a negative-frequency wave-packet can indeed be analytically continued to complex spacetime points of the form $x^{\mu} + i y^{\mu}$, with  $y^{\mu} \in C^+$.
\\

Now we are ready to attack the calculation of the transition amplitude in the Schwarzschild background. The first step is to eliminate the ``singularity'' at the horizon $r=2M$ by adopting the Eddington-Finkelstein coordinates introduced above (\ref{EF}). Near the horizon, an outgoing mode from $\mathcal{H}$, which will have  a positive frequency at infinity, is given by expression (\ref{modes}) with a minus sign in front of $r_*$, and with  $\omega > 0$. Introducing Eddington-Finkelstein coordinates, this mode function can be re-written as
\begin{equation}
\label{posmode}
[ \varphi_{\omega}^{\rm out} (v,r)]_{{\rm near}\,\mathcal H} \propto e^{-i \omega v} \, 
e^{+2 i \omega r} \left( \frac{r-2M}{2M} \right)^{i 4 M \omega} \, ,
\end{equation}
where we used
\begin{equation}
t - r_* = t + r_* - 2 r_* = v - 2 r_* = v - 2r - 4M \ln \left( \frac{r-2M}{2M} \right) \ .
\end{equation}
The outgoing mode (\ref{posmode}) appears to be singular on the horizon where oscillations have shorter and shorter wavelengths as $r \to 2M$, and not to be defined inside the horizon.

The local, $x$-space criterion for characterizing positive and negative frequency modes, discussed above, can be applied - in a local frame near the horizon - to the wave packet (\ref{posmode}). Since the infinitesimal displacement $r \to r - \varepsilon$, $v \to v$ is seen to be future directed and null, this  criterion finally tells us that the following new, extended wave packet, defined by  applying the analytic continuation $r \to r - i {\varepsilon}$ to (\ref{posmode}),
\begin{equation}
n_{\omega} (v,r) \equiv N_{\omega} \, \varphi_{\omega}^{\rm out} (v,r - 
i {\varepsilon}) \propto e^{-i \omega v} \, e^{2i \omega r} \left( \frac{r - 2M - 
i {\varepsilon}}{2M} \right)^{i 4 M \omega}
\end{equation}
is, when Fourier analyzed in the vicinity of ${\mathcal H}$, a negative  frequency wave packet. Note that the new wave packet $n_{\omega} (v,r)$ is defined also in the interior of the black hole and that we have included a new normalizing factor  $N_{\omega}$ in its definition in terms of the analytic continuation of the ``old'' mode $\varphi_{\omega}^{\rm out}$, which had its own normalization.

The normalisation factor $N_{\omega}$ is relevant for the computation of the Hawking radiation. Initially modes in (\ref{modes}) were normalised as
\begin{equation}
(\varphi_{\omega_1 \ell_1 m_1}, \varphi_{\omega_2 \ell_2 m_2}) = + \delta (\omega_1 - \omega_2) \, \delta_{\ell_1 \ell_2} \, \delta_{m_1 m_2} .
\end{equation}
The analytic continuation $r \to r - i {\varepsilon}$ introduces, via the rotation by $e^{-i\pi}$ of $r-2M$ in $(r-2M)^{i4M\omega}$, a factor $(e^{-i\pi})^{i4M\omega} = e^{+4\pi M \omega}$ in the left part ($r < 2M$) of $n_{\omega}$, that is
\begin{equation}
\label{neg}
n_{\omega} (r) = N_{\omega} [\theta (r-2M) \, \varphi_{\omega}^{\rm out} (r-2M) +  e^{4\pi M \omega} \, \theta (2M-r) \, \varphi_{\omega}^{\rm out} (2M-r)] \, ,
\end{equation}
where $\theta (x)$ is the Heaviside step function. The interpretation of eq. (\ref{neg}) is that the  initial negative-frequency mode $n_{\omega}$, localised near the horizon, over time, splits   into an outgoing mode $\varphi_{\omega}^{\rm out} (r-2M)$, that is observed as a positive-frequency mode at infinity, and a second mode $\varphi_{\omega}^{\rm out} (2M-r)$ that is absorbed by the black hole.

Now the calculation of the scalar product $(n_{\omega_1 \ell_1 m_1} , n_{\omega_2 \ell_2 m_2} )$ brings a factor $\vert N_{\omega} \vert^2 \, [1 - (e^{4 \pi M \omega})^2]$, where the minus sign is due to the essentially negative-frequency aspect of $\varphi_{\omega}^{\rm out} (2M-r) \, \theta (2M-r)$.
Therefore the appropriate normalisation factor to get 
\begin{equation}
(n_{\omega_1 \ell_1 m_1} ,  n_{\omega_2 \ell_2 m_2}) = - \delta (\omega_1 - \omega_2) \, \delta_{\ell_1\ell_2} \,  \delta_{m_1m_2}  
\end{equation}
for a negative-frequency mode is 
\begin{equation}
\label{N}
\vert N_{\omega} \vert^2 = \frac{1}{e^{8 \pi M \omega} - 1} \, .
\end{equation}

Leaving aside, for the moment the effect of potential $V_{\ell} (r)$, the transition amplitude $T_{ij}$ - the scalar product between $n_j^{\rm in} = 
n_{\omega_1 \ell_1 m_1}$ and $p_i^{\rm out} \propto \varphi_{\omega \ell m}^{\rm  out}$ - would be
\begin{equation}
\label{transampl}
(\varphi_{\omega \ell m}^{\rm out} , n_{\omega_1 \ell_1 m_1}) = N_{\omega} \, \delta 
(\omega - \omega_1) \, \delta_{\ell \ell_1} \, \delta_{mm_1} \, .
\end{equation}
Therefore, the number of created particles (\ref{bog}) will contain  $\vert N_{\omega} \vert^2$ times the square of $\delta (\omega - \omega_1)$, which, by Fermi's Golden 
Rule, is simply $\delta (\omega - \omega_1) \times \int dt / 2\pi$.

To take into account for the gravitational potential $V_{\ell} (r)$, it is necessary to add the {\it grey body} factor $\Gamma_{\ell} (\omega)$, giving the fraction of the flux 
of $\varphi_{\omega}^{\rm out}$ which ends up at infinity because of the effect of $V_{\ell} (r)$. The final result for Hawking radiation in a Schwarzschild black hole background is
\begin{equation}
\label{Hawking}
\frac{d \langle N \rangle}{dt} = \sum_{\ell , m} \int \frac{d\omega}{2\pi} \, \vert N_{\omega} \vert^2 \, \Gamma_{\ell} (\omega) = \sum_{\ell , m} \int  \frac{d\omega}{2\pi} \, \frac{\Gamma_{\ell} (\omega)}{e^{8\pi M\omega} - 1} \, .
\end{equation}
A black hole radiates as if it were a black body of temperature $T_{\rm BH} = 1 / (8\pi M)$, screened by a gray body factor $\Gamma_{\ell} (\omega)$.

From the Planck factor in (\ref{Hawking}), it is possible to find the Hawking temperature, 
\begin{equation}
T=\hbar \frac{g}{2 \pi} .
\end{equation}  
This result determines the $\hat{\alpha}$ coefficient to be  $\hat{\alpha} = \frac{1}{4}$. Therefore the Bekenstein entropy is
\begin{equation}
S_{BH}=\frac{A}{4 G \hbar}  \; .
\end{equation}
\\

The extension of the analysis proposed to more general classes of black holes backgrounds leads essentially to the substitution of the factor $4M$ by the inverse 
of the surface gravity of the black hole. The Planck spectrum contains a factor $(e^{2\pi(\omega-p_{\varphi} \Omega - e\Phi) / \kappa} - 1)^{-1}$ summarising the 
combined effect of the general temperature and of the couplings of the conserved angular momentum $p_{\varphi}$ and electric charge $e$ to, respectively, the 
angular velocity $\Omega$ and electric potential $\Phi$ of the black hole.\\

Hawking's radiation is not astrophysically relevant for stellar-mass or larger black holes, nevertheless it has been observed \cite{Damour:1976jd} that the combined effect of the grey body factor and of the zero-temperature limit of  $(e^{\frac{2 \pi (\omega - \omega_0)}{g}}-1)^{-1}$ could yield potentially relevant particle creation phenomena in Kerr-Newman BHs, associated to the ``superradiance'' of modes with frequencies $\mu < \omega < \omega_0$, where $\mu$ is the mass of the created particle.

\section{Surface electrodynamic properties}

The modern study of the {\it local dynamics} of black hole horizons originates from a ``holographic''  description of a black hole's properties: all of the physics of a black hole is condensed in the description of a set of surface quantities on the horizon - the surface of the black hole - and a set of bulk properties outside of the horizon. This approach is called the ``membrane paradigm'' \cite{Thorne:1986iy}).\\

A first interesting example of this holographic description pertaining to the electromagnetic properties of charged black holes. It is possible to modify the field equations of the electromagnetic field $F_{\mu \nu}= \partial_{\mu} A_{ \nu}- \partial_{\nu} A_{ \mu}$ - that in principle is expected to permeate the whole spacetime - so as to replace the internal electrodynamics of the black hole by surface effects. Maxwell's equations
\begin{align}
\displaystyle \nabla_{\nu} F^{\mu \nu} &=  4 \pi J^{\mu},\\
\displaystyle \nabla_{\mu}J^{\mu} &= 0,
\end{align}
can be modified by introducing a fictitious field $F_{\mu \nu}(x) \Theta_{\mathrm{H}}$, where $\Theta_H$ is a Heaviside-like step function, equal to $1$ outside 
the BH and $0$ inside. The new equation for $F_{\mu \nu}$ is of the form
\begin{align}
\nabla_{\nu}\left( F^{\mu \nu} \Theta \right)= \left( \nabla_{\nu} F^{\mu \nu} \right) \Theta+F^{\mu \nu}\nabla_{\nu}\Theta
\end{align}
where it is possible to recognise two source terms, that we can re-interpret introducing a surface current $j_{H}^{\mu}$ defined as
\begin{equation}
 j_{H}^{\mu}=\frac{1}{4 \pi}F^{\mu \nu}\nabla_{\nu}\Theta \ ,
\end{equation}
where the term $\nabla_{\nu}\Theta$ contains a Dirac $\delta$-function which restricts it to the horizon. If we consider a generic scalar function $\varphi \left( x \right)$ such that $\varphi \left( x \right)=0$ on the horizon, with $\varphi \left( x \right)< 0$ inside the horizon, and $\varphi \left( x \right)> 0$ outside it, by the properties of the Heaviside step function we get
\begin{equation}
\partial_{\mu}\Theta_H = \partial_{\mu} \theta\left( \varphi \left( x \right) \right)=\delta \left( \varphi \left( x \right) \right) \partial_{\mu} \varphi \ ,
\end{equation}
where $\delta$ is the a Dirac delta function with support on the horizon.
The modified field equation, with the new source term, now reads
\begin{equation}
\label{currentconser}
\nabla_{\nu}\left( F^{\mu \nu} \Theta \right) =  4 \pi \left( J^{\mu} \Theta+j_{H}^{\mu} \right) .
\end{equation}
\\

To complete the description of black hole electrodynamics it is necessary to modify the current conservation equations. In particular, it is useful to define a surface current density on the black hole. For this purpose it is important to stress that there is an important subtlety in the definition of the normal vector to the horizon, due to the fact that, as discussed before, the horizon is a null hypersurface. Indeed, the covariant vector $\ell_{\mu}$ such that  $\ell_{\mu}\mathrm{d}x^{\mu}$ for any infinitesimal displacement $\mathrm{d}x^{\mu}$ on the horizon has norm zero $\ell_{\mu} \ell^{\mu}=0$ and therefore cannot be normalised in the usual way adopted in the Euclidian space. In stationary axisymmetric  spacetimes,  $\ell_{\mu}$ can be uniquely normalised by demanding that the corresponding directional gradient  $\ell^{\mu}\partial_{\mu}$ be of the form
\begin{equation}
 \frac{\partial}{\partial t} + \Omega \frac{\partial}{\partial \phi} \ ,
\end{equation}
with a coefficient one in front of the time-derivative term. For this reason it is possible assume that  $\ell_{\mu}$ is normalized in such a way to that its normalization is compatible with the usual normalization when considering the limiting case of stationary-axisymmetric spacetimes. Therefore given any normalization, there exists a scalar $\omega$ such that
\begin{equation}
\ell_{\mu}=\omega \partial_{\mu} \varphi,
\end{equation}
hence it is possible to define a $\delta$ function on the horizon
\begin{equation}
\delta_{H}=\frac{1}{\omega}\delta \left( \varphi \right),
\end{equation}
such that
\begin{equation}
\partial_{\mu} \Theta_H=\ell_{\mu}\delta_H.
\end{equation}
\\

At this point it is possible to re-write the surface current in the form
\begin{equation}
j_{H}^{\mu}=K^{\mu}\delta_{\mathrm{H}},
\end{equation}
where we defined a black hole surface current density as
\begin{equation}
\label{bhscd}
K^{\mu}=\frac{1}{4 \pi}F^{\mu \nu}\ell_{\nu} \ ,
\end{equation}
 making evident that the surface current is due to the presence of the electromagnetic field on the horizon.
The conservation of electric current now spells out
\begin{equation}
\nabla_{\mu}\left( \Theta_H J^{\mu}+K^{\mu}\delta_H \right) = 0,
\end{equation}
which is manifestly the conservation of the sum of the bulk current $\Theta_H J^{\mu}$ and of the boundary current $K^{\mu}\delta_H$. In this way the description of  electromagnetic phenomena for black holes has be rephrased in terms of surface quantities defined locally on the horizon.\\

It is useful to introduce some more technical tools to describe the surface physics of black holes. Performing a change of coordinates using (advanced) Eddington-Finkelstein-like time coordinates for which $t=x^0$,  $x^1$ is equal to zero on the horizon (like $r - r_+$  in the Kerr-Newman case), and $x^{A}$ for $A=2,3$ denotes some angular-like coordinates. In the new coordinates the normalisation of the normal vector to he horizon takes the form
\begin{equation}
\label{velocity}
\ell^{\mu}\partial_{\mu}=\frac{\partial}{\partial t}+v^A\frac{\partial}{\partial x^A}.
\end{equation}
Expression (\ref{velocity}) suggests the interpretation of $v^A$ as the velocity of some ``fluid particles'' on the horizon, which can be seen as the ``constituents'' of a null  membrane.
Following this analogy, we can - as is usually done in the study of fluids - introduce the gradient of the velocity field, splitting it into its  symmetric and antisymmetric parts, where the antisymmetric part is interpreted as a local rotation and has no consequence on the physics. The symmetric part can be further divided into its trace and trace-free parts, namely
\begin{equation}
\frac{1}{2}\left( \partial_i v_j+\partial_j v_i \right) =\sigma_{ij}+ \frac{1}{d}\partial \cdot v \delta_{ij}
\end{equation}
where $d$ is the spatial dimension of the considered fluid - which will be $d=2$ for black holes in $4$-dimensions. The usual description interprets the first term as the shear, and the second as the rate of expansion of the fluid. Analogous quantities will be defined for black holes.\\

The local spacetime geometry on the horizon is described by the degenerate metric 
\begin{equation}
\label{degmetric}
\mathrm{d}s^2|_{x^1=0}=\gamma_{AB}\left( t,x^C \right) \left( \mathrm{d}x^A-v^A\mathrm{d}t \right) \left( \mathrm{d}x^B-v^B\mathrm{d}t \right) \ ,
\end{equation}
where $v^A=\mathrm{d}x^A/\mathrm{d}t$.  $\gamma_{AB}\left( t,\vec{x} \right)$  is a symmetric rank 2 tensor, {\it i.e.} a time-dependent 2-metric such that the horizon may by viewed as a 2-dimensional brane. The element of area of a spatial sections $S_t$ therefore can be expressed as
\begin{equation}
\mathrm{d}A=\sqrt{\mathrm{det}\gamma_{AB}}\mathrm{d}x^2\wedge \mathrm{d}x^3.
\end{equation} 
One can decompose the current density $K^{\mu}$ into a time component $\sigma_H=K^0$, and two spatial components $K^A$ tangent to
 the spatial slices  $S_t$ ($t =$ const.) of the horizon,
\begin{equation}
K^{\mu}\partial_{\mu}=\sigma_H\partial_t+K^A\partial_A
\end{equation}
in which $\partial_t=\ell^{\mu}\partial_{\mu}-v^A\partial_A$ so that
\begin{equation}
K^{\mu}\partial_{\mu}=\sigma_H \ell^{\mu}+(K^A-\sigma_H v^A)\partial_A .
\end{equation}

The total electric charge of the spacetime  can be defined as the integral 
\begin{equation}
Q_{\rm tot}=\frac{1}{4 \pi} {\oint}_{S_{\infty}}{\frac{1}{2}F^{\mu \nu}\mathrm{d}S_{\mu\nu}} \ ,
\end{equation}
which can be rewritten, using Gauss' theorem, as the sum of a surface integral on the horizon and a volume integral in between the horizon and  $\infty$ - that can be seen as the charge  $Q_H$ contained in space - of the form
\begin{equation}
Q_H=\frac{1}{4 \pi} {\oint}_H{\frac{1}{2}F^{\mu \nu}\mathrm{d}S_{\mu\nu}} .
\end{equation}
The surface element can be expressed in terms of the null vector orthogonal to the horizon $\ell_\mu$ and of a new null vector $n^{\mu}$, transverse  to the horizon and orthogonal to the spatial sections $S_t$, that is normalised as $n^{\mu} \ell_{\mu}= + 1$,
\begin{equation}
\label{surfelem}
\mathrm{d}S_{\mu \nu}=\frac{1}{2}\varepsilon_{\mu \nu \rho \sigma} \mathrm{d}x^{\rho}\wedge 
\mathrm{d}x^{\sigma}=\left( n_{\mu} \ell_{\nu}- n_{\nu} \ell_{\mu} \right) \mathrm{d}A \ .
\end{equation}
Using the definition of the surface current, the black hole's charge is written as 
\begin{equation}
Q_H= {\oint}_H \sigma_H  dA,
\end{equation}
where $\sigma_H $ was introduced before as the time component of the surface current.\\

The above definition of the charge for a black hole forces one to consider the density $\sigma_H $  as a charge distribution on the horizon. Therefore one can interpret 
\begin{equation}
\sigma_H = K^{\mu} n_{\mu} = \frac{1}{4 \pi} F^{\mu \nu} n_{\mu} l_{\nu}
\end{equation}
as the analog of the expression 
\begin{equation}
\sigma = \frac{1}{4 \pi} E^i n_i
\end{equation}
giving the electric charge distribution on a metallic object. Extending the reasoning by analogy to the spatial currents flowing along the surface one can rewrite the current conservation equation (\ref{currentconser}) as
\begin{equation}
\frac{1}{\sqrt{\gamma}}\frac{\partial}{\partial t}\left( \sqrt{\gamma} \sigma_H \right)+\frac{1}{\sqrt{\gamma}}\frac{\partial}{\partial x^A}\left( 
\sqrt{\gamma} K^A \right)=-J^{\mu}\ell_{\mu}.
\end{equation}
This form renders manifest the role of the surface current in ``closing'' the external current injected ``normally'' into the horizon in combination with an an increase in the local horizon charge density.\\

The electromagnetic 2-form restricted to the horizon takes the form
\begin{equation}
\frac{1}{2}F_{\mu \nu}\mathrm{d}x^{\mu} \wedge \mathrm{d}x^{\nu}|_{\mathrm{H}} = E_A \mathrm{d}x^A \wedge \mathrm{d}t+B_{\perp}\mathrm{d}A \ .
\end{equation}
Taking the exterior derivative of the left-hand side of the above expression one gets
\begin{equation}
\nabla \times \vec{E} = -\frac{1}{\sqrt{\gamma}}\partial_t \left( \sqrt{\gamma} B_{\perp} \right).
\end{equation}
which relates the electric and magnetic fields on the horizon.\\

An analog of the Ohm's law relating the electric field to the current can be written in the form 
\begin{equation}
E_A+\epsilon_{AB}B_{\perp}v^B=4 \pi \gamma_{AB}\left( K^B-\sigma_H v^B \right),
\end{equation}
that can be recast as 
\begin{equation}
\vec{E}+\vec{v}\times \vec{B_{\perp}}=4 \pi \left( \vec{K}-\sigma_H \vec{v} \right).
\end{equation}
From this expression it is natural to define the {\it surface electrical resistivity} for black holes to be equal to $\rho=4 \pi=377\,$ Ohm \cite{D1978,Z1978}.

\section{Black hole surface viscosity}

The holographic approach has a second interesting application to Einstein's equation on the surfaces of black holes. Indeed defining appropriate surface quantities related to the spacetime connection one is able to obtain from the equations of gravity a ``surface hydrodynamics'' described by a Navier-Stokes-like equation. The path followed in defining surface electrodynamical quantities must be amended due to the non-linear nature of Einstein's equations\footnote{For a detailed derivation of the results reported in this section one can refer to \cite{D1979,D1982,Gourgoulhon:2005ch,Gourgoulhon:2005ng}}.\\

To address the issue it is necessary to make a few more technical remarks about the near horizon geometry. Given any hypersurface, the parallel transport along some  {\it tangent} direction, which we denote by $\vec{t}$,  of the  (normalized) vector $\vec{\ell}$ normal to the hypersurface yields another tangent vector. In general $\vec{\ell} \cdot \vec{\ell} = \epsilon$, where $\epsilon$ will be $-1$ for a time-like hypersurface, $+1$ for a space-like one and zero for a  null hypersurface. The  directional gradient  of the norm of the vector $\ell$ along the arbitrary tangent vector $\vec{t}$ will give
\begin{equation}
\small( \nabla_{\vec{t}} \, \vec{\ell} \small) \cdot \vec{\ell}=0 \ .
\end{equation}
Therefore the vector $\small( \nabla_{\vec{t}}\, \vec{\ell} \small)$ must be tangent to the hypersurface. More generally there exists a linear map $K$, acting in the tangent plane to the hypersurface, such that $ \nabla_{\vec{t}} \,\vec{\ell} = K(\vec{t})$. In the case of time-like or space-like hypersurfaces, this map is called {\it Weingarten map} and is given by the mixed-component $K_j^i$ version of the {\it extrinsic curvature} of the hypersurface $K_{i j}$. The case of a null hypersurface is more involved since the extrinsic curvature $K_{i j}$ is not uniquely defined. In any case, there exists a mixed-component tensor $K_j^i$, intrinsically defined as the Weingarten map $K$ in $\nabla_{\vec{t}} \vec{\ell} = K(\vec{t})$.
 
In the coordinate system we defined above $x^0, x^1, x^A$ ($A =2,3$), where the horizon is located at $x_1=0$, a basis of vectors tangent to the horizon can be defined containing 
the null vector $\vec{\ell}$, and the two space-like  vectors $\vec{e}_A = \partial_A$. With this basis, the Weingarten map $K$ is fully described by the set of equations
\begin{align}
\nabla_{\vec{\ell}}\, \vec{\ell} & = g \, \vec{\ell},\\
\nabla_A\vec{\ell} &=  \displaystyle \Omega_A \vec{\ell}+D_A^B \vec{e}_B.
\end{align}
The first equation spells the fact that $\vec{\ell}$ is tangent to a null geodesic lying within the horizon, and can be seen as a very general definition of the {\it surface gravity} $g$. In the second equation, a two-vector $\Omega_A$, and the mixed component  $ D_A^B$ of a symmetric two-tensor $ D_{A B}$, which measures the ``deformation'' in time of the geometry of the horizon, are present.  Tensor $D_{A B}$ is defined as 
\begin{equation}
D_{AB}=\gamma_{B C} D_A^C=\frac{1}{2} \frac{D\gamma_{AB}}{\mathrm{d}t}
\end{equation}
where where $D/dt$ denotes the {\it Lie derivative} along $\vec{\ell}= \partial_t + v^A \partial_A$ of the degenerate metric $\gamma_{B C}$ defined in (\ref{degmetric}). In explicit form the deformation tensor can be written as
\begin{align}
\displaystyle D_{AB} &= \frac{1}{2}\left( \partial_t \gamma_{AB}+v^C \partial_C \gamma_{AB}+ \partial_A v^C\gamma_{CB}+\partial_B v^C \gamma_{AC}\right)\\
\displaystyle &= \frac{1}{2} \left( \partial_t \gamma_{AB}+v_{A|B}+v_{B|A}\right) .
\end{align}
Here the symbol `$_|$' indicates a covariant derivative with respect to the Christoffel symbols of the 2-geometry $\gamma_{AB}$. The tensor is made of two parts,  the ordinary time derivative of $\gamma_{AB}$, and that from the variation of the generators of velocity $v^A$ along the horizon. To interpret this object, it  is useful to split the deformation tensor  $D_{AB}$ into a trace-less part and a trace, 
\begin{equation}
D_{AB}=\sigma_{AB}+\frac{1}{2}\theta  \ ,
\gamma_{AB}
\end{equation}
 where the traceless part $\sigma_{AB}$ is interpreted as a ``shear tensor'', while the trace
 \begin{equation}
 \theta=D_A^A=\frac{1}{2}\gamma^{AB}\partial_t \gamma_{AB}+v^A_{|A} \ ,
 \end{equation}
is an ``expansion'' term. The remaining component of the Weingarten map, namely the 2-vector $\Omega_A$, is defined as 
\begin{equation}
\label{omega}
\Omega_A=\vec{n}\cdot \nabla_A \vec{\ell} \ 
\end{equation}  
with $\vec{\ell}\cdot \vec{n}=1$. To give this component a physical meaning it is useful to introduce an angular momentum for the black hole.\\

Let us consider, as before, an axisymmetric spacetime. There will exist a Killing vector related to this symmetry,
\begin{equation}
\vec{k}=k^{\mu} \frac{\partial}{ \partial x^{\mu}} =\frac{\partial}{ \partial \varphi} \ ,
\end{equation}
to which the Noether's theorem will associate a conserved total angular momentum, which can be written as a surface integral over the 2-sphere, $S_{\infty}$
\begin{equation}
J_{\infty}=-\frac{1}{8 \pi}\int_{S_{\infty}} \frac{1}{2}\nabla^{\nu}k^{\mu}\mathrm{d}S_{\mu \nu},
\end{equation}
where the surface element  $\mathrm{d}S_{\mu \nu}$ was defined above in the case of the charge distribution. As before, one can transform the surface integral into the sum of a volume integral containing the angular momentum of the matter present outside the horizon and a surface integral over the horizon
\begin{equation}
 J=J_{\mathrm{matter}}+J_H \ .
\end{equation}
The second term can be interpreted as the black hole angular momentum.

Using the expression (\ref{surfelem}) for the surface element in terms of the vectors $\ell_\mu$ and $n_\mu$ and observing that  we can perform the exchange  $\ell^{\nu} \nabla_{\nu}k^{\mu}=k^{\nu}\nabla_{\nu}\ell^{\mu}$ since the two vectors have zero commutator 
\begin{equation}
\small[ \vec{\ell}, \vec{k} \small] = 0 \ ,
\end{equation}
the black hole angular momentum $J_H$ takes the integral form 
\begin{equation}
J_H=-\frac{1}{8 \pi} \int_{S_H} n_{\mu}k^{\nu}\nabla_{\nu}\ell^{\mu}\mathrm{d}A.
\end{equation}
We can re-write the above expression as the projection of $\Omega_A$ on to the direction of the rotational Killing vector $\vec{k} = \partial_{\varphi}$, so that we have
\begin{equation}
J_H=-\frac{1}{8 \pi}\oint_S k^A\Omega_A \mathrm{d}A 
\end{equation}
where $k^A\Omega_A$ is the $\varphi$-component of $\Omega_A$.  It is natural at this point to define  a ``surface density of linear momentum'' as 
\begin{equation}
\pi_A=-\frac{1}{8 \pi}\Omega_A=-\frac{1}{8 \pi}\vec{n} \cdot \nabla_A\vec{\ell}\ . 
\end{equation}
Finally we get for the total angular momentum of the black hole the expression
\begin{equation}
 J_H=\int_S \pi_{\varphi}\mathrm{d}A \ .
\end{equation}
\\

It is possible to find a dynamical equation for the local quantities defined above by contracting Einstein's equations with the normal to the horizon. In this way one relates surface quantities to the flux of the  energy-momentum tensor $T_{\mu \nu}$ into the horizon.  By projecting Einstein's equations along $\ell^{\mu}e_A^{\nu}$, one finds
\begin{equation}
\frac{D\pi_A}{\mathrm{d}t}=-\frac{\partial}{\partial x^A} \left( \frac{g}{8 \pi} \right)+\frac{1}{8 \pi}\sigma_{A\,|B}^B -\frac{1}{16 \pi}\partial_A
 \theta-\ell^{\mu}T_{\mu A}
\end{equation}
where
\begin{align}
 \frac{D\pi_A}{\mathrm{d}t} &= \left( \partial_t+\theta \right) \pi_A+v^B \pi_{A|B}+v_{\,|A}^B\pi_B, \\
\sigma_{AB} &= \frac{1}{2} \left( \partial_t \gamma_{AB}+v_{A|B}+v_{B|A} \right) 
-\frac{1}{2} \theta \gamma_{AB}, \\
\theta &= \frac{\partial_t \sqrt{\gamma}}{\sqrt{\gamma}}+v_{\, |A}^A
\end{align}
correspond to a convective derivative, a shear and an expansion rate respectively. 

The equation found can be compared to the Navier-Stokes equation for a viscous fluid
\begin{equation}
\left( \partial_t+\theta \right) \pi_i+v^k\pi_{i,k}=-\frac{\partial}{\partial x^i}p+2\eta \sigma_{i\, ,k}^k+\zeta \theta_{,i}+f_i,
\end{equation}
where $\pi_i$ is the momentum density, $p$ the pressure, $\eta$ the shear viscosity, 
$\sigma_{ij}=\frac{1}{2}\left( v_{i,j}+v_{j,i}\right)-\mathrm{Trace}$,
 the shear tensor, $\zeta$ the bulk viscosity, $\theta=v_{\, ,i}^i$ the expansion rate, and $f_i$ the external force density. The striking analogy between these two equations suggests the description of a black hole's surface as a brane with positive surface pressure $p= + \frac{g}{8 \pi}$, 
external force-density $f_A=-\ell^{\mu}T_{\mu A}$ which corresponds to the flow of external linear momentum, surface shear viscosity  $\eta= +\frac{1}{16 \pi}$, and surface bulk viscosity $\zeta=-\frac{1}{16 \pi}$.\\

It is important to remark the non-relativistic character of the black hole hydrodynamical-like equations, a quite surprising feature, in spite of the ``ultra-relativistic'' nature of black holes.

\section{Local thermodynamics of black holes}

In previous sections, starting from Maxwell's equations and Einstein's equations, a surface resistivity and a surface shear viscosity of black holes were defined. On the same grounds, the existence of a local version of the second law of thermodynamics for black holes is of particular interest. Indeed, physical systems verifying Ohm's law and the Navier-Stokes equation are expected to be also endowed with thermodynamic dissipative equations (the equivalent to Joule's law). Na\"ively, the expected ``heat production rate'' in each surface element $\mathrm{d}A$ would be of the form 
\begin{equation}
\label{naive}
\dot{q}=\mathrm{d}A\left[ 2 \eta \sigma_{AB}\sigma^{AB}+\zeta \theta^2+\rho \left( \vec{\mathcal{K}}-\sigma_H \vec{v} \right) ^2 \right],
\end{equation}
where $\rho$ is the surface resistivity, and $\eta$ and $\zeta$ the shear and bulk viscosities defined above. The corresponding local increase of local entropy $s = \hat{\alpha} \mathrm{d}A$ can be found to be 
\begin{equation}
\frac{\mathrm{d}s}{\mathrm{d}t} = \frac{\dot{q}}{T}.
\end{equation}
where $T=\frac{g}{8 \pi \hat{\alpha}}$ is the local temperature on the surface of the black hole\footnote{Even though, as consequence of the zeroth law of black holes thermodynamics, the surface gravity is uniform on the horizon of stationary black holes, it is, in general, non-uniform for evolving black holes.}.

A precise result can be obtained by contracting Einstein's equations with $\ell^{\mu} \ell^{\nu}$. The projection gives the equation
\begin{equation}
\label{entrvar}
\frac{\mathrm{d}s}{\mathrm{d}t}-\tau
\frac{\mathrm{d^2}s}{\mathrm{d}t^2}=\frac{\mathrm{d}A}{T}\left[2 \eta
\sigma_{AB}\sigma^{AB}+\zeta \theta^2+\rho \left(
\vec{\mathcal{K}}-\sigma_H \vec{v}\right) ^2\right] \ .
\end{equation}
The similarity of the evolution law for the entropy found with eq. (\ref{naive}) is quite striking, but there is a relevant difference due to the unexpected second derivative of local entropy on the left hand side appearing with a minus sign and a coefficient $\tau=g^{-1}$, that can be regarded as a time scale. This term is a correction to usual near-equilibrium thermodynamics, which involves only the first order time derivative of the entropy\footnote{An interesting observation is that for $\hat{\alpha}=1/4$ one gets  $\tau=\frac{1}{2 \pi T}$, that corresponds to the inverse of the lowest ``Matsubara frequency'' associated to the temperature $T$.}.

In the limit of an adiabatically slow evolution of the black-hole state, eq. (\ref{entrvar}) reduces to the usual thermodynamical law giving the local increase of the entropy of a fluid element heated by the dissipation associated to viscosity and the Joule's law. 

In the approximation of constant  $\tau$, the rate of increase of entropy is
\begin{equation}
\frac{\mathrm{d}s}{\mathrm{d}t}=\int_t^{\infty}\frac{\mathrm{d}t'}{\tau}e^{-\frac{\left( t'-t \right)}{\tau}}\left( \frac{\dot{q}}{T}\right) \left( t'\right) .
\end{equation}
In other words, the rate of increase of entropy  is defined as integral of the heat dissipation over the future. This fact points to a very peculiar feature of black holes: their {\it acausal} nature. Indeed, black holes are defined as null hypersurfaces which are evolving towards a stationary state in the far future.\\

An important value for the black holes physics is the ratio of the shear viscosity $\eta= 1/(16 \pi)$ to the entropy density $ \hat{s} = s/dA = \hat{\alpha}$, that for the Bekenstein-Hawking value of $\hat{\alpha}= 1/4$, is
\begin{equation}
\frac{\eta}{\hat{s}}= \frac{1}{\hat{\alpha} 16 \pi}= \frac{1}{4 \pi} \ .
\end{equation}
This value of the ratio $\eta/\hat{s}$ has been found Kovtun, Son and Starinets in the gravity duals of strongly coupled gauge theories, using the AdS/CFT correspondence \cite{Kovtun:2004de,Son:2007vk}.

%_____________________________________________
% Second Chapter
\chapter{Elements of fluid dynamics}
%_____________________________________________

Fluid dynamics is  the low energy effective description of any interacting quantum field theory, in the regime in which fluctuations have sufficiently  long wavelengths. Hydrodynamics provides a statistical description on macroscopic scales of the collective physics of a large number of microscopic constituents. A classical reference on the subject is \cite{Landau:1965pi}, while for a more specific discussion of relativistic fluids one can also refer to \cite{Andersson:2006nr}. %An introductory discussion of the matter can be found also in \cite{Weinberg}. 

Quantum systems in equilibrium can be described by the grand canonical ensemble, given the temperature and chemical potentials for various conserved charges. Observables of the system are given by correlation functions in the grand canonical density matrix. When the system is perturbed away from equilibrium, for fluctuations whose wavelengths are large compared to the scale set by the local energy density or temperature, the system can be described at a macroscopic level in terms of fluid dynamics. The variables of such a description are the local densities of all conserved charges together with the local fluid velocities  

The intuition is that sufficiently long-wavelength fluctuations correspond to variations that are slow on the scale of the local energy density/temperature. Therefore, the system can be considered at equilibrium patchwise and locally the temperature can be seen as constant. Therefore while the grand canonical ensemble remains a valid approach to describe the physics locally,  fluid dynamics describes macroscopically how different local domains roughly at equilibrium  interact and exchange thermodynamic quantities. 

Fluid dynamics is a quite surprising field already as a description of classical fluids. Although classical fluid dynamics equations have been intensively studied for almost two centuries, their extremely rich phenomenology remains not completely understood. Well known open problems are the so called ``Navier�Stokes existence and smoothness problems'' \cite{Fefferman:2000wo}, that is the demonstration, for non-relativistic incompressible viscous fluids described by the Navier-Stokes equations, of the existence of globally regular solutions. Moreover interesting phenomena such as turbulence remain to be completely clarified. The holographic mapping of the fluid dynamical system into classical gravitational dynamics, that will be the topic of the next chapter, could be a useful tool to deal with open problems in both fields. 

In this chapter we will start reviewing relativistic fluid dynamics \cite{Andersson:2006nr, Weinberg}, then we will summarise some aspects of conformally invariant fluids \cite{Rangamani:2009xk}. Finally, we will discuss the non-relativistic limit of fluid dynamics \cite{Landau:1965pi, Bhattacharyya:2008kq}.

\section{Relativistic ideal fluids}

Let us start by defining in a more formal way the hydrodynamic regime. In any interacting system there is an intrinsic length scale, the ``mean free path length'' $\ell_{\text{mfp}}$, which constitutes the characteristic length scale of the system. In the kinetic theory context, $\ell_{\text{mfp}}$ is the mean free motion of the constituents between successive interactions.  The hydrodynamic regime is therefore archived when considering description of the system on at length scales which are large compared to $\ell_{\text{mfp}}$. At this scale, microscopic inhomogeneities are sufficiently smeared out.\\

The equations of relativistic fluid dynamics can be summarised as 
\begin{equation}
\nabla_\mu T^{\mu\nu} = 0 \  ,\qquad \nabla_\mu J^\mu_I = 0 \ 
\label{cons-hydr}
\end{equation}
where the first states the conservation of the stress tensor $T^{\mu\nu}$ and the second one refers to the conservation of charge currents $J^\mu_I$, where $I = \{1,2, \cdots\}$ indexes the set of conserved charges characterizing the system. To  characterize a system it is necessary to specify the stress tensor and charge currents in terms of the basic thermodynamic variables.

Let us consider a QFT living on a $d$-spacetime dimensional background ${\mathcal B}_d$ with coordinates $x^\mu$ and non-dynamical metric $g_{\mu\nu}$. The thermodynamic variables of the system are the fluid velocity $u_\mu$, normalized as
\begin{equation}
u_\mu \, g^{\mu \nu} \, u_\nu = -1 \ ;
\end{equation}
the local energy density $\rho$ and charge densities $q_I$, that are seen as extrinsic quantities; pressure $p$, temperature $T$,  and chemical potentials $\mu_I$ that are intrinsic quantities determined by the equation of state.

Let us focus first on the volume element of an ideal fluid which has no dissipation, as seen in the reference frame where it is at rest. In this frame Pascal's law is expected to hold. The pressure exerted by a given element of fluid is the same in all directions and is perpendicular to the surface on which it acts. 
The $i$-th component of the force acting on a surface element $d{\bf f}$ can be written as $T^{ij}df_j$. Therefore in the {\it local rest frame} we can write $T^{ij}df_j = p df_j$, that is  $T_{ij}= p \delta_{ij}$. The component $T^{00}$ is the energy density $\rho$, while components $T^{0\mu}$, that refer to the momentum density, are zero in the local rest frame. 

Summarising, in the frame in which the ideal fluid is at rest, the energy-momentum tensor has the diagonal form
 \begin{equation}
T^{\mu \nu}=
\begin{bmatrix}
\rho & 0 & 0  & 0 & \dots\\
0 & p & 0  & 0 & \dots \\
0 & 0 & p  & 0 & \dots \\
0 & 0 & 0  & p & \dots \\
\vdots & \vdots & \vdots & \vdots & \ddots
\end{bmatrix}
\end{equation}

In a generic reference frame the energy-momentum tensor that we wrote in the rest frame will be 
\begin{equation}
\left(T^{\mu \nu}\right)_{\text{ideal}} = \rho\, u^\mu \, u^\nu + p\,\left(g^{\mu \nu} + u^\mu \, u^\nu\right)  \ ,  \\
\end{equation}
where $u^\mu$ is the fluid velocity. Similarly, in the local rest frame the components of the charge current are the charge density itself .The particle flux will have a time component given by the density of particles, while the spatial components will form the spatial flux vector. Therefore we can write the charge current as   
\begin{equation}
\left(J^\mu_I\right)_{\text{ideal}} = q_I \, u^\mu  .
\end{equation}
Another relevant current to be defined is the entropy current, that similarly to the charge current can be written as
\begin{equation}
\left(J^\mu_S\right)_{\text{ideal}} = s\, u^\mu  , 
\end{equation}	
where $s$ is the local scalar entropy density. The entropy current describes the variations of entropy in the fluid. It is well known that from the conservation equations and standard thermodynamic relations one finds that the entropy current is conserved for an ideal fluid 
\begin{equation}
\nabla_\mu \left(J^\mu_S\right)_{\text{ideal}}  = 0 .
\end{equation}

A convenient notation can be formulated by observing that  the $d$-velocity $u^\mu$ is oriented along the temporal direction. Therefore, it is possible to use $u^\mu$ to decompose the  spacetime into spatial slices with induced metric
\begin{equation}
P^{\mu \nu} = g^{\mu \nu} + u^{\mu} \, u^{\nu}  ,
\end{equation}
where $P^{\mu \nu}$ can be seen as projector onto spatial directions, which has the following properties
\begin{equation}
P^{\mu \nu} \, u_{\mu} = 0 \ , \qquad
P^{\mu \rho} \, P_{\rho \nu} = P^{\mu}_{\ \nu} = P^{\mu \rho} \, g_{\rho \nu} \ , \qquad
P_{\mu}^{\ \mu} = d-1  .
\end{equation}

The ideal fluid stress tensor may be re-written in terms of the new projector as 
\begin{equation}
\left(T^{\mu \nu}\right)_{\text{ideal}}= \rho\, u^\mu \, u^\nu + p \, P^{\mu \nu}  . 
\label{ideal2}
\end{equation}	

\section{Relativistic dissipative fluids}

Real fluids manifest dissipative effects, resulting in the creation of entropy consistently with the second law of thermodynamics. Indeed, in general the conservation of entropy current does not hold. These dissipative effects operate in the fluid to equilibrate it when perturbed away from a given initial equilibrium configuration. From a microscopic point of view, dissipative effects are due to interaction terms in the underlying QFT of which fluid dynamics is a macroscopic description. Therefore, we expect the terms incorporating dissipative effects in the  the stress tensor and charge currents to depend on the coupling constants of the underlying quantum field theory.\\

Following \cite{Rangamani:2009xk} we will introduce dissipative terms with a procedure inspired by the usual way in which effective field theories are modeled, taking into account all possible terms that can appear in the effective Lagrangian consistent with the underlying symmetry at the order required. In the same way, we will introduce in the hydrodynamic approximation all possible operators - constructed as derivatives of the velocity field and thermodynamic variables -,  consistent with the symmetries. This procedure is quite natural in that the hydrodynamics can be seen as an effective theory for an underlying microscopic quantum field theory.

To clarify the approach we will follow, in preparation for the construction to first order in the gradient expansion of the dissipative terms of the stress tensor, we will look for all possible symmetric two tensors built out of the gradients of the velocity field and thermodynamic variables. Clearly the conservation equations (\ref{cons-hydr}) -  which are first order in derivatives - can be used to simplify the expression for the first order stress tensor. In this way correction terms can be expressed as derivatives of the velocity field alone. This process can clearly be iterated to higher orders. \\

Let us start introducing generic dissipative terms to the stress tensor,  $\Pi^{\mu\nu}$, and charge currents, $\Upsilon ^\mu$, that we will have to determine 
\begin{eqnarray}
\left(T^{\mu \nu}\right)_{\text{dissipative}} &=& \rho\, u^\mu \, u^\nu + p\,\left(g^{\mu \nu} + u^\mu \, u^\nu\right)  + \Pi^{\mu\nu} \ ,  \nonumber \\
\left(J^\mu_I\right)_{\text{dissipative}} &=& q_I \, u^\mu + \Upsilon ^\mu .
\end{eqnarray}	

To proceed in a rather systematic way to the enumeration of dissipative terms, it will be necessary - as we have done in the case of the ideal fluid - to define a reference frame in which new terms will be constructed. The choice of the frame is clearly related to the choice of the fluid velocity.\\

In the discussion of  ideal fluids we defined a velocity field such that, in the local rest frame of a fluid element, the stress tensor components longitudinal to the velocity gave the local energy density in the fluid. One can make the same requirement in the study of dissipative fluids. The corresponding gauge choice is known as  {\em Landau frame} and is defined by demanding that the dissipative contributions be orthogonal to the velocity field, {\it i.e.}
\begin{equation}
\Pi^{\mu\nu} \, u_\mu = 0 \ , \qquad \Upsilon ^\mu\, u_\mu =0 \ .
\label{landauframe}
\end{equation}	
The idea in adopting the Landau frame is to define the velocity field $u^\mu$ to be given by the unique (normalised) timelike eigenvector of $T_{\mu\nu}$, so that the definition of the velocity field is tied to the energy-momentum transport in the system. In what follows, we will work in the Landau frame and we will use eq. (\ref{landauframe}) to constrain the dissipative terms of the stress tensor and charge currents.\\

To warm up, we may start by considering the decomposition of the velocity gradient $\nabla^{\nu}u^{\mu}$ into a part along the velocity - given by the {\it acceleration} $a^{\mu}$ -, and a transverse part. The transverse part can be decomposed into a trace, the {\it divergence} $\theta$, and a traceless part. Symmetric components are to be identified as the {\it shear} $\sigma^{\mu \nu}$ of the fluid, while the antisymmetric components can be seen as the {\it vorticity} $\omega^{\mu \nu}$. Therefore we can write the velocity gradient as
\begin{equation}
\nabla^{\nu}u^{\mu}  = - a^{\mu} \, u^{\nu}  + \sigma^{\mu \nu} + \omega^{\mu \nu} + \frac{1}{d-1} \, \theta \, P^{\mu\nu} \ .
\end{equation}
The divergence, acceleration, shear, and vorticity, have been defined\footnote{In a more compact form we can write the projectors as
\begin{eqnarray}
P^{\mu \alpha} \, P^{\nu \beta} \, \nabla_{(\alpha} \, u_{\beta)} = P^{\rho(\mu} \nabla_ \rho u^{\nu)} \ , \nonumber  \\
P^{\mu \alpha} \, P^{\nu \beta} \, \nabla_{[\alpha} \, u_{\beta]} = P^{\rho[\mu} \nabla_ \rho u^{\nu]} \nonumber \ .
\end{eqnarray} 
} 
as
\begin{align}
\label{theta}
\theta &= \nabla_{\mu}u^{\mu} = P^{\mu\nu} \, \nabla_{\mu}u_{\nu}  \\
\label{a}
a^{\mu}&= u^{\nu} \, \nabla_{\nu}u^{\mu} \equiv {\mathscr D} u^{\mu} \\
\label{sigma}
\sigma^{\mu\nu} &= \nabla^{(\mu}u^{\nu)} + u^{(\mu} \, a^{\nu)}  - \frac{1}{d-1} \, \theta \, P^{\mu\nu}
=  P^{\mu \alpha} \, P^{\nu \beta} \, \nabla_{(\alpha} u_{\beta)}  - \frac{1}{d-1} \, \theta \, P^{\mu\nu} \\
\label{omega}
\omega^{\nu\mu} &= \nabla^{[\mu}u^{\nu]} + u^{[\mu} \, a^{\nu]} 
= P^{\mu \alpha} \, P^{\nu \beta} \, \nabla_{[\alpha} u_{\beta]} \ ,
\end{align}
and satisfy by construction the following properties:
\begin{align}
& a_{\mu}\,u^{\mu} = 0 \, \qquad \,\,\,\,\,\, P_{\mu\nu}\,a^{\mu} = a_{\nu} \ , \\
& \sigma^{\mu \nu} \, u_{\mu} = 0 \ , \qquad
\sigma^{\mu \rho} \, P_{\rho \nu} = \sigma^{\mu}_{\ \nu} \ , \qquad
\sigma_{\mu}^{\ \mu} = 0 \ , \\
& \omega ^{\mu \nu} \, u_{\mu} = 0 \ , \qquad
\omega ^{\mu \rho} \, P_{\rho \nu} = \omega ^{\mu}_{\ \nu} \ , \qquad
\omega_{\mu}^{\ \mu} = 0 \ .
\end{align}

Note that above and in what follow we adopt standard symmetrization and anti-symmetrization conventions. For any tensor $F_{ab}$ we define the symmetric part $F_{(ab)} = \frac{1}{2}\, \left(F_{ab} + F_{ba}\right) $ and the anti-symmetric part $F_{[ab]} = \frac{1}{2}\left( F_{ab} - F_{ba}\right)$ respectively. Moreover we indicate with ${\mathscr D} $ the velocity projected covariant derivative: ${\mathscr D}  \equiv u^\mu \, \nabla_\mu$.\\

As mentioned above, the conservation equation (\ref{cons-hydr}) can be used to simplify the expression of the first order stress tensor. Useful relations can be derived projecting  the conservation equation, along the velocity field and transversally
\begin{eqnarray}
\label{iden}
u^\nu \, \nabla_\mu \left(T^{\mu\nu}\right)_{\text{ideal}} &=& 0 \;\; {\Longrightarrow}\;\; 
(\rho + p)\, \nabla_\mu u^\mu + u^\mu \nabla_\mu \rho =0 \\ 
\qquad P_{\nu \alpha} \,\nabla_\mu \left(T^{\mu\nu}\right)_{\text{ideal}} &=& 0 \;\; {\Longrightarrow} \;\; P_\alpha^{\ \mu} \nabla_\mu P + (\rho+p)\, P_{\nu\alpha} \,u^\mu\,\nabla_\mu u^\nu =0 \ . \nonumber
\end{eqnarray}	
Given the identities (\ref{iden}) that we derived, it is possible in the definition of dissipative terms for the stress-tensor to take into account for only symmetric two tensors built from the  velocity gradients. Using  the Landau frame condition, one is able to single out two candidate dissipative terms: 
\begin{equation}
\Pi^{\mu\nu}_{(1)} = -2\, \eta \, \sigma^{\mu \nu} - \zeta\, \theta\, P^{\mu\nu} \ ,
\end{equation}	
that appear associated with two new parameters the {\it shear viscosity}, $\eta$, and the {\it bulk viscosity}, $\zeta$. 

In the same way, it is possible spoiling the conservation of charge current to simplify the search for new terms. In particular it is possible to use the conservation equations to eliminate the acceleration term, limiting the new contribution to ones which are first order in the derivatives of the thermodynamic variables $\rho$ and $q_I$ and also the velocity field. Another potential contribution arises from the  pseudo-vector  
\begin{equation}
\ell^\mu = \epsilon_{\alpha \beta \gamma}^{\hspace{5.5mm}\mu} \, u^\alpha\, \nabla^\beta u^\gamma .
\label{elle}
\end{equation}	
This possible term will be responsible for mixing contributions with different parity structure at first order\footnote{This kind of contribution for charge current appears in fluid dynamics derived from the AdS/CFT correspondence for the ${\mathcal N} =4$ Super Yang-Mills fluid and is linked to Chern-Simons couplings of the bulk gauge field in the gravitational description \cite{Erdmenger:2008rm, Banerjee:2008th}.}. 

The new terms, at first order, for the charge current are
\begin{equation}
\Upsilon ^\mu_{(1)I} = -\widetilde{\varkappa}_{IJ} \, P^{\mu\nu} \, \nabla_{\nu} q_J - \widetilde{\gamma}_I\,  P^{\mu\nu} \, \nabla_{\nu} \rho - \mho_I\, \ell^\mu \ , 
\end{equation}	
where  $\widetilde{\varkappa}_{IJ}$ is the matrix of charge diffusion coefficients, $\widetilde{\gamma}_I$ indicates the contribution of the energy density to the charge current, and $\mho_I$ which are the pseudo-vector transport coefficients. in terms of the temperature $T$ and of the chemical potential $\mu_I$, the current reads 
\begin{equation}
\Upsilon ^\mu_{(1)I} = -\varkappa_{IJ} \, P^{\mu\nu} \, \nabla_{\nu} \left(\frac{\mu_J}{T}\right) - \mho_I\, \ell^\mu - \gamma_I\, P^{\mu\nu}\,\nabla_\nu T \ .
\end{equation}	
\\

Summing up, the stress tensor and the current of charge for a dissipative fluid at leading order in gradient expansion are
\begin{eqnarray}
T^{\mu \nu}&=& \rho\, u^\mu \, u^\nu + p\,\left(g^{\mu \nu} + u^\mu \, u^\nu\right)  -2\, \eta \, \sigma^{\mu \nu} - \zeta\, \theta\, P^{\mu\nu}\ ,  \nonumber \\
J^\mu_I &=& q_I \, u^\mu -\varkappa_{IJ} \, P^{\mu\nu} \, \nabla_{\nu} \left(\frac{\mu_J}{T}\right) -  \mho_I\, \ell^\mu - \gamma_I\, P^{\mu\nu}\,\nabla_\nu T \ .
\label{dissipative fluid}
\end{eqnarray}	
where the set of transport coefficients $\{\eta, \zeta, \varkappa_{IJ}, \gamma_I, \mho_I\}$ are to be determined to completely specify the relativistic viscous fluid. In principle transport coefficients  can be worked out from the fundamental QFT underlying the fluid dynamic description.\\

The dissipative fluid's entropy is not conserved, therefore the second law implies that the entropy current should have non-negative divergence, {\it i.e.}
\begin{equation}
\nabla_\mu J^\mu_S \ge 0 \ .
\end{equation}	

In principle,  the discussion can be extended to determine higher order terms in derivative expansion. However, usually these terms are irrelevant in the hydrodynamic regime. Indeed, these terms are less and less important moving towards longer wavelengths and are therefore suppressed in low energy description. On the contrary terms at first order are important since their presence defines a channel for the fluid to relax back to equilibrium after a perturbation.

\section{Conformal hydrodynamics}\label{confluid}

In this section we will introduce conformal fluids, that can be thought of as the low energy macroscopic description of field theories which are conformally invariant. We will start by discussing the restriction on first order terms in the derivative expansion of the stress-tensor for a generic dissipative fluid arising from the requirement of conformal invariance. In the next section, we will review the useful Weyl covariant formalism \cite{Loganayagam:2008is}, before singling out operators at second order that can appear in conformal hydrodynamics. This discussion is motived by the construction of gravitational duals to hydrodynamics that will be the topic of the next chapter.

Having in mind a relativistic fluid on a background manifold ${\mathcal B}_d$ with metric $g_{\mu \nu}$, we can consider a Weyl transformation of the metric 
\begin{equation} 
 g_{\mu\nu} = e^{2\phi}\, \widetilde{g}_{\mu\nu} \ , \qquad \qquad
  g^{\mu\nu} = e^{-2\phi}\widetilde{g}^{\mu\nu} .
\label{wtr}
\end{equation}
Remembering that the velocity field is normalised as $u^\mu \,u_\mu = -1$, it will scale under a Weyl transformation (\ref{wtr}) as
\begin{equation}
 u^\mu = e^{-\phi}\, \widetilde{u}^\mu \ .
\end{equation}	
Combining these facts one finds thate the spatial projector also transforms homogeneously
\begin{equation}
P^{\mu\nu}=g^{\mu\nu}+u^{\mu}u^{\nu}=e^{-2\phi}\, \widetilde{P}^{\mu\nu} \ .
\end{equation}

A generic tensor ${\mathcal Q}_{\mu_1 \cdots \mu_n}^{\nu_1 \cdots \nu_m}$ is said to be conformally invariant if it transforms homogeneously under Weyl rescalings of the metric  (\ref{wtr}) , i.e., 
\begin{equation}
{\mathcal Q} = e^{-w\,\phi} \, \widetilde{{\mathcal Q}} \ ,
\end{equation}
where $w$ is called {\it conformal weight of the tensor} and depends on the index positions. To have a proper conformally invariant operator in the theory, it is also necessary that the dynamical equations for ${\mathcal Q}$ remain invariant under conformal transformations.\\

Now we can address the question about additional constraints on the stress tensor, at first order, due to conformal invariance. From the conservation equations for the energy-momentum tensor (\ref{cons-hydr}), one finds that it transforms homogeneously under 
Weyl rescalings of the metric with weight $d+2$
\begin{equation}
T^{\mu \nu} = e^{-(d+2) \,\phi}\, {\widetilde T}^{\mu \nu} \ ,
\label{Wtrstrsstensor}
\end{equation}	
provided that its trace is zero $T^{\mu}_{\,\,\,\,\mu} = 0$.

The null trace condition implies for an ideal fluid that 
\begin{equation}
T_\mu^\mu =0 \;\;\;\; \Longrightarrow \;\; \;\; p = \frac{1}{d-1} \, \rho \ .
\end{equation}	
This relation between pressure and energy density fixes the speed of sound in conformal fluids as a function of the spacetime dimension: 
$$c_s = \frac{1}{\sqrt{d-1}}\ .$$

The charge current transforms homogeneously with weight $d$ under conformal transformations, {\it i.e.}
\begin{equation}
J^\mu= e^{-d \,\phi}\, {\widetilde J}^\mu  . 
\end{equation}	

The last pieces to recollect are the scaling dimensions of the thermodynamic variables. It is possible to see that the temperature  scales under a Weyl transformation with weight $1$
\begin{equation}
T  = e^{-\phi} \, \tilde{T} \ ,
\end{equation}
and this implies from the thermodynamic Gibbs-Duhem relation, $p+\rho = s\, T + \mu_I\, q_I$, that the chemical potentials $\mu_I$ also have weight $1$. Moreover from eq. (\ref{Wtrstrsstensor}) it follows that the energy density transforms under a Weyl transformation as
\begin{equation}
\rho  = e^{-d \phi} \, \tilde{\rho} \ .
\end{equation}	

Using the properties of transformation listed above, one is lead to the following expression for the stress tensor of an ideal fluid
\begin{equation}
\left(T^{\mu\nu}\right)_{\text{ideal}} = \alpha \, T^d\, \left(g^{\mu \nu} + d \, u^\mu \ u^\nu\right) \ , 
\end{equation}	
where $\alpha$ is a dimensionless normalization constant fixed by the underlying microscopic conformal field theory.\\

The next step is to discuss dissipative corrections at first order to the ideal fluid stress tensor. The most convenient approach, is to single out all of the operators at order one in the derivative expansion with the right symmetries transforming homogeneously under conformal transformations. With a bit of work, it is possible to show that the covariant derivative of $u^{\mu}$ transforms inhomogeneously, {\it i.e.}
\begin{equation}
 \nabla_{\mu}u^{\nu}=\partial_{\mu}u^{\nu}+\Gamma_{\mu\lambda}^{\nu}\, u^{\lambda}
  =e^{-\phi}\, \left[\widetilde{\nabla}_{\mu}\, \widetilde{u}^{\nu}+\delta^{\nu}_{\mu}\, \widetilde{u}^{\sigma}\, \partial_{\sigma}\phi- \widetilde{g}_{\mu\lambda}\, \widetilde{u}^{\lambda}\, \widetilde{g}^{\nu\sigma}\, \partial_{\sigma}\phi\right] \ ,
\label{covderivtrans}
\end{equation}
where it is useful as an intermediate result that the Christoffel symbols transform as 
\begin{equation}
 \Gamma_{\lambda\mu}^{\nu}= \widetilde{\Gamma}_{\lambda\mu}^{\nu} + \delta^{\nu}_{\lambda}\, \partial_{\mu}\phi+ \delta^{\nu}_{\mu}\, \partial_{\lambda}\phi- \widetilde{g}_{\lambda\mu}\,\widetilde{g}^{\nu\sigma}\, \partial_{\sigma}\phi\ .
\end{equation}

From eq. (\ref{covderivtrans}) it is possible to obtain the transformation laws of fluid dynamical quantities
\begin{align}
&\theta = \nabla_{\mu}u^{\mu}
  =e^{-\phi}\, \left({\widetilde\nabla}_{\mu} {\widetilde u}^{\mu}+(d-1)\, \widetilde{u}^{\sigma}\, \partial_{\sigma}\phi\right) = e^{-\phi} \, \left(\widetilde{\theta} + (d-1)\, \widetilde{{\mathscr D}} \phi \right) \ , \nonumber \\
& a^{\nu}= {\mathscr D} u^\nu = u^{\mu}\nabla_{\mu}u^{\nu}
  =e^{-2\phi}\left(\widetilde{a}^{\nu}+\widetilde{P}^{\nu\sigma}\, \partial_{\sigma}\phi\right), \nonumber \\
& \sigma^{\mu\nu} = P^{\lambda(\mu} \nabla_\lambda u^{\nu)}   - \frac{1}{d-1} \, P^{\mu\nu}\,\nabla_{\lambda}u^{\lambda} 
= e^{-3\,\phi} \; \widetilde{\sigma}^{\mu\nu}, \nonumber \\
& \l^{\mu} =  u_{\alpha}\,\epsilon^{\alpha \beta \gamma \mu}
\nabla_\beta u_{\gamma} = e^{-2 \,\phi}\, {\widetilde \l^\mu} \ .
\label{transdyquant}
\end{align}

In the last equation one uses the fact that all epsilon symbols should be treated as tensor densities in curved space\footnote{It is possible to show \cite{Weinberg} that the epsilon symbols scale as metric determinants {\it i.e.},    $\epsilon_{\alpha \beta \gamma \delta}  \propto \sqrt{-g}$, and  $\epsilon^{\alpha \beta \gamma \delta}\propto \frac{1}{\sqrt{-g}}$. Therefore the correct scaling behavior  under conformal transformations will be $\epsilon_{\alpha \beta\gamma \delta} = e^{4\phi} \; {\widetilde \epsilon}_{\alpha \beta \gamma \delta}$ and  $\epsilon^{\alpha \beta\gamma \delta} = e^{-4\phi} \;  {\widetilde \epsilon}^{\,\alpha \beta \gamma \delta}$.}.\\

From eq. (\ref{transdyquant}) we can deduce the restrictions that one must impose on a conformal dissipative fluid at first order. Since  $\theta$ (and $a^\mu$) transform  inhomogeneously under Weyl transformations, the bulk viscosity should vanish for a conformal fluid $\zeta = 0$. With the same procedure, for the charge current, it turns out that the contribution from the chemical potential and temperature should appear in the combination $\mu_I/T$. Moreover since the gradient of the temperature $P^{\mu\nu}\,\nabla_\nu T$ transforms inhomogeneously, the coefficient of the term in which it appears will be zero $\gamma_I = 0$.

Summing up, one finds that the stress tensor and the charge current, at first order, for a viscous fluid are
\begin{eqnarray}
T^{\mu \nu}&=& \alpha\, T^d \, \left(g^{\mu \nu} +d\,  u^\mu \, u^\nu\right)  -2\, \eta \, \sigma^{\mu \nu} \ ,  \nonumber \\
J^\mu_I &=& q_I \, u^\mu -\varkappa_{IJ} \, P^{\mu\nu} \, \nabla_{\nu} \left(\frac{\mu_J}{T}\right) -  \mho_I\, \ell^\mu \ ,
\end{eqnarray}	
where we have used the generalized Stefan-Boltzmann expression for the pressure. 

\section{Weyl covariant formulation of conformal hydrodynamics}

The Weyl covariant formulation of conformal hydrodynamics is a quite powerful tool to explore second order corrections to the stress tensor \cite{Loganayagam:2008is}. 

On the conformal class of metrics ${\mathcal C}$, defined on  the background manifold ${\mathcal B}_d$, it is possible to define a derivative operator that transforms covariantly under Weyl transformations. Let us start by defining a torsionless connection $\nabla^{\text{Weyl}}$, called the Weyl connection, such that  for every metric in the conformal class ${\mathcal C}$ there exists a connection one-form ${\mathcal A}_\mu$ for which
\begin{equation}
\nabla^{\text{Weyl}}_\alpha g_{\mu \nu}  = 2 \, {\mathcal A}_\alpha \,g_{\mu\nu} \ .
\label{wcmet}
\end{equation}	
It is possible to define a  Weyl covariant derivative  ${\mathcal D}_\mu$ as
\begin{equation}
{\mathcal D}_\mu = \nabla^{\text{Weyl}}_\mu + w \, {\mathcal A}_\mu
\label{wderiv}
\end{equation}
or in a more explicit form
\begin{eqnarray}
{\mathcal D}_\lambda {\mathcal Q}^{\mu\cdots}_{\nu \cdots} &\equiv& \nabla_\lambda {\mathcal Q}^{\mu\cdots}_{\nu \cdots} + w\, {\mathcal A}_\lambda \,{\mathcal Q}^{\mu\cdots}_{\nu \cdots}\nonumber \\
&&  +\left( g_{\lambda\alpha}\, {\mathcal A}^\mu - \delta^\mu_\lambda \, {\mathcal A}_\alpha - \delta^\mu_\alpha \, {\mathcal A}_\lambda\right) \,  {\mathcal Q}^{\alpha\cdots}_{\nu \cdots} + \cdots \nonumber \\
&&- \left( g_{\lambda\nu}\, {\mathcal A}^\alpha - \delta^\alpha_\lambda \, {\mathcal A}_\nu - \delta^\alpha_\nu \, {\mathcal A}_\lambda\right) \, {\mathcal Q}^{\mu\cdots}_{\alpha \cdots} -\cdots \ .
\label{wderiv2}
\end{eqnarray}	
The Weyl covariant derivative  ${\mathcal D}_\mu$ transforms in a covariant way under conformal transformations, when acting on a conformally invariant tensor. In particular 
\begin{equation}
 {\mathcal D}_\lambda  {\mathcal Q}^{\mu\cdots}_{\nu \cdots} = 
 e^{-w\, \phi} \, \widetilde{{\mathcal D}}_\lambda \widetilde{{\mathcal Q}}^{\mu\cdots}_{\nu \cdots} \ ,
\end{equation}
in other words, the Weyl covariant derivative of a conformally invariant tensor transforms homogeneously with the same weight as the tensor itself.

From the definition, it follows that the Weyl connection is  metric compatible 
\begin{equation}
{\mathcal D}_\alpha g_{\mu\nu} = 0 \ ,
\end{equation}
since $w=-2$ for $g_{\mu\nu}$.

Requiring that the new derivative of the fluid velocity is transverse and traceless
\begin{equation}
u^\alpha \, {\mathcal D}_\alpha u^\mu = 0 \ , \qquad {\mathcal D}_\alpha u^\alpha = 0 \ .
\end{equation}	
the connection one-form ${\mathcal A}_\mu$ is uniquely determined as 
\begin{equation}
{\mathcal A}_\mu = u^\lambda\, \nabla_\lambda u_\mu - \frac{1}{d-1}  \, u_\mu 
\, \nabla^\lambda u_\lambda \equiv a_\mu - \frac{1}{d-1}\, \theta \, u_\mu  \  .
\label{wconn}
\end{equation}	\\

Once the Weyl covariant derivative has been defined, one is lead to define an associated curvature as the commutator between two covariant derivatives. In this case, the standard procedure, can be followed keeping in mind some subtleties that we are going to describe.  For a given covariant vector field $V_\mu=e^{-w\phi}\widetilde{V}_\mu\ $, it is possible to define a Weyl-invariant Riemann curvature tensor ${\mathcal R}_{\mu\nu\lambda\sigma}$ and a Weyl-invariant field strength ${\mathcal F}_{\mu \nu}$  as
\begin{equation}
[\mathcal{D}_\mu,\mathcal{D}_\nu]V_\lambda = w\ \mathcal{F}_{\mu\nu}\ V_\lambda - \mathcal{R}_{\mu\nu\lambda}{}^\alpha\  V_\alpha\ \ ,
\end{equation}
where
\begin{equation}
\mathcal{F}_{\mu\nu} = \nabla_\mu \mathcal{A}_\nu - \nabla_\nu \mathcal{A}_\mu \ ,
\end{equation}
and
\begin{eqnarray}
\mathcal{R}_{\mu\nu\lambda}{}^\alpha = R_{\mu\nu\lambda}{}^\alpha &+& \nabla_\mu \left[{g}_{\lambda\nu}\mathcal{A}^{\alpha} - \delta^{\alpha}_{\lambda}\mathcal{A}_\nu  - \delta^{\alpha}_{\nu}\mathcal{A}_{\lambda}\right] - \nabla_\nu 
\left[{g}_{\lambda\mu}\mathcal{A}^{\alpha} - \delta^{\alpha}_{\lambda}\mathcal{A}_\mu  - \delta^{\alpha}_{\mu}\mathcal{A}_{\lambda}\right]\nonumber\\
&+&\left[{g}_{\lambda\nu}\mathcal{A}^{\beta} - \delta^{\beta}_{\lambda}\mathcal{A}_\nu  - \delta^{\beta}_{\nu}\mathcal{A}_{\lambda}\right]
\left[{g}_{\beta\mu}\mathcal{A}^{\alpha} - \delta^{\alpha}_{\beta}\mathcal{A}_\mu  - \delta^{\alpha}_{\mu}\mathcal{A}_{\beta}\right] \nonumber\\
&-& \left[{g}_{\lambda\mu}\mathcal{A}^{\beta} - \delta^{\beta}_{\lambda}\mathcal{A}_\mu  - \delta^{\beta}_{\mu}\mathcal{A}_{\lambda}\right]
\left[{g}_{\beta\nu}\mathcal{A}^{\alpha} - \delta^{\alpha}_{\beta}\mathcal{A}_\nu  - \delta^{\alpha}_{\nu}\mathcal{A}_{\beta}\right] \ .
\end{eqnarray}
The  Weyl-invariant Riemann curvature tensor can be rewritten in the more compact form
\begin{equation}
{\mathcal R}_{\mu\nu\lambda\sigma}= R_{\mu\nu\lambda\sigma} +4\, \delta^\alpha_{[\mu}g_{\nu][\lambda}\delta^\beta_{\sigma]}\left(\nabla_\alpha {\mathcal A}_\beta + {\mathcal A}_\alpha {\mathcal A}_\beta - \frac{{\mathcal A}^2}{2} g_{\alpha\beta} \right) - {\mathcal F}_{\mu\nu}\, g_{\lambda\sigma} \ ,
\label{Reimannconf}
\end{equation}	
where ${\mathcal F}_{\mu \nu}$ is the field strength for the Weyl connection defined above.

Both the Riemann tensor  ${\mathcal R}_{\mu\nu\lambda\sigma}$ and the field strength ${\mathcal F}_{\mu \nu}$  are invariant under Weyl tranformations
\begin{equation}
{\mathcal R}_{\mu\nu\lambda}{}^\alpha=\widetilde{{\mathcal R}}_{\mu\nu\lambda}{}^\alpha \ ,  \qquad \qquad {\mathcal F}_{\mu\nu}=\widetilde{{\mathcal F}}_{\mu\nu} \ .
\end{equation}
There are relevant differences between the Weyl covariant Riemann tensor and the ordinary one in the symmetries they have:
\begin{align}
&{\mathcal R}_{\mu\nu\lambda\sigma}+{\mathcal R}_{\mu\nu\sigma\lambda} = -2\, {\mathcal F}_{\mu\nu} \,g_{\lambda\sigma} \ , \nonumber \\
&{\mathcal R}_{\mu\nu\lambda\sigma}-{\mathcal R}_{\lambda\sigma\mu\nu} = \delta^\alpha_{[\mu}g_{\nu][\lambda}\delta^\beta_{\sigma]} \,{\mathcal F}_{\alpha\beta} - {\mathcal F}_{\mu\nu}\, g_{\lambda\sigma} + {\mathcal F}_{\lambda\sigma} \,g_{\mu\nu}  \ , \nonumber \\
&{\mathcal R}_{\mu\alpha\nu\beta}\,V^\alpha \,V^\beta-{\mathcal R}_{\nu\alpha\mu\beta}\,V^\alpha \,V^\beta = - {\mathcal F}_{\mu\nu}\ V^\alpha \,V_\alpha \ .
\end{align}

In analogy with the ordinary case, given the Riemann tensor one can define a Ricci tensor 
\begin{eqnarray}
{\mathcal R}_{\mu\nu}&\equiv & {\mathcal R}_{\mu\alpha\nu}{}^{\alpha}  \\
&=& R_{\mu\nu} -(d-2)\left(\nabla_\mu {\mathcal A}_\nu + {\mathcal A}_\mu {\mathcal A}_\nu -{\mathcal A}^2 g_{\mu\nu}  \right)-g_{\mu\nu}\nabla_\lambda{\mathcal A}^\lambda - {\mathcal F}_{\mu\nu} \ , \nonumber
\label{ricciconf}
\end{eqnarray}
and a Ricci scalar
\begin{align}
{\mathcal R} \equiv {\mathcal R}_{\alpha}{}^{\alpha} = R -2(d-1)\nabla_\lambda{\mathcal A}^\lambda + (d-2)(d-1) {\mathcal A}^2 \ ,
\label{ricciscalconf}
\end{align}
that transform as
\begin{equation}
{\mathcal R}_{\mu\nu} = \widetilde{{\mathcal R}}_{\mu\nu} \ , \qquad \qquad  {\mathcal R} = e^{-2\phi} \widetilde{{\mathcal R}} \ .
\end{equation}

A general class of tensor that appears in the study of conformal manifolds, are the  the Weyl-covariant tensors that are independent of the background fluid velocity. A relevant example of this class of operators is the Weyl curvature $C_{\mu\nu\lambda\sigma}$, that can be defined as the trace free part of the Riemann tensor. For $d\geq3$ it is
\begin{equation}
{\mathcal C}_{\mu\nu\lambda\sigma} \equiv {\mathcal R}_{\mu\nu\lambda\sigma}+4\,\delta^\alpha_{[\mu}g_{\nu][\lambda}\delta^\beta_{\sigma]}\mathcal{S}_{\alpha\beta} = C_{\mu\nu\lambda\sigma} - {\mathcal F}_{\mu\nu} g_{\lambda\sigma} = e^{2\phi}\,\widetilde{{\mathcal C}}_{\mu\nu\lambda\sigma}
\label{Weyltensor}
\end{equation}
where we introduced ${\mathcal S}_{\mu\nu}$,  the {\it Schouten tensor}, defined as
\begin{align}
{\mathcal S}_{\mu\nu} &\equiv \frac{1}{d-2}\left({\mathcal R}_{\mu\nu}-\frac{{\mathcal R} g_{\mu\nu}}{2(d-1)}\right) \nonumber \\ 
&= S_{\mu\nu}-\left(\nabla_\mu {\mathcal A}_\nu + {\mathcal A}_\mu {\mathcal A}_\nu - \frac{{\mathcal A}^2}{2} g_{\mu\nu} \right) -\frac{{\mathcal F}_{\mu\nu}}{d-2} =\widetilde{\mathcal{S}}_{\mu\nu} \ .
\label{Schouten}
\end{align}
The Weyl Tensor $C_{\mu\nu\lambda\sigma}=\mathcal{C}_{\mu\nu\lambda\sigma} +{\mathcal F}_{\mu\nu} g_{\lambda\sigma} $ defined in eq. (\ref{Weyltensor}) is a conformal tensor. Moreover, it turns out that $C_{\mu\nu\lambda\sigma}$ has the same symmetry properties as the Riemann Tensor $R_{\mu\nu\lambda\sigma}$, {\it i.e.}
\begin{align}
C_{\mu\nu\lambda\sigma} = -C_{\nu\mu\lambda\sigma} &= -C_{\mu\nu\sigma\lambda}=C_{\lambda\sigma\mu\nu} \\
C_{\mu\alpha\lambda}{}^\alpha & =0 \ .
\label{proprC}
\end{align}
From equation (\ref{proprC}), it follows that $C_{\mu\alpha\nu\beta}\,u^\alpha\, u^\beta$ is a symmetric traceless and transverse tensor.

Given the Schouten and the Weyl curvature tensors, one can recast other curvature tensors in terms of these new tensors:
\begin{equation}
\begin{split}
{\mathcal R}_{\mu\nu\lambda\sigma}&=\mathcal{C}_{\mu\nu\lambda\sigma} -\delta^\alpha_{[\mu}g_{\nu][\lambda}\delta^\beta_{\sigma]}\mathcal{S}_{\alpha\beta}, \\
 {\mathcal R} &= 2(d-1)\mathcal{S}_\lambda{}^\lambda \\
{\mathcal R}_{\mu\nu}&= (d-2) \mathcal{S}_{\mu\nu} + \mathcal{S}_\lambda{}^\lambda g_{\mu\nu} ,\\
%\mathcal{G}_{\mu\nu} &= (d-2)(\mathcal{S}_{\mu\nu}-\mathcal{S}_\lambda{}^\lambda g_{\mu\nu}) .\\
\end{split}
\end{equation}

\section{Conformal dissipative fluids}

Now we can use the Weyl covariant formalism to provide a classification, up to the second order in the derivative expansion, of the operators that could appear in the stress tensor of a conformal dissipative fluid. Second order term will be relevant in the discussion of holographic duals of conformal fluids.

Let us start by writing the conservation equations for the stress tensor in terms of  the new covariant derivative, {\it i.e.}
\begin{eqnarray}
{\mathcal D}_\mu T^{\mu \nu} &=& \nabla_{\mu}\ T^{\mu \nu} +w \, {\mathcal A}_\mu \, T^{\mu \nu} + 
\left( g_{\mu\alpha}\, {\mathcal A}^\mu - \delta^\mu_\mu\, {\mathcal A}_\alpha - \delta^\mu_\alpha \, {\mathcal A}_\mu \right) \, T^{\alpha \nu} \nonumber\\
&\,& +  \left(g_{\mu\alpha}\, {\mathcal A}^\nu - \delta^\nu_\mu \, {\mathcal A}_\alpha - \delta^\nu_\alpha \, {\mathcal A}_\mu\right) \, T^{\mu \alpha} \nonumber \\
&=& \nabla_\mu \, T^{\mu \nu}+ (w -d -2)\, {\mathcal A}_\mu \, T^{\mu \nu} - {\mathcal A}^\nu\, T^\mu_{\,\,\,\,\mu}
\nonumber \\
&=& \nabla_\mu T^{\mu \nu}  \  ,
\end{eqnarray}	
where inhomogeneous terms cancel out, remembering that $T^\mu_{\,\,\,\,\mu}=0$ and $w=d+2$. In the presence of a conformal anomaly, that can be present for 
CFTs on curved manifolds in even spacetime dimensions, the dynamical equation for $T_{\mu \nu}$ spells out
\begin{equation}
{\mathcal D}_\mu T^{\mu\nu} = 
\nabla_\mu T^{\mu\nu}  + {\mathcal A}^\nu \, \left(T^\mu_{\,\,\,\,\mu} - W \right) =0 \ ,
\end{equation}
where $W$ is the trace anomaly.

Given the definition of the Weyl covariant derivative, it is possible also to rewrite the building blocks of the stress tensor at first order in the covariant formulation 
\begin{equation}
\sigma^{\mu\nu} = {\mathcal D}^{(\mu} u^{\nu)} \ , \qquad 
\omega^{\mu\nu} = -{\mathcal D}^{[\mu} u^{\nu]} \ ,
\end{equation}
which have weight $w =3$.

To introduce second derivative terms, we need to pin down all of the possible two derivative operators that transform homogeneously under conformal transformations.  The fist kind of term we can have are those that involve the squares of the first derivative operators, {\it i.e.}
\begin{equation}
\sigma^\mu_{\;\alpha} \, \sigma^{\nu\alpha}  = e^{-4\,\phi}\, \widetilde{\sigma} ^\mu_{\;\alpha} \, \widetilde{\sigma} ^{\nu\alpha} \ ,\qquad 
\omega^\mu_{\;\alpha} \, \omega^{\nu\alpha} =  e^{-4\,\phi}\,\widetilde{\omega}^\mu_{\;\alpha} \, \widetilde{\omega}^{\nu\alpha} \ , \qquad 
\sigma ^\mu_{\;\alpha}\,\omega^{\nu\alpha} = e^{-4\,\phi}\,\widetilde{\sigma} ^\mu_{\;\alpha}\,\widetilde{\omega}^{\nu\alpha} \ .
\end{equation}	
Moreover there exist also two derivative terms built from the fluid velocity. The possible terms of this kind are
\begin{align}
& {\mathcal D}_{\mu}{\mathcal D}_{\nu} u^\lambda = {\mathcal D}_{\mu} {\sigma}_{\nu}{}^{\lambda} + {\mathcal D}_{\mu}{\omega}_{\nu}{}^{\lambda} = e^{-\phi}\widetilde{{\mathcal D}}_{\mu}\widetilde{{\mathcal D}}_{\nu}\tilde{u}^{\lambda}\ ,\\
& {\mathcal D}_{\lambda} {\sigma}_{\mu\nu} =  e^{\phi}\, \widetilde{{\mathcal D}}_{\lambda} \widetilde{\sigma}_{\mu\nu} \ , \\
& {\mathcal D}_{\lambda} {\omega}_{\mu\nu} =e^{\phi} \,\widetilde{{\mathcal D}}_{\lambda} \widetilde{\omega}_{\mu\nu}  \ . 
\end{align}

An other kind of second order terms is one which involves the temperature $T$ and various chemical potentials $\mu_I$. After some effort, it turns out that all of the possible operators in this group are
\begin{equation}
\begin{split}
&{\mathcal D}_\mu\left(\frac{\mu_I}{T}\right) = \widetilde{{\mathcal D}}_\mu  \left(\frac{\widetilde{\mu}_I}{\widetilde{T}}\right) \ ,\\ 
&{\mathcal D}_\mu T = e^{-\phi} \widetilde{{\mathcal D}}_\mu \widetilde{T} \\
&{\mathcal D}_\lambda{\mathcal D}_\sigma\left(\frac{\mu_I}{T}\right) = \widetilde{{\mathcal D}}_\lambda\widetilde{{\mathcal D}}_\sigma\left(\frac{\widetilde{\mu}_I}{\widetilde{T}}\right) \, \\
&{\mathcal D}_\lambda{\mathcal D}_\sigma T = e^{-\phi}\, \widetilde{{\mathcal D}}_\lambda\widetilde{{\mathcal D}}_\sigma\widetilde{T}  \ .
\end{split}
\end{equation}

In addition to the terms listed above, there is a one last term to consider, obtained by contracting the curvature tensors in the Weyl  tensor (\ref{Weyltensor}) with two velocity vectors
\begin{equation}
C_{\mu\alpha\nu\beta}\,u^\alpha \,u^\beta  = \widetilde{C}_{\mu\alpha\nu\beta}\,\tilde{u}^\alpha \,\tilde{u}^\beta 
\end{equation}	
that is a symmetric, traceless and transverse tensor.\\

Having classified all possible operators that can enter in the construction of dissipative terms for a viscous conformal fluid, it is possible to proceed in isolating the terms that will be useful later on. In particular, for sake of simplicity, we will construct a dissipative generalisation of a conformal fluid with no conserved charges. For this choice, there are five possible symmetric traceless tensors which transform homogeneously under Weyl rescalings at second order:
\begin{align}
&{\mathfrak T}_1^{\mu\nu} =2\, u^\alpha \, {\mathcal D}_\alpha \sigma_{\mu\nu}, \qquad {\mathfrak T}_2^{\mu\nu} =C_{\mu\alpha\nu\beta}\,u^\alpha \,u^\beta, \\
&{\mathfrak T}_3^{\mu\nu}  =4\,\sigma^{\alpha\langle\mu}\, \sigma^{\nu\rangle}_{\ \alpha}, \qquad \,\,\,\, {\mathfrak T}_4^{\mu\nu} =  2\, \sigma ^{\alpha\langle\mu}\,\omega^{\nu\rangle}_{\ \alpha}, \\
&{\mathfrak T}_5^{\mu\nu}=   \omega ^{\alpha\langle\mu}\, \omega ^{\nu\rangle}_{\ \alpha} .
\end{align}
We summarise all the symmetric traceless tensors which transform homogeneously under Weyl rescalings up to the second order in table 2.1.\\

\begin{table}[tb]
\begin{center}
\begin{tabular}{ll}
\toprule
First order:&  \qquad $\sigma^{\mu\nu}$  \\
\midrule
Second order:& \qquad ${\mathfrak T}_1^{\mu\nu} =2\, u^\alpha \, {\mathcal D}_\alpha \sigma_{\mu\nu}$  \\
\, & \qquad ${\mathfrak T}_2^{\mu\nu} =C_{\mu\alpha\nu\beta}\,u^\alpha \,u^\beta$ \\
\, & \qquad ${\mathfrak T}_3^{\mu\nu}  =4\,\sigma^{\alpha\langle\mu}\, \sigma^{\nu\rangle}_{\ \alpha}$ \\
\, & \qquad ${\mathfrak T}_4^{\mu\nu} =  2\, \sigma ^{\alpha\langle\mu}\, \omega^{\nu\rangle}_{\ \alpha} $\\
\, & \qquad $ {\mathfrak T}_5^{\mu\nu}=   \omega ^{\alpha\langle\mu}\, \omega ^{\nu\rangle}_{\ \alpha} $ \\
\bottomrule
\label{table}
\end{tabular}
\end{center}
\caption{Symmetric and homogeneous traceless operator for a conformal viscous non-charged fluid up to the second order.}
\end{table}

Having completed the exploration of possible dissipative terms up to the second order, we are finally able to write the correction to the stress tensor for a conformal viscous fluid that, for simplicity, we will assume to have no conserved charge. The general contributions to the stress tensor are:
\begin{eqnarray}
T_{(0)}^{\mu \nu} &=& \alpha\, T^d \, \left(g^{\mu \nu} +d\,  u^\mu \, u^\nu\right)  -2\, \eta \, \sigma^{\mu \nu} \nonumber \\
\Pi_{(1)}^{\mu\nu} &=& -2\, \eta\,\sigma^{\mu\nu} \nonumber \\
\Pi_{(2)}^{\mu\nu} &=&  \tau_\pi\,\eta\, {\mathfrak T}
_1^{\mu\nu} + \kappa\,{\mathfrak T}
_2^{\mu\nu} + \lambda_1\, {\mathfrak T}
_3^{\mu\nu} + \lambda_2\, {\mathfrak T}
_4^{\mu\nu} + \lambda_3\, {\mathfrak T}
_5^{\mu\nu} \  .
\end{eqnarray}	
Where $\{\eta, \tau_\pi, \kappa, \lambda_i\}$ for $i = \{1,2, 3\}$ is a set of six transport coefficients characterizing the viscous fluid.

\section{Nonrelativistic limit of fluid dynamics}

In this section we will discuss, following \cite{Bhattacharyya:2008jc}, the non-relativistic scaling limit of the fluid dynamical equations of a non-charged fluid (we will not assume, for the moment, conformal invariance),  {\it i.e}
\begin{equation}
\nabla_\mu T^{\mu\nu}=0 \ ,
\end{equation}
with 
\begin{equation}
T^{\mu \nu}= \rho u^\mu u^\nu + {\rm p} P^{\mu\nu} -2 \eta \sigma^{\mu\nu} -\zeta \theta P^{\mu\nu} + \ldots
\end{equation}
where ${\rm p}$ is the pressure\footnote{In this section ${\rm p}$ will indicate the relativistic pressure and $p$ the nonrelativistic one.}, $\rho$ the energy density, $\eta$ the shear viscosity, $\zeta$ the bulk viscosity of the fluid, to the incompressible non-relativistic Navier-Stokes equations 
\begin{align}
\label{NS} 
\frac{\partial {\vec v}}{\partial t}+ {\vec v}\cdot \nabla  {\vec v}& = -{\vec \nabla} p
+ \nu \nabla^2 {\vec v} + {\vec f} \\
{\vec \nabla} \cdot {\vec v}&=0
\end{align}
where ${\vec v}$ is the fluid velocity, $p$ the fluid pressure, $\nu$ the shear viscosity and ${\vec f}$ an external forcing function.

In particular, we will discuss the non-relativistic limit 
\begin{equation}
\begin{split}
\delta x& \sim \frac{1}{T \epsilon}\\
\delta t&\sim  \frac{1}{T \epsilon^2}\\
v^i & \sim   \epsilon\\
\delta {\rm p} & \sim  T^d \epsilon^2\\
\epsilon & \to 0
\end{split}
\end{equation}
where $\delta x$ is a spatial length scale, $\delta t$ the temporal scale while  $v^i$ and $\delta {\rm p}$ have to be seen as estimates of the magnitude 
of velocity and pressure fluctuations on a configuration of fluid at rest and in equilibrium.

The meaning of this limit can be understood by remembering that it is necessary to scale to long distances to be in the fluid dynamical regime. Time intervals should scale like spatial intervals squared, as it is easy to realise looking at the dispersion relation for shear waves,  $\omega = i \nu k^2$ .
These two scalings determine the scaling law for velocities. Finally the pressure fluctuations are to be scaled such that  they cannot accelerate the fluid velocities outside this scaling limit. 

Let us start by considering the flow of a fluid on a spacetime with a metric that can be cast in the form
\begin{equation}
G_{\mu\nu}=g_{\mu\nu}+H_{\mu\nu} \ ,
\end{equation}
where $H_{\mu\nu}$ is an arbitrary small fluctuation of a  background metric $g_{\mu\nu}$ that we want to write as\footnote{This form of the metric $g_{\mu\nu}$ can be seen as a simply a choice of coordinate system, for a large class of metrics.} 
\begin{equation} 
g_{\mu\nu}dx^\mu dx^\nu= -dt^2 + g_{ij}dx^i dx^j.
\end{equation}

The fluid flow on the background space equipped with the metric $G_{\mu\nu}$ can be reinterpreted as a forced flow on the space with metric $g_{\mu \nu}$  \cite{Bhattacharyya:2008jc}, mapping the velocity field ${\tilde u}^\mu$ on the space $G_{\mu\nu}$ to a velocity field $u^\mu$ on $g_{\mu\nu}$. In constructing the map the normalisation of the velocity field $u^2={\tilde u}^2 =-1$ must be maintained. 

The fluid velocity ${\tilde u}^\mu$ on the space $G_{\mu\nu}$ can be written as
\begin{equation}
{\tilde u}^\mu  = \frac{1}{\sqrt{V^2}} \left( 1, {\vec V} \right) \ ,
\end{equation}
where ${\vec V}$ is the $d-1$ spatial vector with components $V^i$. $V^\mu$ is a $d$ component object with components $(1, {\vec V} )$ and 
$V^2$ is $G_{\mu \nu} V^\mu V^\nu$.

A perturbative expansion of the velocity ${\tilde u}^\mu$  at the first order in the small fluctuation $H_{\mu\nu}$ gives
\begin{align} 
{\tilde u}^\mu& = u^\mu + \delta u^\mu + \ldots \nonumber \\ 
u^\mu & = \frac{1}{\sqrt{1 - g_{ij} V^i V^j}} \left( 1, {\vec V} \right)\\
\delta u^\mu & = -u^\mu \frac{u^\alpha u^\beta H_{\alpha \beta}}{2} \nonumber 
\end{align}
where we have expressed the series in terms of $u^\mu$, the new velocity of the fluid referring to the metric $g_{\mu\nu}$. The idea that we want to follow is to consider terms in which $H_{\alpha \beta}$ appears as contributions to an effective forcing function in the non-relativistic Navier-Stokes  equation (\ref{NS}).

We want to consider $H_{\alpha \beta}$ as a small amplitude and long distance fluctuation on a uniform fluid at rest - it is always a solution  to the equations of fluid dynamics on the space with metric $g_{\mu\nu}$.  The fluid at rest will be described as a fluid with pressure ${\rm p}_0$, energy density $\rho_0$  on a manifold that is ``close'' to $g_{\mu\nu}$. We set the components of the fluctuation to
\begin{align}
H_{00}& = \epsilon^2 h_{00}(\epsilon x^i, \epsilon^2 t)\nonumber \\
H_{0i}& = \epsilon A_i (\epsilon x^i, \epsilon^2 t)\nonumber\\
H_{ij}&= \epsilon^2 h_{ij} (\epsilon x^i, \epsilon^2 t) \nonumber
\end{align}
and we assume consistently that the velocity and pressure satisfy
\begin{equation} 
\begin{split}
V^i&=\epsilon v^i(\epsilon x^i, \epsilon^2 t)\\
\frac{{\rm p}-{\rm p}_0}{\rho_0 + {\rm p}_0}&= \epsilon^2 p(\epsilon x^i, \epsilon^2 t) 
\end{split}
\end{equation}
where $\epsilon$ is a parameter arbitrarily small and one can define a covariant vector $v^\mu=(1, v^i)$. The normalisation of the pressure fluctuations, $\rho_0 + P_0$, has been introduced for future convenience. Energy density $\rho$ and viscosity $\nu$ also scale in a similar fashion. 

The conservation equations are expected to reduce to the classical  Navier-Stokes equations in this limit. Let us start by considering the temporal component of $\nabla_\mu T^{\mu \nu}$, we get
\begin{equation}
\begin{split}
\nabla_\mu T^{\mu 0}& = \epsilon^2 \left[\rho_e \left( \nabla_i v^i\right) \right] +  {\cal O}(\epsilon^4)\\
\rho_e&= \rho_0+p_0\\
\end{split}
\end{equation}
therefore, for $\epsilon \to 0$, this equation reduces to 
\begin{equation}
\nabla_i v^i=0 ,
\end{equation} 
where $\nabla_i$ is the covariant derivative with respect to the purely spatial metric $g_{ij}$. 
Spatial components of the conservation equations, after some manipulation, can be cast in the form
\begin{multline}
\nabla_\mu T^{\mu i} = \epsilon^3  \left[ \rho_e \nabla^i p + \rho_e \nabla_\mu \left( v^i v^\mu \right) \right.\\
\left.  -2 \eta \nabla_j  \left(\frac{ \nabla^j v^i + \nabla^i v^j}{2}-g^{ij}\frac{\vec{\nabla} \cdot {\vec v}} {d-1} \right) 
-\zeta \nabla_i {\vec \nabla}\cdot {\vec v}  -f^i \right] + {\cal O}(\epsilon^5)  
\label{spns}
\end{multline}
where the forcing function $f$ is defined as
\begin{equation}
f^i=\rho_e\left( \frac{\partial_i h_{00}}{2} - \partial_0 A_i  - \frac{\partial_j (\sqrt{g} A_i v^j)}{\sqrt{g}} + v^j\partial^i A_j \right) \ .
\end{equation}

The coefficient of $\epsilon^3$ can be rewritten in a more transparent form. We begin using the expression 
\begin{equation}
\nabla_i v^i= \frac{\partial_i(\sqrt{g} v^i)}{\sqrt{g}}=0 \nonumber \ ,
\end{equation}
to get
\begin{equation}
 \nabla^i p + \partial_0 v^i  + {\vec v}. \nabla v^i - \nu \left(  \nabla^2 v^i  +   R^i_j v^j  \right) 
= \frac{\partial^i h_{00}}{2} - \partial_0 A^i + F^i_j v^j
\end{equation}
where we defined a field strength $F_{ij}=\partial_i A_j-\partial_j A_i$ associated to the vector field $A_i$, and a {\it kinematical viscosity} $\nu=\eta/\rho_e$ of the fluid. It this then possible to redefine variables, splitting the gauge field $A_i$ into its pure curl and pure divergence parts, {\it i.e.}
\begin{equation}
A_i= a_i +\nabla_i \chi \ ,
\end{equation}
where $\nabla_i a^i=0$, and one finds
\begin{equation}
f_{ij}\equiv \partial_i a_j-\partial_j a_i= F_{ij}.
\end{equation}
It is also useful to define an effective pressure 
\begin{equation}
p_e=p-\frac{1}{2}  h_{00} + {\dot \chi}.
\end{equation}
In terms of the new variables, equation (\ref{spns}) becomes
\begin{equation}
 \nabla_i p_e + \partial_0 v_i + {\vec v}. \nabla v_i - \nu \left( \nabla^2 v_i +  R_{ij} v^j \right)   = -\partial_0 a^i- v^j f_{ji} \ ,
 \label{cond-spnv}
 \end{equation}
which is indeed  the Navier-Stokes equation with a forcing function generated by an effective background electromagnetic field on the effectively charged fluid. \\ 

In the case of a charged fluid, the discussion follows the same track, adding the set of conservation equations for conserved charges (\ref{cons-hydr}). It turns out that, in the non-relativistic limit, these equations lead to supplementary conditions under which the fluid velocity is divergence-free.\\

A point that it is useful to assess regards how Cauchy conditions for the relativistic fluid are transmitted to the non-relativistic fluid. The Cauchy data of the relativistic dynamical equations consist of $d$ real functions of space: the value of the pressure field and the  value of the $d-1$ independent velocity fields defined on an initial timeslice \footnote{Since the fluid conservation equations are of first order in time, the Cauchy data of the problem does not include the time derivatives of all these fields.}. In the case of the non-relativistic incompressible Navier-Stokes equations, one can observe that the derivative of eq.(\ref{cond-spnv}) results in the condition
\begin{equation}\label{press}
\nabla^2 p_e = -\nabla_i v^j \nabla_j v^i - v^i v^j R_{ij}  \ + \nabla_i \left[ \left( -\nu R^i_j + f^i_j \right) v^j \right] \ ,
\end{equation}
that determines the pressure of the fluid as a function of the fluid velocity. Therefore, the set of independent data necessary to determine the flow of the fluid is given by $d-2$ real functions that parameterize an arbitrary divergence free velocity field. Indeed,  two of the degrees of freedom of the equations of relativistic fluid dynamics are projected away in the scaling limit. One can see that these two degrees of freedom correspond the fluctuations of the pressure and the divergence of the velocity\footnote{As observed in \cite{Bhattacharyya:2008jc}, at the linearized level these two degrees of freedom combine together in sound mode fluctuations. Hence the physical fact behind the reduction of Cauchy data in the non-relativistic limit is that sound waves are projected out in the scaling limit}. 

\section{Nonrelativistic symmetries for conformal fluids}

As we have discussed in the previous section, the Navier-Stokes equations may be obtained as the scaling limit of any relativistic equations of fluid 
dynamics. For the purpose of our work we are interested in particular in conformal fluids  due to their connection with gravity. Therefore it is interesting to assess in which form the conformal symmetry descends to a symmetry of the Navier-Stokes equations, in the non-relativistic scaling limit we proposed above.\\

A dilatation consists of a diffeomorphism 
\[
(x')^\mu=\frac{x^\mu}{\lambda}, ~~~ T'=T\ , ~~~ (u')^\mu=\frac{u^\mu}{\lambda}\ , ~~~ (g')_{\mu\nu}=  \lambda^2 g_{\mu\nu}\ ,
\]
associated with a Weyl transformation 
\[
{\tilde x}^\mu= (x')^\mu\ , ~~~ {\tilde g}_{\mu\nu}= \frac{(g')_{\mu\nu}} {\lambda^2}\ , ~~~ {\tilde u}^{\mu}= \lambda (u')^{\mu}\ , 
~~~ {\tilde T}= \lambda T \ .
\]
The compound action is therefore:
\begin{equation}
{\tilde x}^\mu = \frac{x^\mu}{\lambda}, ~~~\tilde{g}_{\mu\nu}= g_{\mu\nu}, ~~~ \tilde{u}^\mu= u^\mu, ~~~\tilde{T}=\lambda T .
\end{equation}
This action is a symmetry of the  equations of conformal relativistic fluid dynamics, but it is not a symmetry of the Navier-Stokes equations. Indeed, kinematical viscosity $\nu$ in the Navier-Stokes equations, is proportional to $\frac{1}{T}$, hence it is not invariant under dilatation transformations.  However as was shown in \cite{Bhattacharyya:2008jc}, it is possible to modify dilatation transformations into a true symmetry of the Navier-Stokes equations by ``absorbing'' the transformation of $\nu$ into the ``anomalous'' transformations of time and velocity. \\

Let us start by considering special conformal transformations. The scaling law for the velocity and temperature fields, under a special conformal transformation, may be obtained by combining  a  diffeomorphism with the appropriate Weyl transformation. For an infinitesimal conformal transformation one gets 
\begin{align}
\label{spconf}
\delta x^\mu& = -2 c.x x^\mu +x^2 c^\mu \nonumber \\
\delta u^\mu&= -2\left[ x^\mu c^\nu-x^\nu c^\mu\right] u_\nu - \delta x^\nu \partial_\nu u^\mu \\
\delta T&= 2 c.x T - \delta x^\nu \partial_\nu T . \nonumber 
\end{align}
To verify the covariance of local equations under these symmetry transformations, one has to omit the terms proportional 
to $\delta x^\mu \partial_\mu$, using the following expression for the derivative's transformation
\[
\delta \left( \partial_\beta \right) = 2 \left[ c_\beta  x.\partial -x_\beta c.\partial + x.c \partial_\beta \right] .
\]

Under transformation (\ref{spconf}), the conformal stress tensor becomes
\begin{equation}
\delta T^{\mu\nu}= 2 d (c.x) T^{\mu\nu} + 2 (x^\lambda c^\mu -x^\mu c^\lambda)  T^\nu_\lambda +  2 (x^\lambda c^\nu -x^\nu c^\lambda)T^\mu_\lambda - \delta x^\lambda \partial_\lambda T^{\mu \nu} \ .
 \end{equation}
 Therefore, given the derivative's transformation, any identically traceless stress tensor is invariant under special conformal transformations.\\
 
Special conformal transformations (\ref{spconf}) give rise to an additive shift in the temperature fluctuation, $\delta T$, proportional 
to $x\cdot c T_0$. In order that this shift respects the $\epsilon^2$ scaling of $\delta T$ one has to scale $c_0 \propto \epsilon^4$ and $c_i \propto \epsilon^3$. In this way the transformations (\ref{spconf}) read
\begin{align}
\delta t & = 0 \nonumber \\
\delta x^i& = - t^2 c^i  \nonumber \\
\delta v^i &= -2 c^i t  + t^2  c_j \partial_j v^i  \\
\delta T&= 2 (-c^0 t + c^i x^i)  T  + t^2 c_j \partial_j T . \nonumber \\
\end{align}
To verify the covariance of the conservation equations under these transformations, one has to omit the terms proportional to $t^2 c_j \partial_j$ and to supplement their action with the following derivative transformation
\[
\delta \left( \partial_t \right) = 2 t c^i \partial_i \,  \qquad  \delta \left( \partial_i \right)  =0 \ . 
\]

The symmetry generated by $c^0$ acts trivially. Indeed it does  not act on coordinates or velocities, but merely generates a shift, linear 
in time, of the pressure. On the contrary, the symmetries generated by $c^i$ act nontrivially. Under this transformation one gets
\[
 \delta p_e= \frac{\delta p}{d P_0}= 2 c\cdot x \ ,
 \]
and therefore
\[
\delta \partial_i p_e= 2 c^i \ .
\] 
Finally $\delta( {\dot v^i} + v\cdot\nabla v^i ) = -2 c^i$. Hence, the viscous term - which was responsible for the change in Navier-Stokes equations - is unchanged under the redefined transformation. The Navier-Stokes equations are invariant under a conformal symmetry group.

We can conclude this section, for sake of completeness, by giving few more details on the conformal symmetry group. 

The group generators of this group are the dilatation $D$,  special conformal symmetries $K_i$, Galilean boosts $B_i$, the generator 
of time translations (energy) $H$, momenta $P_i$ and spatial rotations $M_{ij}$. The action of these generators on velocity fields is given by 
\begin{equation} 
\begin{split}
&D v^j  = \left( -2 t \partial_t - x^m \partial_m - 1 \right) v^j  \\
&K_i v^j= -2 t \delta_{ij}  + t^2  \partial_i v^j \\
&B_i v^j=  \delta_{ij} -t \partial_i v^j\\
&H v^j =  - \partial_t v^j \\
&P_i v^j= - \partial_i v^j \\
&M_{ik} v^j= \delta_{ij} v^k - \delta_{kj} v^i - (x^k \partial_i - x^i 
\partial_k) v^j \\
\end{split}
\end{equation}
While the following commutation relations among generators hold:
\begin{equation}\label{comrel} 
\begin{split}
&[D, K_i]= -3 K_i \qquad \,\,  [D, B_i]= -B_i \qquad [D, H]= 2 H\\
&[D, P_i]= P_i  \qquad \qquad [D, M_{ij}]=0  \qquad \,\,\,\,\, [M_{ij}, P_k]= -\delta_{ik} P_j + \delta_{jk} P_i \\
&[ M_{ij}, K_k] = -\delta_{ik} K_j + \delta_{jk} K_i \qquad \qquad \,\,\,\,\,\,\, [M_{ij}, B_k]= -\delta_{ik} B_j + \delta_{jk} B_i \\
&[M_{ij}, H]= 0 \qquad \,\,\,\,\,\,\,\,\,\, [K_i, P_j]= 0 \,\,\,\,\,\,\, \qquad [K_i, B_j]= 0 \\
&[K_i, H]= -2 B_i \,\, \qquad [H, B_j]= -P_i\\
\end{split}
\end{equation}

In addition to the relevant symmetries we enumerated above, the Navier-Stokes equations have, as noted before, an infinite dimensional group of trivial symmetries that act as a shift of the pressure by an arbitrary function of time. These are indeed symmetries since only gradients of the pressure 
enter the Navier-Stokes equations and trivial because the pressure is not really an independent variable, as we discussed above. The action of symmetry generators on the pressure is 
\begin{equation}
\begin{split}
D p_e  =& \left( -2 t \partial_t - x^m \partial_m - 2 \right) p_e  \\
K_i p_e =&  2 x^i  + t^2  \partial_i p_e \\
B_i p_e = & -t \partial_i p_e\\
H p_e = & - \partial_t p_e \\
P_i p_e =& - \partial_i p_e \\
M_{ik} p_e =& - (x^k \partial_i - x^i 
\partial_k) p_e . \\
\end{split}
\end{equation}

These generators acting on the pressure do not yield the commutation relations,  indeed they have additional terms. However spatial derivatives of the pressure field correctly transform according to the algebra (\ref{comrel}).  For this reason, the symmetry algebra (\ref{comrel}) is not represented on the pressure field  itself, but only on its spatial derivatives.

%\section{Boundary conditions}

%_____________________________________________
% Third Chapter
\chapter{Fluid dynamics from gravity}
%_____________________________________________

In this chapter we give a short review of the construction of gravity solutions dual to arbitrary fluid flows proposed in \cite{Bhattacharyya:2008jc}. A detailed and comprehensive exposition of the subject can be found in \cite{Rangamani:2009xk}. As pointed out in the first chapter, the application of  hydrodynamic concepts  to black holes is part of the core intuition of the membrane paradigm \cite{Thorne:1986iy, Damour:2004kwa}, wherein one modeled the black hole horizon by a membrane equipped with fluid like properties. However, in this approach the connection to fluid dynamics is an analogy, and therefore provides a qualitative understanding of  the physical behaviour of black holes via a simpler fluid model. On the contrary, the fluid-gravity correspondence - an example of the  AdS/CFT correspondence - provides a duality between the hydrodynamic description and the  gravitational dynamics.  Indeed, the  precise quantitative connection allows one to map fluid solutions of the boundary field theory into a black hole solutions in the  bulk geometric description with regular event horizons.

The AdS/CFT correspondence conjectures a deep connection between quantum field theories and theories of gravity \cite{Maldacena:1997re,Gubser:1998bc,Witten:1998qj}. The correspondence has proved to be a powerful tool for the large $N$ limit in which the gravitational theory turns classical,  and in a simultaneous strong `t Hooft coupling limit that suppresses  $\alpha'$ corrections to gravitational dynamics.  In this limit the conjecture asserts the equivalence between the effectively classical  large $N$ dynamics of the local single trace operators $\rho_n= N^{-1}\text{Tr}\mathbf{O_n}$ of gauge theory and the classical two derivative equations of Einstein gravity interacting with other fields. The  AdS/CFT correspondence provides a one to one map from  the local and relatively simple bulk equations into unfamiliar and extremely nonlocal equations for the boundary trace operators $\rho_n(x)$. The authors of  \cite{Bhattacharyya:2008jc} proposed a construction which can simplify the extremely complicated equations on the boundary.

Two derivative bulk theories of gravity, containing AdS$_{d+1} \times M_I$ as a solution, admit a consistent truncation to the Einstein equations with a negative cosmological constant.The only fluctuating field in this truncation is the Einstein frame graviton; all other bulk fields are simply set to their background AdS$_{d+1} \times M_I$ values. This fact implies the existence of a sector of decoupled and universal dynamics of the stress tensor in the corresponding dual field theories. Indeed, all single trace operators other than the stress tensor may consistently be set to zero as the stress tensor undergoes its dynamics. The dynamics of the stress tensor happens to be universal since it is governed by the same equations of motion in each member of this infinite class of strongly coupled CFTs. 

A further simplification in the dynamics of the dual theory can be obtained by considering the regime in which the local stress tensor varies on a length scale that is large, at any point,  compared to a local equilibration length scale - intuitively, a ``mean free path'' - which is set by the ``rest frame'' energy density at the same point. Under this condition, it is expected that boundary configurations should be locally thermalized, and therefore well described by the equations of boundary fluid dynamics. In particular, it has been proven that the complicated nonlocal $T_{\mu\nu}$ dynamics reduces to the familiar boundary Navier Stokes equations of fluid dynamics in this long wavelength limit  \cite{Bhattacharyya:2008jc}. 

In \cite{Bhattacharyya:2008jc} a perturbative procedure was proposed,  in a boundary derivative expansion, to construct a large class of asymptotically AdS$_5$ long wavelength solutions to Einstein's equations with a negative cosmological constant. These solutions are parameterized by a four-velocity field $u^\mu(x^\mu)$  and a temperature field $T(x^\mu)$ that have to obey the four dimensional generalized Navier Stokes equations $\nabla_\mu T^{\mu\nu}=0$ where the stress tensor $T^{\mu\nu}(x^\mu)$ is a local functional of the velocity and the temperature fields. 

The explicit map from the space of solutions of a distinguished set of Navier Stokes equations to the space of long wavelength solutions of asymptotically AdS$_{d+1}$ gravity was established up to second order for $d= 4, 5$ in \cite{Bhattacharyya:2008jc} and than generalized and studied in more detail in \cite{Bhattacharyya:2008kq, VanRaamsdonk:2008fp, Bhattacharyya:2008xc, Bhattacharyya:2008ji, Dutta:2008gf, Loganayagam:2008is}. In particular, in \cite{Bhattacharyya:2008xc} it was proven that, subject to mild assumptions, these spacetimes have regular event horizons. Successively, the construction was extended to spacetimes of arbitrary dimensionality \cite{Bhattacharyya:2008mz, Haack:2008cp}.

\section{Black branes in AdS$_{d+1}$}

Let us start by analysing the universal sector of a string background of the form $\text{AdS}_{d+1} \times X$ where $X$ is a compact internal manifold, suitable for obtaining a consistent string vacuum  and whose properties will be irrelevant in the following. Einstein's equation of gravity with a negative cosmological constant $\Lambda$ follow from the Einstein-Hilbert action
\begin{equation}
S_{\text{bulk}} = \frac{1}{16\pi\,G_N^{(d+1)}}\,\int\, d^{d+1}x\, \sqrt{-G} \,\left(R - 2\,\Lambda\right) \ .
\end{equation}	
Choosing $R_{AdS} = 1$, Einstein's equations are given by
\begin{align}
\label{ein} 
&E_{M N}= R_{M N} - \frac{1}{2} \, G_{M N} R- \frac{d(d-1)}{2}\, G_{MN}=0\\
&R_{M N} + d\, G_{M N}=0, \qquad R=- d(d+1) , 
\end{align}
where $G_{M, N}$ is the bulk metric, $E_{M N}$ is the Einstein tensor and $M, N = 1, \dots , d+1$\footnote{We adopt the convention of using upper case Latin indices $\{M,N, \cdots\}$ to denote bulk directions, while lower case Greek indices $\{\mu ,\nu, \cdots\}$ refer to field theory or boundary directions. Lower case Latin indices are used, instead, $\{i,j,\cdots\}$ to denote the spatial directions in the boundary. Therefore we have $M, N = 1, \dots , d+1$, and $\mu, \nu=1 \ldots d$.}.

An obvious solution of these equations is AdS spacetime with unit radius  
\begin{equation}
\mathrm{d}s^2=\frac{\mathrm{d}r^2}{r^2} +r^2 \left( \eta_{\mu\nu} \mathrm{d}x^\mu \mathrm{d}x^\nu \right), 
\end{equation}
where now $\mu, \nu=1 \ldots d$.

A different solution to Einstein's equations is the planar Schwarzschild-$AdS_{d+1}$ black brane which in standard coordinates is written as
\begin{align}
\label{AdSSw}
\mathrm{d}s^2 &= -r^2\, f(b\,r) \, \mathrm{d}t^2 + \frac{\mathrm{d}r^2}{r^2\, f(b\,r) }+ r^2\, \delta_{ij} \, \mathrm{d}x^i\, \mathrm{d}x^j \ ,\\
f(r) &= 1-\frac{1}{r^d} \ .
\end{align}	
Eq. (\ref{AdSSw}) can be seen as one-parameter family of solutions labeled by the horizon size $r_+$, which sets the temperature of the black hole
\begin{equation}
T= \frac{d}{4\pi\,b} \  .
\end{equation}	

From the black brane solution (\ref{AdSSw}) one can generate a $d$ parameter family of  {\it boosted black brane} solutions by boosting the solution along the 
translationally invariant spatial directions $x^i$. In Schwarzschild like coordinates, the new solution can be written as 
\begin{align}
\label{blackbrane}
\mathrm{d}s^2 &=\frac{\mathrm{d}r^2}{r^2 f(r) } +r^2 \left( -f(r) u_\mu u_\nu 
+ {\mathcal P}_{\mu\nu} \right)  \mathrm{d}x^\mu \mathrm{d}x^\nu \\
f(r) &= 1-\frac{1}{(br)^d}, \qquad \eta_{\mu \nu} u^\mu u^\nu=-1, \nonumber \\
{\mathcal P}_{\mu\nu} &=\eta_{\mu\nu}+u_\mu u_\nu, \qquad b=\frac{d}{4 \pi T} \ , \nonumber
\end{align}
where
\begin{align}
u^v&=\frac{1}{\sqrt{1-\beta^2}} \\
u^i&=\frac{\beta_i}{\sqrt{1-\beta^2}} \ 
\end{align}
and the velocities $\beta_i$ are all constants with $\beta^2 = \beta_j \, \beta^j$, and $ P^{\mu \nu}= u^\mu u^\nu +\eta^{\mu \nu} $
is the projector onto spatial directions. The temperature $T$  associated to this solution is constant. The parameters which characterize the bulk solution are now the basic hydrodynamical degrees of freedom, temperature and the velocity of the black hole. 

More generally, one can consider solution of the form (\ref{blackbrane}) where $g_{\mu\nu}$ is an arbitrary constant boundary metric of signature $(d-1, 1)$, and $u^\mu$ is a generic constant unit normalized $d$ velocity, {\it i.e.}
\begin{equation}
 g_{\mu \nu} u^\mu u^\nu=-1, \qquad  {P}_{\mu\nu}= g_{\mu\nu}+u_\mu u_\nu, \qquad b=\frac{d}{4 \pi T} \ .
\end{equation}
Clearly the new metric is quite redundant, since a $d(d+3)/2$ parameter set of metrics (\ref{blackbrane}) are all coordinate equivalent. Indeed $g_{\mu\nu}$ can be sent in $\eta_{\mu\nu}$ by an appropriate linear coordinate  transformation $x^\mu \rightarrow \Lambda^\mu_\nu x^\nu$, and $u^\mu$ can subsequently be set to $(1, 0 \ldots 0)$ by a boundary Lorentz transformation. Moreover $b$ can be set to one by uniform rescaling of boundary coordinates  coupled with a rescaling of $r$. 

For our purposes  it is useful, to introduce an even more redundant formulation of solution (\ref{blackbrane}), obtained with via a transformation  $\tilde{r}\to e^{-\phi} r$, {\it i.e.}
 \begin{equation} 
\label{blackbrane2}
\mathrm{d}s^2=\frac{ \left( \mathrm{d}\tilde{r} + \tilde{r}  \mathcal A_\nu \mathrm{d}x^\nu \right)^2 }{\tilde{r}^2 f(r) } +\tilde{r}^2 \left( -f(\tilde{b}\tilde{r}) 
\tilde{u}_\mu \tilde{u}_\nu \mathrm{d}x^\mu \mathrm{d}x^\nu 
+ {\tilde P}_{\mu\nu} \mathrm{d}x^\mu \mathrm{d}x^\nu \right) 
\end{equation}
where $g_{\mu\nu}$, $u^\mu$, and $b$ are as defined in the previous equation, while 
\begin{equation}
\label{weyldef}
\tilde{g}_{\mu\nu}=e^{2\phi(x^\mu)} g_{\mu\nu}, \quad 
\tilde{u}_\mu =e^{\phi(x^\mu)} u_\mu, \quad \tilde{b} = 
e^{\phi(x^\mu)} b, 
\end{equation}
and $\phi(x^\mu)$ is an arbitrary function. 

Metrics of the form (\ref{blackbrane2}) describe the same bulk geometry but can be interpreted as distinct,  though Weyl equivalent, boundary description by regulating them inequivalently near the boundary. In particular, following the references above, spacetimes described by eq. (\ref{blackbrane2}) will be regulated on slices of constant ${\tilde r}$ and consequently seen as states in a conformal field theory on a space with metric ${\tilde g}_{\mu\nu}(x)$.

The non-anomalous part of the boundary stress tensor dual to metric (\ref{blackbrane2}) is given by 
\begin{equation}
T_{\mu\nu} = \frac{1}{16\pi G_{\text{AdS}} \tilde{b}^d} \left(\tilde{g}_{\mu\nu}+d \tilde{u}_\mu \tilde{u}_\nu \right) \ ,
\end{equation}
which corresponds to the stress tensor of an ideal conformal fluid with a pressure $p=1/(16\pi G_{\text{AdS}} b^d)$ 
and without any vorticity or shear strain rate. 

It is useful to observe that the velocity field is the unique time-like eigenvector of the stress tensor
\begin{equation} 
T^\mu_\nu u^\nu = \frac{K}{b^d} u^\mu, ~~~K=-\frac{(d-1)}{16 \pi 
G_{\text{AdS}}}
\end{equation}
and the inverse temperature field $b$ is simply related to its eigenvalue.

\section{Zeroth order {\it ansatz} for the metric}

In the previous section, we described locally asymptotically AdS$_{d+1}$ exact solutions to Einstein's equations (\ref{ein}) which have, as a holographic boundary stress tensor, the ideal conformal fluid stress tensor. This is an expected result as the solution is stationary and therefore corresponds to the global thermal equilibrium. 

Now, to describe hydrodynamics we should perturb the system away from global equilibrium. The natural way to do this is to consider  thermodynamic variables varying along the boundary directions, that is promoting the parameters $b$ ({\it viz.} the temperature $T$), $\beta_i$ ({\it viz.} the velocity $u_\mu$) to functions of the boundary coordinates. In particular, we want discuss solutions to Einstein equations  with slowly varying boundary stress tensors on a boundary manifold that has a weakly curved  boundary metric. These solutions should be locally patch-wise equilibrated, therefore corresponding bulk solution should approximately be given by patching together tubes of the uniform black brane solutions, extending from local patches on the boundary into the bulk.\\

Let us consider a locally asymptotically AdS$_{d+1}$ solution to Einstein's equations, whose corresponding boundary stress tensor everywhere has a unique time-like eigenvector. Let us promote this eigenvector to a function $u^\mu(x)$  of the boundary coordinates, and let us, in the same way, define the inverse of the temperature field as $b(x)$. Now it is useful to define in a precise way in which sense we can consider our solution as ``slowly varying'': denoting by $\delta x(y)$ the smallest length scale of variation of the stress tensor of the corresponding solution at the point $y$, we shell consider  $\delta x (y) \gg  b(y)$. Actually $b(y)$ may be interpreted as the effective length scale of equilibration of the field theory at $y$. Similarly, we say that the boundary metric is weakly curved if $b(y)^2 R(y) \ll 1$ (where $R(y)$ is the curvature scalar, or more generally an estimate of the largest curvature scale in the problem.)

An important issue to address is the shape of the radial curves that tubes - extending from local patches of uniform black brane on the boundary into the bulk - will have. A seeming natural choice is to let tubes follow lines $x^\mu= \rm{const.}$ in the Schwarzschild coordinates, writing the approximate bulk metric as
\begin{equation} 
\label{wronguess}
\mathrm{d}s^2=\frac{\left( \mathrm{d}\tilde{r} + \tilde{r} \mathcal A_\nu \mathrm{d}x^\nu \right)^2}
{\tilde{r}^2 f(r) } +\tilde{r}^2 \left( -f(\tilde{b}\tilde{r}) 
\tilde{u}_\mu \tilde{u}_\nu \mathrm{d}x^\mu \mathrm{d}x^\nu 
+ {\tilde{\mathcal P}}_{\mu\nu} \mathrm{d}x^\mu \mathrm{d}x^\nu \right) 
\end{equation}
The problem with this choice is that it has been shown that it leads, in general, to a geometry with non-regular future horizon\footnote{Examples of metrics of the form (\ref{wronguess}) have been discussed in  \cite{Chamblin:1999by,Benincasa:2007tp,Buchel:2008kd}.}. Furthermore, the solution (\ref{wronguess}) has a relevant issue with causality. Let us consider a variation in the boundary metric at an arbitrary point $y^\mu$, inducing an effective force on the fluid. The future evolution of the fluid is affected only in the  the future boundary light cone of $y^\mu$, that we will denote $C(y^\mu)$\footnote{To be precise this would be strictly true only summing all orders in the fluid expansion. Truncation at any finite order could lead to apparent violations of causality over length scales of order $1/T$.}. The bulk will be affected by the variation only in the region of the space-time - $B(y^\mu))$ - that is the union of all the tubes, that originate in the boundary region $C(y^\mu)$. The natural requirement for bulk causality is that  $B(y^\mu)$ must lie entirely within the future bulk light cone of $y^\mu$. This is not the case in the geometry described by metric (\ref{wronguess}), where the tubes run along lines of constant $x^\mu$ in Schwarzschild coordinates.\\

\begin{figure}
\begin{center}
\includegraphics[scale=0.75]{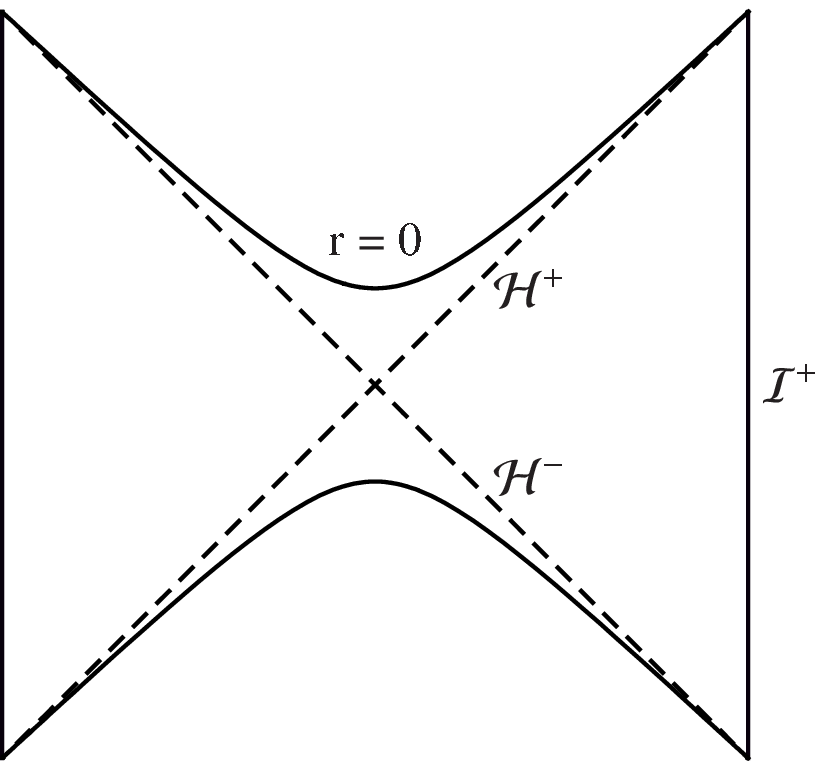}
\hspace{1.5cm}
\includegraphics[scale=0.32]{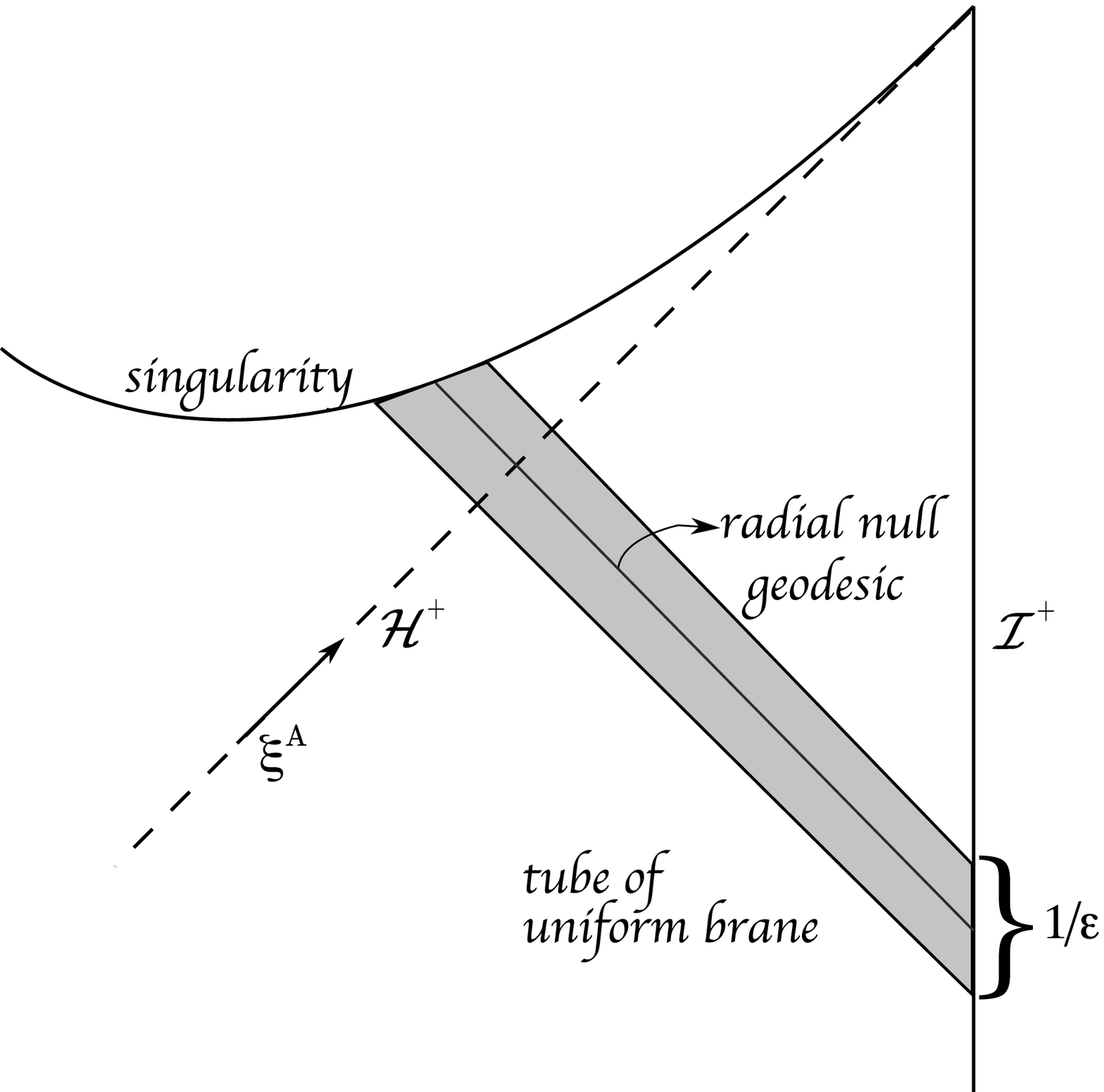} 
\end{center}
\caption{(a) Penrose diagram of the uniform black brane where the dashed lines denote the future event horizons. (b) The shaded tube extending from a local patches of uniform black brane on the boundary into the bulk, and running along ingoing null geodesics, indicates the region of spacetime over which the solution (\ref{guess}) is well approximated by a tube of the uniform black brane. The two figures are taken from Ref.~\cite{Rangamani:2009xk}.}
\label{tubes}
\end{figure}

The appropriate choice to deal with issues of causality turns out to be to promote the parameters of a uniform brane solution, rewritten in Eddington Finkelstein coordinates, to slowly varying boundary functions $u_\mu(x)$ and $b(x)$. In this way one gets the {\it ansatz} for the metric
\begin{equation}
\label{guess}
\mathrm{d}s^2 = -2 u_\mu \mathrm{d}x^\mu \left( \mathrm{d}r + r\ \mathcal{A}_\nu \mathrm{d}x^\nu \right) + r^2 g_{\mu\nu} \mathrm{d}x^\mu \mathrm{d}x^\nu+\frac{r^2}{(br)^d} u_\mu u_\nu \mathrm{d}x^\mu \mathrm{d}x^\nu
\end{equation}
where as before
\begin{equation}
\tilde{g}_{\mu\nu}=e^{2\phi} g_{\mu\nu}(x), \quad \tilde{u}_\mu= 
e^{\phi} u_\mu(x), \quad \tilde{b} = e^{\phi} b(x), 
\end{equation}
and $g_{\mu\nu}(x)$ is a weakly curved boundary metric. The reason for switching to Eddington-Finkelstein coordinates, apart from making issues of regularity more transparent, is that for this choice  tubes are chosen to run along ingoing null geodesics and therefore $B(y^\mu)$ lies entirely within the future bulk light cone of the generic point $y^\mu$ (fig. \ref{tubes}). 

An important observation is that for $g_{\mu\nu}$, $u_\mu$ and $b$ all constants the metrics (\ref{wronguess}) and (\ref{guess}) reduce to equivalent - under a coordinate transformation - descriptions of a uniform brane solution. On the contrary, when $g_{\mu\nu}$, $u_\mu$ and $b$ are functions of the boundaries coordinates $x^\mu$, the geometry described by (\ref{wronguess}) and (\ref{guess}) are inequivalent and can differ qualitatively. Indeed, under mild assumptions, the metric (\ref{guess}) presents a regular future horizon that shields all of the boundary from all future singularities in this space.\\

To summarise the general picture, a given boundary patch corresponds to an entire tube of width set by the scale of variation in the boundary. In the  Eddington-Finkelstein coordinates one has to patch together these tubes to obtain a regular solution to Einstein's equations and moreover this patching can be done order by order in boundary derivatives, just as in fluid dynamics. Therefore the metric (\ref{guess}) will be the first term in a systematic perturbative expansion of a regular solution to Einstein's equations. The perturbative expansion parameter is $1/b \delta x$. The curvature scale in the metric will be assumed to be of the same order as $1/\delta x$.

\section{The perturbative expansion in gravity}

The logic of the perturbative expansion of the metric has been described in detail in \cite{Bhattacharyya:2008jc}. Let us review the reasoning starting with the {\it ansatz} for the bulk metric
\begin{equation}
G_{MN} = G_{MN}^{(0)}+G_{MN}^{(1)}\varepsilon+G_{MN}^{(2)}\varepsilon^2+{\mathcal O}\left(\varepsilon^3\right) \ ,
\label{pertansatz}
\end{equation}
where $\varepsilon$ is the parameter of the derivative expansion and $G_{MN}^{(k)}$ is the correction to the bulk metric at order $k$ that is to be determined with the help of the bulk Einstein equation. It will be shown that perturbative solutions to the gravitation equation  exist only when the velocity and temperature fields obey certain equations of motion that will be determined in a perturbative expansion order by order in $\varepsilon$
\begin{equation}
\beta_i= \beta_i^{(0)}+ \varepsilon \, \beta_i^{(1)}+ {\cal O}\left(\varepsilon^2\right), \qquad b=b^{(0)}+ \varepsilon \, b^{(1)} + {\cal O}\left(\varepsilon^2\right) , 
\label{pertansatz-2}
\end{equation}
where $\beta_i^{(m)}$ and $b^{(n)}$ are all functions of $\varepsilon \,x^\mu$.

In order to give a precise meaning to the coordinates it is necessary to adopt a choice of gauge. In \cite{Bhattacharyya:2008jc} the ``background field'' gauge was adopted, {\it i.e.}
\begin{equation}
G_{rr}= 0 \ , \qquad G_{r \mu }\propto u_\mu \ , \qquad {\rm Tr}\left( (G^{(0)})^{-1} G^{(n)} \right)=0 \qquad  \forall \; n  > 0 . 
\label{gaugecond}
\end{equation}
A more convenient gauge choice was chosen in \cite{Bhattacharyya:2008mz}
\begin{equation}
G_{rr}= 0 \ , \qquad G_{r \mu } =  u_\mu \ . 
 \label{gaugecond2}
 \end{equation}
 The meaning of this choice was discussed in \cite{Bhattacharyya:2008ji}, where it was observed that, under this  gauge fixing, lines of constant $x^\mu$ are ingoing null geodesics, with the radial coordinate $r$ being the affine parameter.
 
Having chosen a gauge, {\it e. g.} the second one, it is possible to substitute the perturbative {\it ansatzs} for the metric (\ref{pertansatz}) and for parameters (\ref{pertansatz-2}) into the bulk Einstein equations (\ref{ein}) and expand it order by order in $\varepsilon$. It is possible to solve the perturbative equation recursively starting form the {\it ansatz} for the order zero metric defined in the previous section. Let us suppose that we have solved the perturbation theory to the
$(n-1)^{{\rm th}}$ order, that is to have determined $G^{(m)}$ for $m\leq n-1$, and have determined the functions $\beta_i^{(m)}$ and $b^{(m)}$ for $m \leq n-2$. The Einstein equation at order $n$ that must be solved can be written as
\begin{equation} 
{\mathbb H}\left[G^{(0)}(\beta^{(0)}_i, b^{(0)})\right] G^{(n)}(x^\mu ) = s_n  ,
\label{pertein}
\end{equation}
where we indicated by ${\mathbb H}$ a linear differential operator of second order in the in the variable $r$ alone. An interesting feature of this operator is its ultralocal nature in the field theory directions. This depends on the fact that $G^{(n)}$ is already of order $\varepsilon^n$, and every boundary derivative appears with an additional power of $\varepsilon$. Indeed, it turns out that ${\mathbb H}$ is a differential operator only in the variable $r$ and is independent of $x^\mu$, that it is independent at order $n$ of the expansion and depends only on the values of $\beta^{(0)}_i$ and $b^{(0)}$ at $x^\mu$. Hence it is possible to solve eq. (\ref{pertein}) point by point on the boundary. What changes order by order is the source term $s_n$ that can be expressed as a local expression of $n^{{\rm th}}$ order in boundary derivatives of $\beta^{(0)}_i$ and $b^{(0)}$, as well as of $(n-k)^{{\rm th}}$ order in $\beta_i^{(k)}$, $b^{(k)}$ for all $k \leq n-1$.

As observed in \cite{Bhattacharyya:2008ji}, it is convenient to organize the $\frac{(d+1)(d+2)}{2}$ equations in (\ref{pertein}) in two classes of equations: $\frac{d(d+1)}{2}$ equations determining the metric, that we call {\em dynamical equations}, and a a second set of $d$ equations which are essentially {\em constraint equations}.\\

The {\em constraint equations} are those of first order in $r$ derivatives and are obtained  by contracting the equations with the one-form ${\mathrm d}r$
\begin{equation}
E^{(c)}_M = E_{MN} \, \xi^N
\end{equation}	
where for our considerations $\xi_N = dr$. Four of the five constraint equations, those with a free index $\mu$ (that is with with legs along the boundary direction), have a simple interpretation: they are the equations of boundary energy momentum conservation:
\begin{equation} 
\nabla_\mu T_{(n-1)}^ {\;\mu \nu}=0 \ ,
\label{eqst}
\end{equation}
where  $T_{(n-1)}^ {\;\mu \nu} $ is the boundary stress tensor dual to the solution expanded up to ${\mathcal O}\left(\varepsilon^{n-1}\right)$. Recalling that all $G_{MN}^{(k)}$ are local functions of $\beta_i$ and $b$, it is clear that  $T_{(n-1)}^ {\;\mu \nu} $ is also a local function (with at most $n-1$ derivatives) that respects all boundary symmetries.

The constraint equations can be used to determine $b^{(n-1)}$ and $\beta_i^{(n-1)}$, this amounts to solving the equations of fluid dynamics at $(n-1)^{\rm th}$ order (\ref{eqst}). There is a non-uniqueness in these solutions given by the zero modes obtained by  linearizing the equations of stress energy conservation at zeroth order. These may be absorbed into a redefinition of the  $\beta^{(0)}_i, b^{(0)}$, and do not correspond to a physical non-uniqueness. \\

The remaining constraints $E_{rr}$ and $E_{\mu \nu}$ are dynamical equations useful to determine the correction that should be added to our initial metric to make it a solution of Einstein equations. By exploiting the underlying symmetries of the zeroth order solution, specifically the rotational symmetry in the spatial sections on the boundary, $SO(d-1)$, it is possible to decouple the system of equations into a set of first order differential operators. Performing this diagonalization of the system of equations one gets a formal solution of the form:
\begin{equation}
G^{(n)} = {\rm particular}(s_n) + {\rm homogeneous}({\mathbb H}) .
\end{equation}	

At this point it is necessary to impose boundary conditions to determine a solution uniquely. A possible choice is to require that the solution is normalizable such that the spacetime is asymptotically AdS$_{d+1}$ and also to demand regularity at all $r \neq 0$. In \cite{Bhattacharyya:2008jc} it was proven that it is possible to require that $G^{(n)}$ is appropriately normalizable  at $r=\infty$  and non-singular at all nonzero $r$, for a generic non-singular and normalizable source $s_n$\footnote{These requirements do not yet completely specify the solution for $G_{MN}^{(n)}$, since H posses a set of zero modes that satisfy both these requirements. A basis for these zero modes is obtained by differentiating the $d$ parameters class of solutions (\ref{blackbrane}) with respect to parameters $\beta_i$ and $b$. These zero modes correspond to infinitesimal shifts of $\beta^{(0)}_1$ and $b^{(0)}$ and therefore may be absorbed by redefining these quantities \cite{Bhattacharyya:2008jc}.}. Furthermore, if the solution at order $n-1$ is non-singular at all nonzero $r$, it is guaranteed to produce a non-singular source at all nonzero $r$. This is particularly important since implies that the non-singular $s_n$ can be inductively constructed.

\section{Outline of the first order computation}

Having described the recursive procedure for constructing a perturbative solution of Einstein equations (\ref{ein}), in this section we can give an outline of the derivation of the metric at first order. 

The zeroth order {\it ansatz} for the metric, $G_{MN}^{(0)}$, has been formulated in (\ref{guess}). The equations that determine $G_{MN}^{(1)}$ at $x^\mu$ are ultralocal, consequently it is sensible to solve the problem point by point. We can pick a point on the boundary $x^\mu = x^\mu_0$, which by exploiting the Killing symmetries of the background can be chosen to be the origin. At $x^\mu_0$ we can use the local scaling symmetry to set $b^{(0)} = 1$ and pass to a local inertial frame so that $\beta^{(0)}_i=0$. 

Making these choices, the metric (\ref{guess}) can be expanded to the first order in the neighbourhood of the origin 
\begin{multline}
{\mathrm d}s_{(0)}^2 = 2\, {\mathrm d}v \, {\mathrm d}r -r^2\, f(r)\, {\mathrm d}v^2 + r^2\, {\mathrm d}x_i \, {\mathrm d}x^i - 2\, x^\mu \,\partial_\mu \beta^{(0)}_i \, {\mathrm d}x^i \, {\mathrm d}r \\
 - 2\, x^\mu \,\partial_\mu \beta^{(0)}_i \, r^2\, (1-f(r))\, {\mathrm d}x^i \, {\mathrm d}v  - d \,\frac{ x^\mu\,\partial_\mu b^{(0)}}{r^{d-2}}\, {\mathrm d}v^2 \  .
\label{expmetric}
\end{multline}	
The metric (\ref{expmetric}) together with $G_{MN}^{(0)}$ has a background part (the first three terms) which is simply the metric of a uniform black brane, plus a correction consisting of first derivatives some of which are known (the last four terms in (\ref{expmetric})) and the reminder of which are to be determined.\\

We now need to make an ansatz for the metric correction at ${\mathcal O}\left({\varepsilon}\right)$, $G^{(1)}_{MN}$, which we wish to determine. As was pointed out in the previous section one can exploit the $SO(d-1)$ spatial rotation symmetry at $x^\mu_0$ to decompose modes into various representations of this symmetry. Modes of $G^{(1)}_{MN}$ transforming under different representations decouple from each other by symmetry. We have  the following decomposition into $SO(d-1)$ irreducible  representations:
\begin{align}
& \text{scalars:} \qquad  G^{(1)}_{vv},  G^{(1)}_{vr}, \sum_i\,  G^{(1)}_{ii},  \nonumber \\
&\text{vectors:}  \qquad  G^{(1)}_{vi}, \nonumber \\
&\text{tensors:} \qquad  G^{(1)}_{ij} .
\end{align}
Einstein's equations can be solved sector by sector. In the scalar sector, one finds that the constraint equations imply that 
\begin{equation}
\frac{1}{d-1}\, \partial_i\beta^{(0)}_i =  \partial_v b^{(0)}  \ , 
\label{scalarcons}
\end{equation}	
 while in the vector sector we have
\begin{equation}
 \partial_i b^{(0)} =  \partial_v \beta^{(0)}_i \ .
\label{veccons}
\end{equation}	
These two equations are equations of energy momentum conservation. 

The differential operators entering in the dynamical equations for the function ${\mathcal F}$ defining the terms  $G^{(1)}_{MN}$ are of the form
\begin{eqnarray}
\text{vector}:&& {\mathbb H}_{{\bf d-1}} {\mathcal F} = \frac{d}{dr}\,\left(\frac{1}{r^{d-1}}\, \frac{d}{dr}\,{\cal O}\right) \nonumber \\
\text{tensor}:&& {\mathbb H}_{{\bf \frac{d(d+1)}{2}}}  {\mathcal F} = \frac{d}{dr}\,\left(r^{d+1}\, f(r)\, \frac{d}{dr}\,{\cal O}\right) 
\end{eqnarray}	
where the index of the operator ${\mathbb H}$ is the representation label of the $SO(d-1)$ rotational symmetry. The scalar sector involves some mixing between different fields and is slightly more involved.  As expected the $\mathbb H$ operators are simple differential operators in the radial variable alone  (the form of the differential operator remains invariant in the course of the perturbation expansion) and can be inverted to find the function ${\mathcal F}$ once the source $s_n$ is specified. 

Extending this procedure, the calculation can in principle be carried out to any desired order in the $\varepsilon$ expansion. Clearly it is necessary to compute at any given order the source terms $s_n$. In addition one always has to ensure that the lower order stress tensor conservation equations are satisfied. For instance, in order for the source terms which appear in the determination of $G^{(2)}_{MN}$  to be ultra-local  at our chosen boundary point $x^\mu_0$, we have to ensure that the first order fluid equations of motion are satisfied. 

\section{Outline of the second order computation}

Once the procedure has been implemented at first order, it is then possible to move on to find a solution to Einstein's equation at the second order. Before proceeding, it is necessary to pause for a moment to notice that the first order computation was performed as a Taylor expansion in the neighbourhood of a special point $x^\mu$ that we chose as the origin. This was enough to obtain $G^{(1)}_{MN}$, while in order to solve the second order problem it is necessary to ensure oneself that the first order constraint is solved to the second order in Taylor expansion of the fields $b^{(0)}$ and $\beta^{(0)}_i$ about the $x_\mu$ that we chose. Therefore the proper requirement is
\begin{equation}
\label{addconstraint}
\partial_\lambda \partial_\mu T^{\mu \nu}_{(0)}(x^\alpha) = 0 \ , 
\end{equation}
this is equivalent to requiring that $T^{\mu \nu}_{(0)}$ satisfies the conservation equation to the order $\varepsilon^2$ before attempting to find the second order stress tensor. In general the right procedure would have been to satisfy the conservation equation globally, however the ultralocality of the set-up allows one to have less stringent requirements, that amounting to checking that conservation holds only at the order at which we are working.\footnote{A good way to rephrase this requirement is to say that before proceed to the second order expansion it is necessary to control that the background defined with the first order calculation solve the first order fluid dynamics.}\\

Once we have checked the requirement (\ref{addconstraint}), it is possible to reexpress Einstein equations schematically as
\begin{equation} 
{\mathbb H}\left[G^{(0)}(\beta^{(0)}_i, b^{(0)})\right] G^{(2)}(x^\mu ) = s_a + s_b  ,
\end{equation} 
where for convenience the source term was written as a sum of the pieces. $s_a$ is a local functional of $\beta^{(0)}_i$ and $b^{(0)}$ up to the second order in field theory derivatives. Terms entering in $s_a$ originated from two field theory derivatives acting on the metric $G^{(0)}_{MN}$ or one field theory derivative acting on $G^{(1)}_{MN}$.  $s_b$ arises from first order derivatives of the velocity and temperature corrections $\beta^{(1)}_i$ and $b^{(1)}$. The second part of the source term is new with respect to the first order computation. \\

Having defined the new source terms, the second order step of the perturbative calculation proceeds similarly to the first order computation\footnote{A detailed derivation of the results for $d=4,5$ at first and second order can be found in the original work  \cite{Bhattacharyya:2008jc}}.

\section{Weyl covariance of the metric}

An important aspect of the bulk metric dual to fluid dynamics is that it has to transform covariantly under the Weyl transformation \cite{Bhattacharyya:2008ji}. Under the gauge choice (\ref{gaugecond2}) the bulk metric can be cast in the general form
\begin{equation}
\label{generalform}
{\mathrm d}s^2 = -2 u_\mu(x) {\mathrm d}x^\mu({\mathrm d}r+\mathcal{V}_\nu(r,x) {\mathrm d}x^\nu)+\mathfrak{G}_{\mu\nu}(r,x){\mathrm d}x^\mu {\mathrm d}x^\nu  
\end{equation}
where  $\mathfrak{G}_{\mu\nu}$ is transverse to the velocity field $u^\mu$, {\it i.e.},  
\begin{equation}
u^\mu \mathfrak{G}_{\mu\nu}=0 . 
\end{equation}
All the Greek indices are raised and lowered using the boundary metric $g_{\mu\nu}$ defined by
\begin{equation}\label{eq:bndmet}
\begin{split}
g_{\mu\nu} = \lim_{r\rightarrow \infty} r^{-2} \left[\mathfrak{G}_{\mu\nu}-  u_{(\mu} \mathcal{V}_{\nu)} \right]
\end{split}
\end{equation}
and $u_\mu$ is the unit time-like velocity field in the boundary, i.e., $g^{\mu\nu}u_\mu u_\nu = -1 $.

Let us consider a bulk diffeomorphism of the form $r=e^{-\phi}\tilde{r}$ supplemented by a scaling in the temperature $b=e^{\phi}\tilde{b}$, where $\phi=\phi(x)$ is defined as a function of the boundary co-ordinates. Components of the metric scale as 
\begin{align}
& u_\mu = e^{\phi}\tilde{u}_\mu, \\ 
&\mathcal{V}_{\mu} = e^{-\phi}\left[\tilde{\mathcal{V}}_{\mu}+\tilde{r}\ \partial_\mu\phi \right],\\
&\mathfrak{G}_{\mu\nu} =\tilde{\mathfrak{G}}_{\mu\nu} \quad \text{and}\quad (\mathfrak{G}^{-1})^{\mu\nu}=(\tilde{\mathfrak{G}}^{-1})^{\mu\nu}\\
&dr +\mathcal{V}_\nu dx^\nu = e^{-\phi} (d\tilde{r}+\tilde{\mathcal{V}}_\nu dx^\nu) .
\label{Weylscaling}
\end{align}

Therefore we conclude that consistency demands that  $\mathcal{V}_{\mu}$ shall transforms like a connection under a Weyl transformation while ${\mathfrak{G}}_{\mu\nu}$ remains invariant. These properties of the metric components yield a prescription for constructing these two objects as appropriate sums of objects with the right transformation properties under Weyl scaling. Therfore ${\mathfrak{G}}_{\mu\nu}$ will be a linear sum of Weyl invariant forms, while $\mathcal{V}_\mu-r\mathcal{A}_\mu$ will be a linear sum of Weyl-covariant vectors with weight unity. The form of the coefficients is not determined by symmetry requirements and have to be fixed by direct calculation.

\section{The metric dual to hydrodynamics}\label{defmetric}

Following the procedure described in the previous sections up to the second order, one find the metric dual to the hydrodynamics on the boundary. In this section we report the results of the calculation that yield, up to second order, a map identifying fluid solutions in the boundary field theory with black brane solutions in the  bulk geometry. 

The metric can be cast in different forms, using different conventions. For the sake of completeness we give in the following a brief overview of some of them. 

A very neat and explicit form is
\begin{align}
ds^2 = &-2 u_\mu dx^\mu \left( dr + r\ A_\nu dx^\nu \right) + \left[ r^2 g_{\mu\nu} +u_{(\mu}\mathcal{S}_{\nu)\lambda}u^\lambda -\omega_{\mu}{}^{\lambda}\omega_{\lambda\nu}\right]dx^\mu dx^\nu \nonumber\\
&+\frac{1}{(br)^d}(r^2-\frac{1}{2}\omega_{\alpha \beta}\omega^{\alpha \beta}) u_\mu u_\nu dx^\mu dx^\nu\nonumber\\
&+2(br)^2 F(br)\left[\frac{1}{b}  \sigma_{\mu\nu} +  F(br)\sigma_{\mu}{}^{\lambda}\sigma_{\lambda \nu}\right]dx^\mu dx^\nu \nonumber\\
&-2(br)^2 \left[K_1(br)\frac{\sigma_{\alpha \beta}\sigma^{\alpha \beta}}{d-1}P_{\mu\nu} + K_2(br)\frac{u_\mu u_\nu}{(br)^{d}}\frac{\sigma_{\alpha \beta}\sigma^{\alpha \beta}}{2(d-1)} \right. \nonumber \\
&\left.  -\frac{L(br)}{(br)^{d}}  u_{(\mu}P_{\nu)}^{\lambda}\mathcal{D}_{\alpha}{\sigma^{\alpha}}_{\lambda}\right] dx^\mu dx^\nu \nonumber\\
&-2(br)^2 H_1(br)\left[u^{\lambda}\mathcal{D}_{\lambda}\sigma_{\mu \nu}+\sigma_{\mu}{}^{\lambda}\sigma_{\lambda \nu} -\frac{\sigma_{\alpha \beta}\sigma^{\alpha \beta}}{d-1}P_{\mu \nu} + C_{\mu\alpha\nu\beta}u^\alpha u^\beta \right]dx^\mu dx^\nu \nonumber \\
&+2(br)^{2} H_2(br)\left[u^{\lambda}\mathcal{D}_{\lambda}\sigma_{\mu \nu}+\omega_{\mu}{}^{\lambda}\sigma_{\lambda \nu}+\omega_\nu{}^\lambda \sigma_{\mu\lambda}\right]  dx^{\mu} dx^{\nu} \ ,
\label{metric-1}
\end{align}
Where the functions appearing in the metric are defined by 
\begin{align}
F(br) &\equiv \int_{br}^{\infty}\frac{y^{d-1}-1}{y(y^{d}-1)}dy ,\\
H_1(br)&\equiv \int_{br}^{\infty}\frac{y^{d-2}-1}{y(y^{d}-1)}dy ,\\
H_2(br)&\equiv \int_{br}^{\infty}\frac{d\xi}{\xi(\xi^d-1)}
\int_{1}^{\xi}y^{d-3}dy \left[1+(d-1)y F(y) +2 y^{2} F'(y) \right] \\
&=\frac{1}{2} F(br)^2-\int_{br}^{\infty}\frac{d\xi}{\xi(\xi^d-1)} \int_{1}^{\xi}\frac{y^{d-2}-1}{y(y^{d}-1)}dy , \\
K_1(br) &\equiv \int_{br}^{\infty}\frac{d\xi}{\xi^2}\int_{\xi}^{\infty}dy\ y^2 F'(y)^2 , 
\end{align}
\begin{align}
K_2(br) &\equiv \int_{br}^{\infty}\frac{d\xi}{\xi^2}\left[1-\xi(\xi-1)F'(\xi) -2(d-1)\xi^{d-1} \right. \nonumber \\
&\left. \quad +\left(2(d-1)\xi^d-(d-2)\right)\int_{\xi}^{\infty}dy\ y^2 F'(y)^2 \right] ,\\
L(br) &\equiv \int_{br}^\infty\xi^{d-1}d\xi\int_{\xi}^\infty dy\ \frac{y-1}{y^3(y^d-1)} \ .
\end{align}
The integral functions have the following asymptotics \cite{Bhattacharyya:2008mz}:
\begin{align}
F(br) & \approx  \frac{1}{br} -\frac{1}{d(br)^d}+ \frac{1}{(d+1)(br)^{d+1}}+\frac{\#}{(br)^{2d}}+\ldots,\\
H_1(br) &\approx \frac{1}{2(br)^2}-\frac{1}{d(br)^d}+ \frac{1}{(d+2)(br)^{d+2}}+\frac{\#}{(br)^{2d}}+\ldots ,\\
H_2(br)&\approx \frac{1}{2(br)^2}-\frac{1}{d(br)^d}\int_{1}^{\infty}\frac{y^{d-2}-1}{y(y^{d}-1)}dy ,\\
K_1(br) &\approx \frac{1}{2(br)^2}-\frac{2}{d(d+1)(br)^{d+1}}\nonumber \\
&\qquad +\frac{2}{(d+1)(d+2)(br)^{d+2}}+\frac{\#}{(br)^{2d}}+\ldots ,\\
K_2(br) &\approx -\frac{(d-3)(d-1)}{2(d+1)(br)^2}+\frac{2(d-2)}{d(br)}\nonumber \\
&\qquad +\frac{1}{d(2d-1)(br)^{d}}+\frac{\#}{(br)^{d+2}}+\ldots ,\\
L(br) &\approx -\frac{1}{d(d+2)(br)^2}+\frac{1}{(d+1)(br)} \nonumber\\
&\qquad-\frac{1}{(d+1)(2d+1)(br)^{d+1}} \nonumber\\
&\qquad -\frac{1}{2(d+1)(d+2)(br)^{d+2}} +\frac{\#}{(br)^{2d}}+\ldots .
\end{align}

The metric (\ref{metric-1}) can be formulated in an equivalent form using the equality
\begin{equation}\label{Sexp:eq}
\mathcal{S}_{\mu\lambda}u^\lambda=-\frac{1}{d-2}\mathcal{D}_\lambda \omega^{\lambda}{}_{\mu}+\frac{1}{d-2}\mathcal{D}_\lambda \sigma^{\lambda}{}_{\mu} -\frac{\mathcal{R}}{2(d-1)(d-2)}u_{\mu}+\ldots \ .
\end{equation}
The resulting expression for the metric is
\begin{align}
ds^2&=-2 u_\mu dx^\mu \left( dr + r\ A_\nu dx^\nu \right) + r^2 g_{\mu\nu} dx^\mu dx^\nu \nonumber \\
&-\left[\omega_{\mu}{}^{\lambda}\omega_{\lambda\nu}+\frac{1}{d-2}\mathcal{D}_\lambda \omega^{\lambda}{}_{(\mu}u_{\nu)}-\frac{1}{d-2}\mathcal{D}_\lambda \sigma^{\lambda}{}_{(\mu}u_{\nu)} \right. \nonumber\\
&\left. +\frac{\mathcal{R}}{(d-1)(d-2)}u_{\mu}u_{\nu}\right] dx^\mu dx^\nu 
+\frac{1}{(br)^d}(r^2-\frac{1}{2}\omega_{\alpha \beta}\omega^{\alpha \beta}) u_\mu u_\nu dx^\mu dx^\nu\nonumber \\
&+2(br)^2 F(br)\left[\frac{1}{b}  \sigma_{\mu\nu} +  F(br)\sigma_{\mu}{}^{\lambda}\sigma_{\lambda \nu}\right]dx^\mu dx^\nu 
-2(br)^2 \left[K_1(br)\frac{\sigma_{\alpha \beta}\sigma^{\alpha \beta}}{d-1}P_{\mu\nu} \right. \nonumber\\
&\left. + K_2(br)\frac{u_\mu u_\nu}{(br)^{d}}\frac{\sigma_{\alpha \beta}\sigma^{\alpha \beta}}{2(d-1)}-\frac{L(br)}{(br)^{d}}  u_{(\mu}P_{\nu)}^{\lambda}\mathcal{D}_{\alpha}{\sigma^{\alpha}}_{\lambda}\right] dx^\mu dx^\nu \nonumber\\
&-2(br)^2 H_1(br)\left[u^{\lambda}\mathcal{D}_{\lambda}\sigma_{\mu \nu}+\sigma_{\mu}{}^{\lambda}\sigma_{\lambda \nu} -\frac{\sigma_{\alpha \beta}\sigma^{\alpha \beta}}{d-1}P_{\mu \nu} + C_{\mu\alpha\nu\beta}u^\alpha u^\beta \right]dx^\mu dx^\nu \nonumber\\
&+2(br)^{2} H_2(br)\left[u^{\lambda}\mathcal{D}_{\lambda}\sigma_{\mu \nu}+\omega_{\mu}{}^{\lambda}\sigma_{\lambda \nu}-\sigma_{\mu}{}^{\lambda}\omega_{\lambda \nu}\right]  dx^{\mu} dx^{\nu} \ .
\label{metric-2}
\end{align}

We will instead use the much more compact form (\ref{generalform}) that renders explicit the Weyl symmetries, adopting the Weyl covariant formalism - with a slight redefinition of the integral functions - that is
\begin{align} 
ds^2 &=G_{MN} \,dX^M \,dX^N  \nonumber \\
&= - 2 \,   u_\mu(x) \, dx^\mu \,\left( dr + {\mathcal V}_\nu(r,x)\,\,dx^\nu\right)+ {\mathfrak G}_{\mu \nu}(r,x) \, dx^\mu\, dx^\nu \ , 
\label{metric-3}
\end{align}
where the fields ${\mathcal V}_\mu$ and ${\mathfrak G}_{\mu\nu}$ are functions of $r$ and $x^\mu$ which admit an expansion in the boundary derivatives
\begin{align}
{\mathcal V}_\mu & = r\, {\cal A}_\mu - {\cal S}_{\mu\lambda}\,u^\lambda - {\mathfrak v}_1(b\,r)\, P^{\; \nu}_\mu\, { \cal D}_\lambda\sigma^{\lambda}_{\ \nu} +u_\mu \, \left[\frac{1}{2}\, r^2 \, f( b\,r) \right. \nonumber \\
&\qquad  \left.+ \frac{1}{4}\, \left(1-f(b\,r)\right) \, \omega_{\alpha\beta}\, \omega^{\alpha\beta} + {\mathfrak v}_2(b\,r) \, \frac{\sigma_{\alpha\beta}\, \sigma^{\alpha\beta}}{d-1}\right] \ , \\
{\mathfrak G}_{\mu\nu} &= r^2\, P_{\mu\nu}  - \omega_{\mu}^{\ \lambda}\,\omega_{\lambda\nu}+ 2\, (b\,r)^2\, {\mathfrak g}_1(b\,r)\, \left[\frac{1}{b}\, \sigma_{\mu\nu} + {\mathfrak g}_1(b\,r) \, \sigma_\mu^{\ \lambda}\,\sigma_{\lambda\nu} \right] \nonumber\\
&\qquad \,- {\mathfrak g}_2(b\,r)  \,\frac{\sigma_{\alpha\beta}\, \sigma^{\alpha\beta}}{d-1} \, P_{\mu\nu} - {\mathfrak g}_3(b\,r) \, \left[{\mathfrak T}_{1\mu\nu} + \frac{1}{2}\, {\mathfrak T}_{3\mu\nu}   + 2\,{\mathfrak T}_{2\mu\nu} \right]\nonumber\\
& \qquad \,+ {\mathfrak g}_4(b\,r)\,  \left[{\mathfrak T}_{1\mu\nu} +  {\mathfrak T}_{4\mu\nu}   \right] .
\label{metricfunctions}
\end{align}
The tensor ${\mathfrak G}_{\mu\nu}$ is transverse, since it is built out of operators that are orthogonal to the velocity, and it can be inverted via the relation 
$\left({\mathfrak G}^{-1}\right)^{\mu\alpha} \, {\mathfrak G}_{\alpha\nu}  = P^\mu_{\;\nu}$. The new integral functions are defined in terms of the old ones as
\begin{align}
{\mathfrak g}_1(br) &= F(br) , \qquad \qquad \, \, \, \, \, \,{\mathfrak g}_2(br) = 2\,(rb)^2 K_1(br), \nonumber \\
{\mathfrak g}_3(br) & = (br)^2 H_1(br) , \qquad \, \, \, {\mathfrak g}_4(br) = (br)^2 H_2(br) , \nonumber \\
{\mathfrak v}_1(br) &=\frac{2}{(br)^{d-2}} L(br) , \qquad{\mathfrak v}_2(br) = \frac{1}{2\, (br)^{d-2}} K_2(br) .
\end{align}
Finally, we recall the following definitions (that are reported in table 2.1)
\begin{align}
{\mathfrak T}_1^{\mu\nu} &=2\, u^\alpha \, { \cal D}_\alpha \sigma_{\mu\nu} \ , \nonumber \\
{\mathfrak T}_2^{\mu\nu} &=C_{\mu\alpha\nu\beta}\,u^\alpha \,u^\beta \ , \nonumber \\
{\mathfrak T}_3^{\mu\nu} &=4\,\sigma^{\alpha\langle\mu}\, \sigma^{\nu\rangle}_{\ \alpha}  \ ,  \nonumber \\
{\mathfrak T}_4^{\mu\nu} &=  2\, \sigma ^{\alpha\langle\mu}\, \omega^{\nu\rangle}_{\ \alpha}  \ ,  \nonumber \\
{\mathfrak T}_5^{\mu\nu} &=   \omega ^{\alpha\langle\mu}\, \omega ^{\nu\rangle}_{\ \alpha} \ .
\end{align}

\section{The boundary stress tensor}

The dual stress tensor corresponding to the metric at the second order presented in the previous section can be determined quite straightforwardly using the holographic prescription in \cite{Henningson:1998gx,Balasubramanian:1999re}. To perform the computation one has to define a regularisation  of the asymptotically AdS$_{d+1}$ spacetime at some cut-off hypersurface $r  = \Lambda_\text{c}$ and consider the induced metric on this surface, which up to a scale factor involving $\Lambda_{\text{c}}$ is our boundary metric $g_{\mu\nu}$. The holographic stress tensor is given in terms of the extrinsic curvature $K_{\mu\nu}$ and metric data of this  cut-off hypersurface
\begin{equation}
K_{\mu\nu} = g_{\mu\rho}\,\nabla^\rho n_\nu
\end{equation}	
where $n^\mu$ is the outward normal to the surface. Now, for asymptotically AdS$_{5}$ spacetimes the prescription in \cite{Balasubramanian:1999re} is
\begin{equation}
T^{\mu\nu} = \lim_{\Lambda_\text{c} \to \infty}\; \frac{\Lambda_\text{c}^{d-2}}{16\pi \, G_N^{(d+1)}} \, \left[ K^{\mu\nu} - K \, g^{\mu\nu} - (d-1)\, g^{\mu\nu} - \frac{1}{d-2}\,  \left(R^{\mu\nu} -\frac{1}{2}\, R \, g^{\mu\nu}\right)\right] .
\end{equation}	

Implementing this procedure, one is able to determine the dual stress tensor corresponding to the metric in the previous subsection
\begin{align}
T_{\mu\nu} &= p\left(g_{\mu\nu}+d u_\mu u_\nu \right)-2\eta \sigma_{\mu\nu} \nonumber \\
&-2\eta \tau_\omega \left[u^{\lambda}\mathcal{D}_{\lambda}\sigma_{\mu \nu}+\omega_{\mu}{}^{\lambda}\sigma_{\lambda \nu}+\omega_\nu{}^\lambda \sigma_{\mu\lambda} \right] \nonumber \\
&+2\eta b\left[u^{\lambda}\mathcal{D}_{\lambda}\sigma_{\mu \nu}+\sigma_{\mu}{}^{\lambda}\sigma_{\lambda \nu} -\frac{\sigma_{\alpha \beta}\sigma^{\alpha \beta}}{d-1}P_{\mu \nu}+ C_{\mu\alpha\nu\beta}u^\alpha u^\beta \right]\\
\end{align}
where
\begin{align}
b &=\frac{d}{4\pi T} ,  \qquad \qquad \qquad \,\,\,\, p=\frac{1}{16\pi G_{\text{AdS}}b^d}, \\
\eta &=\frac{1}{16\pi G_{\text{AdS}}b^{d-1}} ,\quad
\qquad \tau_{\omega} =  b \int_{1}^{\infty}\frac{y^{d-2}-1}{y(y^{d}-1)}dy .
\end{align}
Substituting the standard entropy density, one gets
\begin{align}
\frac{\eta}{s} = \frac{1}{4\pi} \ ,
\end{align}
the celebrated ratio of shear viscosity to entropy density \cite{Kovtun:2004de}.

\section{Horizon location for the dual metric}

A last point to assess is the regularity of the horizon of the black brane metrics that we have derived\footnote{In this section we will follow the derivation proposed in \cite{Bhattacharyya:2008mz}.}. The event horizon ${\mathcal H}^+$ of a given spacetime is defined as the boundary of the past lightcone of future null infinity. This is a formal statement of the physical fact that the spacetime events inside the event horizon of the black hole cannot communicate to the asymptotic region. The future null infinity ${\mathcal I}^+$ is the set of points which are approached asymptotically by null rays which can escape to infinity and it is time-like for asymptotically AdS spacetimes. Since ${\mathcal H}^+$ is the boundary of a causal set, it is a null surface which is, in particular, generated by null geodesics in the spacetime.\\ 

The event horizon of the spacetimes dual to the hydrodynamics on the boundary is the unique null hypersurface that tends, at late times, to the known event horizons of the late time limit of our solutions. As seen above, the metric we found can written, in the gauge $g_{rr}=0$,  $g_{r\mu}=-u_\mu$, in the form (\ref{metric-3}) that makes explicit the invariance under boundary Weyl transformations. 

Let us suppose that the event horizon is given by the equation 
\begin{equation}
\mathcal{S}\equiv r-r_{_H}(x)=0 ,
\end{equation}
where $r_{_H}$ is the radial position of the horizon. The vector $\xi_A$  normal to the horizon is defined by the one-form
 \begin{equation}
 dS= \xi_A dy^A = dr -\partial_\mu r_H dx^\mu \ ,
 \end{equation}
that can be written in a manifestly Weyl covariant form
 \begin{equation}
 \xi_A dy^A = dS= (dr+\mathcal{V}_\lambda dx^\lambda)-\kappa_\mu dx^\mu \ .
 \end{equation} 
Also its dual normal vector can be written in Weyl covariant form as
\begin{align}
\xi^A\partial_A &= {G}^{AB}\partial_{A}\mathcal{S}\partial_{B}= n^\mu(\partial_\mu-\mathcal{V}_\mu\partial_r)-u^\mu\kappa_\mu\partial_r \nonumber \\
&= n^\mu\left[\partial_\mu+\partial_\mu r_{_H} \partial_r\right] =n^\mu\left[\partial_\mu\right]_{r=r_{_H}} \ .
\end{align}
In the above expression we introduced two new Weyl-covariant vectors $\kappa^\mu=e^{-\phi}\tilde{\kappa}^\mu$ and $n^\mu=e^{-\phi}\tilde{n}^\mu$ defined via the expressions 
\begin{equation}
\begin{split}
\kappa_\mu &\equiv \partial_\mu r_{_H}+\mathcal{V}_{\mu H} \\
n^\mu &\equiv u^\mu-(\mathfrak{G}_H^{-1})^{\mu\nu}\kappa_\nu .\\
\end{split}
\end{equation}
The subscript $H$ indicates that the functions are to be evaluated at the event-horizon.

The induced metric on the horizon $\mathcal{H}_{\mu\nu}(x)$ is defined as
\begin{equation}
ds^2_H = \left[G_{AB}(y) dy^A dy^B\right]_{r=r_{_H}(x)} \equiv \mathcal{H}_{\mu\nu}(x) dx^\mu dx^\nu \ ,
\end{equation}
where the boundary coordinates $x^\mu$ have been used as the coordinates on the event horizon. Therefore we have 
\begin{equation}
\mathcal{H}_{\mu\nu}= \mathfrak{G}_{\mu\nu}-u_{(\mu}\kappa_{\nu)}
\end{equation}
and the null-condition on the horizon, $[G_{AB}]_H \xi^A \xi^B= \mathcal{H}_{\mu\nu}n^\mu n^\nu=0$ corresponds to
\begin{equation}
(\mathfrak{G}^{-1})^{\mu\nu}\kappa_\mu\kappa_\nu = 2 u^\mu\kappa_\mu \ .
\label{nullcond}
\end{equation}

Now it is possible to construct a Weyl-covariant derivative expansion{\footnote{As noted in \cite{Bhattacharyya:2008mz} since there is no first order Weyl-covariant scalar, there are no corrections to $r_{_H}$ at the first order in the derivative expansion. A detailed classification of the possible Weyl-covariant tensors is contained in \cite{Bhattacharyya:2008ji}.}} for $r_{_H}$, {\it i.e.}
\begin{align}
r_{_H} &= \frac{1}{b}+b \left(h_1 \sigma_{\alpha\beta} \sigma^{\alpha\beta} + h_2 \omega_{\alpha\beta} \omega^{\alpha\beta} +h_3 \mathcal{R} \right) +\ldots \nonumber \\
&= r_{_H}^{(0)} + r_{_H}^{(2)}+\ldots
\end{align}

Computing $\kappa_\mu$ one gets
\begin{equation}\label{kappasol:eq}
\begin{split}
\kappa_\mu &= \mathcal{D}_\mu b^{-1} - \mathcal{S}_{\mu\lambda} u^\lambda -2 L_H P_{\mu}^{\nu}\mathcal{D}_{\lambda}{\sigma^{\lambda}}_{\nu}   \\
&\,\,\,\,\,\,\,\,  + u_\mu \left[ \frac{1}{4}\omega_{\alpha \beta}\omega^{\alpha \beta} 
+ \frac{K_{2H}}{2(d-1)}\sigma_{\alpha \beta}\sigma^{\alpha \beta}+\frac{d}{2b}r_{_H}^{(2)}
\right] +\ldots\\
 n^\mu &= u^\mu - b^2 P^{\mu\nu}\left[\mathcal{D}_\nu b^{-1} -\mathcal{S}_{\nu\lambda} u^\lambda -2 L_H \mathcal{D}_{\lambda}{\sigma^{\lambda}}_{\nu}\right]\\
 \sqrt{\text{det}_{d-1}\mathfrak{G}_H} &= \frac{1}{b^{d-1}}\left[1+(d-1)br_{_H}^{(2)}+
 \frac{b^2}{2}\omega_{\alpha \beta}\omega^{\alpha \beta}-b^2 K_{1H} \sigma_{\alpha \beta}\sigma^{\alpha \beta}\right]\\
 b^{d-1} n^\mu \sqrt{\text{det}_{d-1}\mathfrak{G}_H} &=\left[1+(d-1)br_{_H}^{(2)}+
 \frac{b^2}{2}\omega_{\alpha \beta}\omega^{\alpha \beta}-b^2 K_{1H} \sigma_{\alpha \beta}\sigma^{\alpha \beta} \right]u^\mu\\
 &\,\,\,\,\,\,\,\, - b^2 P^{\mu\nu}\left[\mathcal{D}_\nu b^{-1} - \mathcal{S}_{\nu\lambda} u^\lambda -2 L_H \mathcal{D}_{\lambda}{\sigma^{\lambda}}_{\nu}\right]
\end{split}
\end{equation}

Now substituting the expression for $\kappa_\mu$ in \eqref{nullcond}, and after some manipulations, one finds the position of the event horizon as
\begin{equation}
\begin{split}
r_{_H} &= \frac{1}{b}+b \left(h_1 \sigma_{\alpha\beta} \sigma^{\alpha\beta} + h_2 \omega_{\alpha\beta} \omega^{\alpha\beta} +h_3 \mathcal{R} \right) +\ldots\\
\end{split}
\end{equation}
where
\begin{equation}
\begin{split}
h_1&=\frac{2(d^2+d-4)}{d^2(d-1)(d-2)} - \frac{K_{2H}}{d(d-1)}\\
h_2&=-\frac{d+2}{2d(d-2)} \ \qquad \text{and}\quad h_3=-\frac{1}{d(d-1)(d-2)}
\end{split}
\end{equation}
and $K_{2H}$ is
\begin{multline}
 K_{2H} = \int_{1}^{\infty}\frac{d\xi}{\xi^2}\left[1-\xi(\xi-1)F'(\xi) -2(d-1)\xi^{d-1} \right.\\
 \left. \quad +2\left((d-1)\xi^d-(d-2)\right)\int_{\xi}^{\infty}dy\ y^2 F'(y)^2 \right] .
\end{multline}

\section{Boundary entropy current}

Dissipative fluids are characterized by entropy production. On the other side of the duality, for black holes at equilibrium there is a natural definition of entropy associated with the area of the event horizon which is the usual Bekenstein-Hawking entropy of a back hole. Unfortunately, in general, when we consider deviations from equilibrium, there is no unambiguous notion of entropy. Indeed, the substantial requirement for the entropy current is to satisfy the second law
\begin{equation}
\nabla_\mu J^\mu_S \geq 0 \ ,
\end{equation}
and it is possible, in principle, to adopt any local function having positive divergence to characterize the irreversibility of the fluid dynamical flow. The only constraint is that the candidate Boltzmann H-function must agree with the thermodynamic notion of entropy in global equilibrium. However since in the case of stationary black holes, as discussed in the first chapter,  the area of the event horizon is associated the entropy of the dual field theory, it seems quite natural to associate the entropy of the field theory with the area of the event horizon\footnote{In \cite{Rangamani:2009xk} it has been noticed that this point is quite delicate. Indeed, a key feature of the event horizon is its ``teleological nature'', {\it i.e.}, the fact that the entire future evolution of the spacetime is needed in order to determine its location. This may clash with the attempts to determine the horizon perturbatively. Associating an entropy current with the event horizon may lead to a non-local and acausal definition of entropy \cite{Chesler:2008hg}. Moreover in  the case of the conformal soliton flow \cite{Friess:2006kw}, a simple hydrodynamic flow on ${\mathbb R}^{d-1,1}$, it was shown that the event horizon area does not capture the entropy of the dual field theory \cite{Figueras:2009iu}. Therefore under certain one should use the area of apparent horizons (more precisely dynamical horizons), to define the entropy current.}.\\
 
In the previous section the geometry of a regular future event horizon ${\mathcal H}^+$ for the  spacetimes dual to boundary fluid dynamics was derived. It is possible, starting from the entropy associated to the area of this horizon, to define an entropy current directly for the field theory in the boundary. To this end, consider spatial sections of the event horizon, which are co-dimension two surfaces in the spacetime, which we label as ${\mathcal H}^+_v$. We are working in a coordinate chart where the coordinates $\alpha^i$ for $i =\{1,\cdots d-1\}$ define a chart on  the spatial section and we use as an affine parameter, the boundary coordinate  $v$, to propagate these surfaces forward along the horizon generator $\xi^A$. On the surface ${\mathcal H}^+_v$  it is natural to define  a $(d-1)$-form whose integral gives the area of the spatial section. The entropy of the black hole will be - assuming that the null energy condition satisfied - proportional to this area. Using the Bekenstein-Hawking formula, in terms of the area of the event horizon, the entropy density is found to be
\begin{equation}
s = \frac{1}{4\,G_N^{(d+1)}} \,\frac{1}{b^{d-1}} \  , \qquad b =  \frac{d}{4\pi\, T} \ . 
\end{equation}\\

To get the entropy current of the field theory, one has to pull the entropy of the black hole back to the boundary. In \cite{Bhattacharyya:2008xc} the authors proposed that one pull-back the area form on the horizon using radially ingoing null geodesics, which provide an isomorphism between the spatial sections on the boundary  and the corresponding ${\mathcal H}^+_v$ on ${\mathcal H}^+$. The general procedure detailed in \cite{Bhattacharyya:2008xc} yields the boundary entropy current
\begin{eqnarray}
J^\mu_S &=& \frac{\sqrt{\text{det}^{(n)}_{d-1}\mathcal{H}}}{4 G_{AdS}}n^\mu \\
&=& \frac{\sqrt{\text{det}^{(n)}_{d-1}\mathcal{H}}}{4 G_{AdS}}\left[u^\mu-(\mathfrak{G}_H^{-1})^{\mu\nu}\kappa_\nu\right] \ .
\end{eqnarray}
To define $\text{det}^{(n)}_{d-1}\mathcal{H}$ one has to split the boundary co-ordinates $x^\mu$ into $(v,x^i)$. After this split, the components of the $n^\mu$ are further split into $(n^v,n^i)$. We denote the $d-1$ dimensional induced metric on the constant $v$ submanifolds of the event horizon by $\mathfrak{h}_{ij}$. Finally, one can define
\begin{equation}
\sqrt{\text{det}^{(n)}_{d-1}\mathcal{H}} = \frac{\sqrt{\text{det}_{d-1}\mathfrak{h}}}{n^v \sqrt{-\text{det} g}} \ 
\end{equation} 
where $g_{\mu\nu}$ is the boundary metric and the expression on the right hand side has been assumed to be pulled back from the horizon to the boundary via the ingoing null-geodesics. \\

Following this procedure one finds that the entropy current takes the form:
\begin{multline}
J^\mu_S = s\,u^\mu + s\, b^2 \, u^\mu \,\left(A_1 \,\sigma_{\alpha\beta}\,\sigma^{\alpha\beta}+A_2 \,\omega_{\alpha\beta}\,\omega^{\alpha\beta} +A_3 \,\mathcal{R}\,\right)\\
 + s\, b^2 \,\left( B_1 \,{\mathcal D}_\lambda \sigma^{\mu\lambda} + B_2 \,{\mathcal D}_\lambda \omega^{\mu\lambda} \right) + \cdots
\end{multline}
where $s$ is the entropy density and $A_{1,2,3}$, $B_{1,2}$ are arbitrary numerical  coefficients. Requiring positivity of the divergence one finds that two of the coefficients must solve the following linear constraint
\begin{equation}
B_1 + 2\, A_3 = 0 \ . 
\end{equation}	
An interesting result is that the entropy current in $d$ dimensions is a Weyl covariant vector of weight $d$. This is consequence of the fact that the entropy density scales like the inverse spatial volume, since the total entropy is dimensionless, and the velocity field scales according to (\ref{Weylscaling}). Using the geometric data of the space-time described by the metric (\ref{metric-3}) the coefficients are fixed to be
\begin{align}
A_1 &= \frac{2}{d^2}\, (d+2) -\left( \frac{1}{2}\, {\mathfrak g}_2(1) +\frac{ 2}{d}\, {\mathfrak v}_2(1)\right) , \qquad A_2 = -\frac{1}{2\,d} \ ,\nonumber\\
B_1 &= -2 \, A_3 = \frac{2}{d\, (d-2)} \ , \qquad B_2 = \frac{1}{d-2}  \ .
\end{align}	
We conclude this section by observing that, writing the divergence of the gravitational entropy current as
\begin{equation}
{\mathcal D}_\mu J^\mu_S = \frac{2\,\eta}{T} \, \left(\sigma_{\mu\nu} + \frac{1}{2}\, \left[ \frac{d}{4\pi\,T}\, (1+ A_1 \,  d) - \tau_\pi\right]\, u^\alpha\, {\mathcal D}_\alpha\sigma_{\mu\nu}\right)^2 + \cdots  \ ,
\end{equation}	
which is accurate up  to third order in the derivative expansion, it is possible to check that it satisfies the requirement of non-negative divergence at this order.

\section[Non-relativistic solutions]{The gravity solution dual to hydrodynamics in the non-relativistic regime}

In chapter $2$ we discussed the non-relativistic scaling limit ofthe  hydrodynamic equations. In particular it was shown that, under this scaling, the incompressible non-relativistic Navier-Stokes equations are equivalent to the relativistic equations of fluid dynamics dual to gravity up to ${\cal O}(\epsilon^3)$.  In this section we will report some results about the gravitational dual to a solution of the non-relativistic Navier-Stokes equations constructed as a small fluctuation about a black brane background that solves all of Einstein's equations up to order ${\cal O}(\epsilon^3)$ \cite{Bhattacharyya:2008jc}.

In computing the bulk metric up to ${\cal O}(\epsilon^3)$, it turns out that only contributions from the zeroth order and the first order in the derivative expansion of the gravitational solutions of are relevant. Keeping in mind the discussion in section $2.7$, the metric up to first order in derivatives is
\begin{equation}
ds^2  = ds_0^2 + ds_1^2
\end{equation}
where
\begin{eqnarray}
ds_0^2 &=& -2 u_\mu dx^\mu dr + \frac{1}{b^d r^{d-2}}u_\mu u_\nu dx^\mu dx^\nu + r^2 g_{\mu\nu}dx^\mu dx^\nu \nonumber \\
ds_1^2 &=& -2 r u_\nu \left(u^\alpha\bar\nabla_\alpha\right) u_\mu dx^\mu dx^\nu \nonumber\\
&+& \frac{2}{d-1}r \left(\bar\nabla_\alpha u^\alpha\right) u_\mu u_\nu dx^\mu dx^\nu + 2 b r^2 F\left(b r\right) \sigma_{\mu\nu}dx^\mu dx^\nu \nonumber 
\end{eqnarray}
and $\sigma_{\mu \nu}$, $b$, and $T$ can be expanded as
\begin{eqnarray}
\sigma_{\mu\nu} &=& \frac{1}{2}\left(\bar\nabla_\mu u_ \nu + \bar\nabla_\nu u_ \mu\right) \\
&\,&+ \frac{1}{2}\left(u_\nu \left(u^\alpha\bar\nabla_\alpha\right) u_\mu 
+ u_\mu \left(u^\alpha\bar\nabla_\alpha\right) u_\nu\right) - \frac{1}{d-1}\left(\bar\nabla_\alpha u^\alpha\right)\left(u_\mu u_\nu + g_{\mu\nu}\right) \nonumber \\
b &=& \frac{d}{4 \pi T} = b_0 + \delta b\\
T &=& T_0 + \delta T .
\end{eqnarray}
$\bar\nabla$ denotes the covariant derivative with respect to the full boundary metric $G_{\mu \nu}$ which is equal to a background $g_{\mu\nu}$ plus perturbation $H_{\mu\nu}$, while $T_0$ is the temperature of the background black brane. 

Contributions of the metric in which the derivative $\bar\nabla$  appears can be re-written as covariant derivatives of the $d-1$ velocity $v_i$ and the metric perturbation $A_i = H_{0i}$ with respect to the spatial part of the background metric $g_{ij}$, {\it i.e.}
\begin{eqnarray}
\bar\nabla_iu_j &=& \nabla_iv_j + {\cal O}(\epsilon^4)\nonumber \\
\bar\nabla_iu_0 + \bar\nabla_0u_i &=& \partial_0(v_i + A_i) - \frac{1}{2}\partial_i h_{00} - \frac{1}{2}\partial_i(v_jv^j) - v^jF_{ij} + {\cal O}(\epsilon^4)\nonumber \\
\bar\nabla_\mu u^\mu &=& \nabla_jv^j + {\cal O}(\epsilon^4)\nonumber \\
u^\mu\bar\nabla_\mu u_0 &=& {\cal O}(\epsilon^4) \\
u^\mu\bar\nabla_\mu u_i &=& \partial_0(v_i + A_i) - \frac{1}{2}\partial_i h_{00} + (v^j\nabla_j)v_i - v^jF_{ij} + {\cal O}(\epsilon^4)\nonumber \\
F_{ij} &=&  \partial_i A_j - \partial_j A_i . \nonumber
\label{derivates}
\end{eqnarray}
Now $\nabla$ is the covariant derivative with respect to the background metric $g_{ij}$. The raising and lowering of the ${i,j}$ indices is to be performed using the metric $g_{ij}$. To simplify the expression of $\sigma_{\mu\nu}$ in \eqref{derivates} the constraint $\nabla_i v^i = 0$ has been used.

Substituting in the metric, the first order part takes the form
\begin{equation} 
\begin{split}
ds_1^2 &= b_0 r^2 F(b_0 r)\left(\nabla_iv_j + \nabla_j v_i\right)dx^i dx^j - 2 b_0 r^2 F(b_0 r)v^j\left(\nabla_iv_j + \nabla_j v_i\right)dt~dx^i\\
& +2r \left(\partial_0(v_i + A_i) - \frac{1}{2}\partial_i h_{00} - v^j F_{ij} + (v^j\nabla_j)v_i\right)dt~dx^i + {\cal O}(\epsilon^4)
\end{split}
\end{equation}
where the first term is of order $\epsilon^2$ and the last two terms are of order $\epsilon^3$. The constraint equation $\nabla_i v^i = 0$, was used to cancel out contributions from the scalar sector. 

Also the zeroth order metric can be expanded in powers of $\epsilon$. One finds that to solve Einstein equation up to order $\epsilon^3$, it is sufficient to expand the zeroth order metric up to order $\epsilon^2$ in the fluctuations. The result is
\begin{equation} 
\begin{split}
ds_0^2 &= \frac{1}{b_0^d r^{d-2}} dt^2 + r^2\left(-dt^2 + g_{ij} dx^i dx^j\right)  + 2 dt~ dr\\
&- \frac{2}{b_0^dr^{d-2}}\left(A_i + v_i\right)dt~ dx^i - 2\left(A_i + v_i\right)dx^i dr\\
&+\frac{1}{b_0^{d+1}r^{d-2}}\left(-d~\delta b + v_jv^j - h_{00}\right)dt^2 \\
&+ \frac{1}{r^{d-2}}\left(A_i + v_i\right)\left(A_j + v_j\right)dx^idx^j  - \left(-v_jv^j + h_{00}\right)dt~dr\\
\end{split}
\end{equation}
where the first line is of order $\epsilon^0$, the second line is of order $\epsilon^1$ and the third and the fourth are of order $\epsilon^2$.

It is possible to show by a direct calculation that the proposed metric solves Einsteins equations up to the third order in $\epsilon$,  provided that the velocity and temperature fields above obey the incompressible Navier-Stokes equations \eqref{NS} \cite{Bhattacharyya:2008jc}.

An important observation arises from the equation that determines $\delta T$ (and hence $\delta b$)
\begin{equation}
\frac{\nabla^2 T}{T_0} = -\nabla_i v^j \nabla_j v^i - v^i v^j R_{ij}  \
+ \nabla_i \left[ \left( -\nu R^i_j + F^i_j \right) v^j \right] +
\frac{1}{2} \nabla^2 h_{00} - \partial_0 (\nabla. A) \ . \nonumber
\end{equation}
$\delta T$ is a spatially nonlocal but temporally ultralocal functional of the velocity fields $v^i$. Therefore, even though the bulk metric at $x^\mu$ is 
determined locally as a function of temperatures and velocities at $x^\mu$, it is not determined locally as a function of velocities at $x^\mu$. This nonlocality is a consequence of the infinite speed of sound in the scaling limit.

% Third Chapter
%_____________________________________________
\chapter{Matter and Gravity}
%_____________________________________________

In this chapter we will discuss some techniques and issues that are relevant to the study of black holes and singular hypersurfaces in gravity.

In particular we will start by reviewing various possible {\it energy conditions} that enter in black holes physics \cite{Hawking:1973uf, Wald:1984rg}. Then we discuss the issue of the violation of the loosest among these condition, the null energy condition, and its physical consequences \cite{Hsu:2004vr, Dubovsky:2005xd, Buniy:2005vh}.

Later on we will introduce some geometric notions useful in the description of hypersurfaces and in particular the {\it extrinsic curvature}. These concepts will be used in applying Israel's seminal work on singular hypersurfaces \cite{Israel:1966rt} to the configuration that we will present in the next chapter. 

\section{Energy conditions}\label{Energysec}

In general relativity, an energy condition is one of various alternative conditions which can be applied to the matter content to define some of its properties. Since almost every spacetime is a possible solution to Einstein's equations for some particular choice of energy-momentum tensor $T_{\mu \nu}$, it is important to understand what general restrictions can hold on the energy-momentum of a physical system and which are their implications. Indeed, energy conditions play a critical role in the study of black holes and cosmological singularities,  entering in important theorems such as the no hair theorem and the laws of black hole thermodynamics \cite{Hawking:1973uf, Wald:1984rg}. Moreover, even though a violation of an energy condition does not necessarily imply a pathology in a physical system, the violation of the null energy condition has been related to the insurgence of instabilities and superluminal propagation (one can see {\it e.g.} \cite{Hsu:2004vr, Dubovsky:2005xd, Buniy:2005vh}).

Let us start  with the weak energy condition. \\

{\it \textbf{Weak energy condition}: The energy-momentum tensor at each point $p$ on a manifold ${\mathcal M}$, $p \in {\mathcal M}$, obeys the inequality 
\begin{equation}
T_{\mu \nu} w^\mu w^\nu \geq 0 \ ,
\end{equation}
for any timelike vector $w$ in the tangent space at $p$, $w \in T_p$. By continuity this will also be true for any null vector $w \in T_p$.}
\\

The weak energy condition is equivalent to the statement that the energy density of any matter distribution, as measured by any observer in spacetime, must be nonnegative. Indeed, for an observer whose world line at $p$ has unit tangent $V$, the local energy density is $T_{\mu \nu} V^\mu V^\nu$. \\

Following \cite{Hawking:1973uf}, we will explore more in depth the meaning of this assumption using the fact that it is possible to cast the $4$-dimensional stress tensor $T_{\mu \nu}$ in four canonical forms, with respect to an orthonormal basis $\{ \bf{e_0, e_1, e_2, e_3}\}$.

\begin{equation}
 \text{Type\,\, I} \qquad \qquad 
T^{\mu \nu}=
\begin{bmatrix}
\rho & 0 & 0  & 0 \\
0 & p_1 & 0  & 0  \\
0 & 0 & p_2 & 0  \\
0 & 0 & 0  & p_3 
\end{bmatrix}
\end{equation} 
In this case, the stress tensor has a unique timelike eigenvector (unless $\rho= -p_i$, ($i = 1,2,3$)). $\rho$ is the energy-density measured by an observer whose world line has unit tangent vector ${\bf e_0}$ at $p$, while the three eigenvectors $p_\alpha$ are the pressures in the three spacelike direction spanned by ${\bf e_i}$ ($i = 1,2,3$)\footnote{As discussed in chapter 2, if the stress tensor is that of a perfect fluid, then the pressures in the spacelike directions are equal $p_1=p_2=p_3=p$.}. The energy-momentum tensor can be cast in this form for fields with non-zero rest mass and for some zero rest mass fields, except those of the Type II.

 \begin{equation}
\text{Type\,\, II} \qquad \qquad
T^{\mu \nu}=
\begin{bmatrix}
\nu +\kappa & \nu & 0  & 0 \\
\nu & \nu -\kappa & 0  & 0  \\
0 & 0 & p_1 & 0  \\
0 & 0 & 0  & p_2 
\end{bmatrix}
 \qquad  \nu = \pm 1
\end{equation} 
In this case, the stress tensor has a double null eigenvector ($\bf{e_0+e_1}$). The stress tensor has this form in the case of radiation traveling in the direction $\bf{e_0+e_1}$. In this case $p_1$, $p_2$ and $\kappa$ are zero.

 \begin{equation}
\text{Type\,\, III} \qquad \qquad 
T^{\mu \nu}=
\begin{bmatrix}
\nu & 0 & 0  & 0 \\
0 & - \nu & 1  & 1  \\
0 & 1 & -\nu & 0  \\
0 & 1 & 0  & p 
\end{bmatrix}
\end{equation}
The stress tensor of this type has a triple null eigenvector ($\bf{e_0+e_1}$). Actually there is no observed field of this kind.

 \begin{equation}
 \text{Type\,\, IV} \qquad \qquad 
T^{\mu \nu}=
\begin{bmatrix}
0 & \nu & 0  & 0 \\
\nu & - \kappa & 0  & 0  \\
0 & 0 & p_1 & 0  \\
0 & 0 & 0  & p_2 
\end{bmatrix}
\qquad \kappa^2 < 4 \nu^2
\end{equation}
This stress tensor corresponds to the case in which there is no timelike or null eigenvector. No field of this type has been observed.

The week energy condition holds for type I under conditions
\begin{equation}
\rho \geq 0 \, \qquad \rho + p_i \geq 0 \qquad \text{for} \qquad  i = 1,2,3 \ ;
\end{equation}
and for type II if
\begin{equation}
p_1 \geq 0 \ , \qquad p_2 \geq 0 \ , \qquad \kappa\geq 0 \ ,\qquad  \nu = +1 \ ;
\end{equation}
while it does not hold for type III and IV.\\

{\it \textbf{Dominant energy condition (A)}: The energy-momentum tensor at each $p \in {\mathcal M}$, obeys the inequality 
\begin{equation}
T_{\mu \nu} w^\mu w^\nu \geq 0 \ ,
\end{equation}
for any future directed timelike vector $w \in T_p$, and $-T^{\mu}_\nu w^\nu$ is a future directed non-spacelike vector.}
\\

The dominant energy condition embodies the notion that matter should flow along timelike or null worldlines. It may also be interpreted stating that to any observer the local energy density should appear nonnegative and the local energy flow should be non-spacelike. It is possible to restate the dominant energy condition as:\\

{\it \textbf{Dominant energy condition (B)}: In any orthonormal basis the energy dominates the other components of the stress tensor
\begin{equation}
T^{00} \geq \left|T^{\mu \nu}\right| \qquad \text{for any} \qquad (\mu, \nu) \ .
\end{equation}
}
\\

This condition holds for type I if
\begin{equation}
\rho \geq 0 \, \qquad  - \rho \leq p_i \leq \rho  \qquad \text{for} \qquad  i = 1,2,3 \ ;
\end{equation}
while it holds for type II under the conditions
\begin{equation}
\nu = +1\ , \qquad  \kappa \geq 0 \ , \qquad 0 \leq p_1 \leq \kappa \qquad  \text{for} \qquad  i = 1,2  \ ;
\end{equation}
and again it does not hold for type III and IV. Therefore the dominant energy condition, in addition to requirement of the weak energy condition, states that the pressure should not exceed the energy density.\\

{\it \textbf{Strong energy condition}: The energy-momentum tensor at each $p \in {\mathcal M}$, obeys the inequality 
\begin{equation}
\left(T_{\mu \nu} -\frac{1}{2}T g_{\mu \nu} \right)w^\mu w^\nu \geq 0 \ ,
\end{equation}
for any future directed normailsed timelike vector $w \in T_p$.}
\\

Since the Einstein equation implies 
\begin{equation}
T_{\mu \nu} -\frac{1}{2} T g_{\mu \nu} = \frac{1}{8\pi} R_{\mu \nu}
\end{equation}
the strong energy condition is a statement about the Ricci tensor.

The strong energy condition holds in type I for
\begin{equation}
\rho + \sum_{i=1}^3 p_i \geq 0 \ , \qquad \rho + p_i \geq 0 \qquad \text{for} \qquad i=1,2,3 \ ,
\end{equation}
and for Type II if
\begin{equation}
\nu = +1, \qquad \kappa \geq 0, \qquad p_1\geq 0,  \qquad p_2\geq 0 \ .
\end{equation}
It is useful to notice that this condition does not imply the weak one. \\

In presence of a cosmological constant term, the strong energy condition is known as {\it the null convergence condition}. In this case the statement is simply modified to 
\begin{equation}
\left(T_{\mu \nu} -\frac{1}{2}T g_{\mu \nu} + \frac{1}{8\pi}\Lambda g_{\mu \nu} \right)w^\mu w^\nu \geq 0 \ .
\end{equation}
The conditions for type one now are 
\begin{equation}
\rho + \sum_{i=1}^3 p_i - \frac{1}{4 \pi}\Lambda \geq 0 \ , \qquad \rho + p_i \geq 0 \qquad \text{for} \qquad i=1,2,3 \ ,
\end{equation}
and for Type II read
\begin{equation}
\nu = +1, \qquad \kappa \geq 0, \qquad p_1\geq 0,  \qquad p_2\geq 0, \qquad p_1+p_2 - \frac{1}{4 \pi}\Lambda \geq 0\ \ .
\end{equation}

The weakest requirements on the stress tensor are formulated in the null energy condition.\\

{\it \textbf{Null energy condition (NEC)}: The energy-momentum tensor at each $p \in {\mathcal M}$, obeys the inequality 
\begin{equation}
T_{\mu \nu} k^\mu k^\nu \geq 0 \ ,
\end{equation}
for any future directed null vector $k \in T_p$.}
\\

The null energy condition holds for type I under the condition
\begin{equation}
\rho + p_i \geq 0 \qquad \text{for} \qquad  i= 1,2,3 \ ,
\end{equation}
and for type II if
\begin{equation}
\nu = +1, \qquad \nu+\kappa+p_i \geq 0 \qquad i=1,2 \ .
\end{equation}

It is easy to convince oneself that the null energy condition is implied by both strong and weak conditions, while the weak one is implied by the dominant energy condition. Relations among different energy conditions are summarised in the following diagram:

\begin{empheq}[box=
\fbox]{align*}
\\
\qquad \left.
\begin{array}{ccc}
\, & \, & \text{Strong E.C.} \\
\, & \, & \, \\
\text{Dominant E.C.} & \Longrightarrow & \text{Weak E.C.} 
\end{array}
\right\} \, \, \, \Longrightarrow \,\,\, \text{Null E.C.} \qquad
\\
\end{empheq}
\\

In table 4.1 we summarise energy conditions and their associated constraints on type I stress tensors. The discussion above - that we carried on in 4 dimensions - can be easily generalised to higher dimensional spaces. 

\begin{table}[tb]
\begin{center}
\begin{tabular}{lcc}
\toprule
Name & Statement & Conditions for Type I \\
\midrule
Weak & $T_{\mu \nu} w^\mu w^\nu \geq 0$ & $\rho \geq 0, \, \rho + p_i \geq 0$ \\
Dominant &$-T^{\mu}_\nu w^\nu$ future directed & $\rho \geq 0, \,   \left| \rho \right| \geq p_i$ \\
Strong & $\left(T_{\mu \nu} -\frac{1}{2}T g_{\mu \nu} \right)w^\mu w^\nu \geq 0$ & $\rho + \sum_{i=1}^3 p_i \geq 0, \, \rho + p_i \geq 0$ \\
Null & $T_{\mu \nu} k^\mu k^\nu \geq 0$ & $\rho + p_i \geq 0$ \\
\bottomrule
\end{tabular}
\caption{Energy conditions.}
\end{center}
\end{table}

\section{Violation of the null energy condition}

Energy conditions are typically fulfilled by classical fields. However it is well known that they can be violated by quantized matter fields. The most common example is the Casimir vacuum energy between two conduction plates separated by a distance $\delta$
\begin{equation}
\rho = - \frac{\pi^2}{720}\frac{\hbar}{\delta^4} \ .
\end{equation}
\\

In this kind of situation ione usually formulates an averaged version of the energy conditions. Indeed, for each of the energy conditions listed in the previous section it is possible to formulate a corresponding averaged version. Average energy condition requirements of non-negativity must hold only on average along the flowlines of the appropriate vector fields. For example  the {\it averaged null energy condition} states that the integral along a null geodesic $\gamma$ of $T_{\mu \nu} k^\mu k^\nu$ must be non-negative
\begin{equation}
\int_\gamma T_{\mu \nu} k^\mu k^\nu d\lambda \geq 0 \ .
\end{equation}
This condition holds for noninteracting scalar and electromagnetic fields in arbitrary quantum states; this is true even though $T_{\mu \nu} k^\mu k^\nu$ may be locally negative.
\\

Even though, in some cases energy conditions are violated ({\it e.g.} the strong energy condition may be violated in some inflation scenarios), it is generally believed that any well behaved physical systems should respect at least the null energy condition. In various settings it has been been shown that the violation of the null energy condition results in the insurgence of instabilities in the system sourcing $T_{\mu \nu}$ \cite{Hsu:2004vr, Dubovsky:2005xd}. \\

For example, in \cite{Dubovsky:2005xd} the authors showed that classical solutions of both minimally and non-minimally coupled scalar-gauge models which violate the null energy condition are unstable. Moreover it was also proven that perfect fluids which violate the NEC are unstable. Finally quantum states in which the expectation of the energy-momentum tensor violates the NEC cannot be the ground state, including models with fermions. On the basis of these results, it seems that physically interesting cases of violation of the null energy condition are likely to be ephemeral.\\
		
In \cite{Buniy:2005vh} the authors studied systems of derivatively coupled scalar fields with coordinate dependent condensates, without taking into account gravitational back-reactions. In this setting it was suggested that a violation of the null energy condition does not strictly imply instabilities - such as ghosts or imaginary frequencies - in the system. Indeed, it was proven that there may exist an acceptable effective field theory with a positive definite Hamiltonian for quadratic perturbations around a background whose energy momentum violates the null energy condition. It turns out that a necessary feature of such a model is the anisotropy of the background and the presence of superluminal modes. More generally it was suggested that  for systems that are either isotropic or do not feature superluminality, a violation of the null energy condition always implies an unescapable instability.  In the next chapter we will be interested in two nonisotropic backgrounds which both violate the (averaged) null energy condition on a hypersurface.

\section{Some notions on hypersurfaces}

Let us introduce some notation and some concepts related to the study of hypersurfaces in General Relativity \cite{Misner:1974qy}. We will introduce concepts in four dimensions. These can be readily generalised to higher dimensions.

Consider a curve parametrised by a parameter $\tau$ on  a four-dimensional Riemannian manifold $\mathcal{M}$ of class $C^4$. The covariant derivative of a smooth vector function $A^\mu$ defined on this curve will be denoted (we adopt the notation of \cite{Weinberg}) as:
\begin{equation}
\frac{D A^\mu(\tau)}{D \tau} = \frac{d A^\mu}{d \tau} + A^\lambda \Gamma^\mu_{\lambda \nu} \frac{d x^\nu}{d \tau} \ .
\end{equation}

If $A^\mu(u, v)$ is defined on a two-space $x^\mu = x^\mu(u, v)$, the standard definition of the Riemann tensor spells out 
\begin{equation}
\left[ \frac{D \,\,}{D v}, \frac{D \,\,}{D u} \right] A^\mu = R^\mu_{\,\,\, \nu \lambda \kappa} A^\nu \frac{\partial x^\lambda}{\partial u} \frac{\partial x^\kappa}{\partial v} \ ,
\end{equation}

Now, let us consider a smooth hypersurface $\Sigma$ in $\mathcal{M}$ with unit normal 
\begin{equation}
n\cdot n = \varepsilon(n) = 
\left\{
\begin{array}{lc}
+1 & \text{spacelike} \, n \\
-1 & \text{timelike} \, n 
\end{array}
\right. \ .
\end{equation}

An infinitesimal displacement on the hypersurface can be expressed in terms of the intrinsic coordinates $\xi^i$ as
\begin{equation}
{\rm d}s = e_{(i)} {\rm d}\xi^i \ ,
\end{equation}
where $e_{(i)}$ are a natural set of independent tangent vectors associated with the intrinsic coordinates $\xi^i$. In terms of the system of coordinates $\{x^\mu\}$ defined on $V$, the vector  $e_{(i)}$ can be expressed as
\begin{equation}
e_{(i)}^\mu = \frac{\partial x^\mu}{\partial \xi^i} .
\end{equation}
In what follows we will write explicitly 4-dimensional indices only when necessary, otherwise leaving them implicit to avoid confusion. 

Given a tangent vector function lying in the hypersurface, it is possible to define intrinsic components as
\begin{equation}
A = A^i \cdot e_{(i)} \ ,
\end{equation}
The scalar product of this vector with the base vector $e_{(j)}$ is
\begin{equation}
(A \cdot e_{(j)}) = A^i (e_{(i)} \cdot e_{(j)}) = A^i g_{ij} = A_j \ ,
\end{equation}
where the metric three tensor $g_{ij}$ was introduced.

The covariant derivative of the vector A in the direction $e_{(i)}$ is
\begin{equation}
^{(4)}\nabla_{e_{(i)}} A = ^{(4)}\nabla_i A = ^{(4)}\nabla_{e_{(i)}} (e_j A^j) = e_j \frac{\partial A^j}{\partial x^i}+ (^{(4)}\Gamma^\mu_{ji} e_\mu) A^j \ 
\end{equation}
where we used the notation ``$^{(4)}$'' to make evident the dimension of the spacetime in which objects are defined. Indeed, the above definition has components  ``out of the hypersurface''. 

To get a covariant derivative defined intrinsically to the 3-dimensional hypersurface it is necessary to project the above derivative orthogonally onto $\Sigma$. In this way one gets
\begin{equation}
A_{h|i} \equiv e_h\cdot ^{(3)}\nabla_{e_{(i)}} A = \frac{\partial A_h}{\partial x^i} - A^m \Gamma_{m,hi} \ .
\end{equation} 

The connection $\Gamma_{m,hi}$ in three dimensions can be expressed in terms of the basis vector and of the metric 3-tensor $g_{ij}=e_{(i)} \cdot e_{(i)}$
\begin{equation}
\label{3gamma}
^{(3)}\Gamma_{m,hi} = e_{(m)} \cdot  ^{(3)}\nabla_i e_{(h)} \ .
\end{equation}
Given this definition, it is evident that intrinsic covariant derivatives and the associated intrinsic Riemann 3-tensor $R^i_{\,\, jmn}$ do not depend on the nature of the embedding and are therefore invariant under changes of the embedding space that preserve the intrinsic metric on $\Sigma$.

We are now in a position to define the notion of {\it extrinsic curvature}, that can be seen as a description of the curvature of a slice of the 3-geometry relative to that of the 4-geometry. This is measured by the covariant variation $^{(4)}\nabla_i n$ of the unit normal. In components the extrinsic curvature is defined\footnote{In the definition of $K_{ij}$ we adopt the sign convention of \cite{Israel:1966rt}, which is opposite to that in  \cite{Misner:1974qy}} by 
\begin{equation}
^{(4)}\nabla_i n = K_i^{\, j} e_{(j)} .
\end{equation}
Taking the scalar product of this definition with the basis vector $e_{(m)}$ and remembering that 
$$e_{(m)}\cdot n=0$$
by definition, we get
\begin{equation}
\label{ksym}
K_{im}=K_i^{\, j}g_{jm} = e_{m}\cdot \frac{\partial n}{\partial \xi^i}= - n \cdot \frac{\partial e_{m}}{\partial \xi^i} =- n \cdot \frac{\partial e_{i}}{\partial \xi^m} = K_{ji} \ ,
\end{equation}
that spells out the symmetry properties of $K_{ij}$.

The Gauss-Weingarten equations can be derived directly form (\ref{3gamma}) and (\ref{ksym})
\begin{equation}
\frac{\partial e_{i}}{\partial \xi^i}  = - \varepsilon(n) K_{ij} n + \Gamma^h_{ij}e_{(h)} \ .
\end{equation}

It is possible to work out a different expression relating the extrinsic curvature to the standard Riemann tensor. Particularly useful are the equations of Gauss and Codazzi:
\begin{align}
& ^{(4)}R_{\alpha \beta \gamma \delta} e_{(a)}^\alpha e_{(b)}^\beta e_{(c)}^\gamma e_{(d)}^\delta = ^{(3)}R_{abcd} + K_{ac} K_{bd} - K_{bc}K_{ad}\ , \\
& ^{(4)}R_{\alpha \beta \gamma \delta} n^\alpha e_{(b)}^\beta e_{(c)}^\gamma e_{(d)}^\delta = K_{bd|c}-K_{bc|d} \ .
\label{codazzo}
\end{align}

Equations (\ref{codazzo}) can be contracted using respectively $g^{bc} g^{ad}$ and $g^{bd}$ and observing that
\begin{equation}
g^{bc} e_{(b)}^\beta e_{(c)}^\gamma = g^{\beta \gamma} - \varepsilon(n) n^\beta n^\gamma \ . 
\end{equation}
The contracted versions of the equations of Gauss and Codazzi are
\begin{align}
- 2 \varepsilon (n) ^{(4)}G_{\alpha \beta} n^\alpha n^\beta &= ^{(3)}R + K_{ab}K^{ab}- K^2 \ , \\
^{(4)}G_{\alpha \beta} e_{(a)}^\alpha n^\beta &= K^b_{a|b}- K_{|a}\ ,
\label{codazzi}
\end{align}
where $K= g^{ab}K_{ab}$ and all terms are independent of the $x^\mu$ coordinates.\\

Consider now a 3-tensor field $S_{ab}$ living on the hypersurface $\Sigma$, it is possible to define an associated 4-dimensional discontinuous vector 4-tensor as
\begin{equation}
S^{\alpha \beta} =
\left\{
 \begin{array}{lc}
 S^{ab} e_{(a)}^\alpha e_{(b)}^\beta & \text{on} \,\,\, \Sigma \\
 0 & \text{off} \,\,\, \Sigma 
 \end{array}
 \right. \ ,
 \label{Sext}
\end{equation}
then we have
\begin{equation}
\label{nablaesse}
\nabla_\mu S^{\alpha \beta} = e_{(a)}^\alpha S^{ab}_{\,\,\,\,\,\,|b}-\varepsilon(n)S^{ab}K_{ab}n^\alpha \ 
\end{equation}
where $\nabla_\mu$ is the standard covariant derivative defined on the coordinates $x^\mu$.

\section[Singular hypersurfaces]{Singular hypersurfaces \\ in General Relativity}\label{Israelsec}

An important issue in gravitational theory is the formulation of correct junction conditions at surfaces of discontinuity. These surfaces can be characterised by a jump discontinuity in the density, such as {\it boundary layers} or  {\it shock waves}; or by the fact that the density become infinity, as happens for {\it surface layers}. In Newtonian gravitation, one has to set a system of coordinates which has to be {\it a priori} well defined, supplemented by the appropriate continuity  and jump conditions connecting the potential and its first derivatives across the surface. In Einstein's theory of gravitation, the issue is more complicated. Indeed, the smoothness of the gravitational potential $g_{\alpha \beta}$ is determined by the smoothness of the physical conditions {\it and} by the smoothness of the coordinates one chooses to describe the space-time manifold \cite{Israel:1966rt}. The problem is therefore made more subtle by the necessity of distinguishing between the physical discontinuity and spurious ``bumps'' that may arise from an unhappy choice of coordinates.\\

Let us define in a more precise form the classification of  surfaces of discontinuity. Let us focus on a hypersurface $\Sigma$ which separates the spacetime in two four-dimensional manifolds $V^-$ and $V^+$. Moreover we assume that $V^-$ and $V^+$ are both  of class $C^4$ and contain $\Sigma$ as part of their boundaries. The normal $n$ to $\Sigma$, that we can choose to be directed from $V^-$ to $V^+$, is assumed to be spacelike everywhere. Let $K^-_{ij}$ and $K^+_{ij}$ be the two extrinsic curvatures of $\Sigma$ associated respectively with embeddings $V^-$ and $V^+$.

Given extrinsic curvatures, we can give a definition of singular hypersurfaces  independent of the choice of four-dimensional coordinates. We define, {\it singular hypersurfaces of higher order} surfaces $\Sigma$ for which everywhere
\begin{equation}
K^-_{ij} = K^+_{ij} \ ,
\end{equation}
and  {\it singular hypersurfaces of order one}  or histories of a {\it surface layer} as those surfaces $\Sigma$ for which
 \begin{equation}
K^-_{ij} \neq K^+_{ij} \ .
\end{equation}
In particular, (the histories of)  {\it boundary layers} are part of the class of the  {\it singular hypersurfaces of higher order} that contain also all of the regular hypersurfaces.\\

The case of boundary surfaces can be rapidly assessed. From eq. (\ref{codazzi}) we find
\begin{equation}
\left. G_{\alpha \beta} n^\alpha n^\beta\right|^-= \left. G_{\alpha \beta} n^\alpha n^\beta\right|^+ \ , \qquad \left. G_{\alpha \beta} e_{(a)}^\alpha n^\beta\right|^-= \left. G_{\alpha \beta} e_{(a)}^\alpha n^\beta\right|^+ \ ,
\label{boudsurf}
\end{equation}
where we considered the limits of the function $G_{\alpha \beta} n^\alpha n^\beta$ and $G_{\alpha \beta} e_{(a)}^\alpha n^\beta$ at $\Sigma$ from $V^+$ and $V^-$, and we used the fact that the expressions (\ref{codazzi}) have a continuous right hand side across $\Sigma$. Equations (\ref{boudsurf}) establish identities between terms that may be evaluated distinctly in two independently chosen 4-dimensional coordinate systems on $V^-$ and $V^+$. It is interesting to notice that there is no requirement about the continuous matching of charts $x^\alpha_-$ and $x^\alpha_+$ on $\Sigma$. The only requirement about the two charts is that, given the set of intrinsic coordinates $\xi^i$, the equations of $\Sigma$
\begin{equation}
x_-^\alpha = f(\xi^1, \xi^2, \xi^3) \ , \qquad x_+^\alpha = g^\alpha(\xi^1, \xi^2, \xi^3)  ,
\end{equation}
must be formulated in terms of two $C^4$ functions $f^\alpha$, $g^\alpha$. \\

The case of a {\it surface layer} is more interesting. We start by defining the non-vanishing 3-tensor
\begin{equation} 
\label{gammaij}
\gamma_{ij} = K^+_{ij} - K^-_{ij}  \ .
\end{equation}
In terms of $\gamma_{ij}$ it is possible to define the {\it surface energy tensor} of the layer, as\footnote{As before we set the Newton constant and the speed of light in vacuum equal to one, $G= c = 1$.}
\begin{equation}
8 \pi S_{ij} =  - \gamma_{ij} + g_{ij}\gamma_k^{\,\,k} \ .
\label{esse}
\end{equation}
Eq. (\ref{esse}) can be reformulated as
\begin{equation}
\label{gamma2}
\gamma_{ij}= - 8 \pi \left(S_{ij} - \frac{1}{2} g_{ij} S \right) \ ,
\end{equation}
where $S= S_k^{\,\,k}$.

For a surface layer in vacuo $G_{\alpha \beta}=0$ both in $V^-$ and $V^+$, therefore equations (\ref{codazzi}) are reduced to constraints on the extrinsic curvature:
\begin{align}
^{(3)}R+ K_{ab}^{\pm}K^{ab}_\pm -K^2_{\pm} &= 0 \ , \\
 (K^\pm)^{b}_{\,\,a|b} - K^\pm_{|a} &= 0 \ .
\label{constK}
\end{align}
 
After some manipulations  eq (\ref{constK}) can be recast as the following group of constraints:
\begin{align}
\label{Szero}
S^{ab}_{\,\,\,\,\,\,|b} &=0 \ , \\
 \widetilde{K}^b_{\,\, a|b} -\widetilde{K}_{|a}  &= 0 \ , \\
 \label{Szero2}
 \widetilde{K}_{ab} S^{ab} &= 0 \ , \\
^{(3)}R+ \widetilde{K}_{ab}\widetilde{K}^{ab} - \widetilde{K}^2 & = - 16 \pi^2 (S_{ab}S^{ab}-\frac{1}{2} S^2) \ ,
\end{align}
where 
\begin{equation} 
 \widetilde{K}_{ab} \frac{1}{2}(K_{ab}^+K_{ab}^-) \ , \qquad   \widetilde{K} = g^{ab}  \widetilde{K}_{ab} \ .
\end{equation}

Given two independent charts on $V^-$ and $V^+$, respectively $x^\alpha_-$ and $x^\alpha_+$, we can define, using eq. (\ref{Sext}), 4-dimensional extensions $S^\pm_{\alpha \beta}$ of $S_{ab}$ in the two domains $V^\pm$. Hence, we can reformulate condition (\ref{Szero}) using relation (\ref{nablaesse}) as
\begin{equation}
\label{Scondizioni}
\left. e_{(a)}^\alpha \nabla_\beta S_\alpha^{\,\, \beta}\right|^\pm = 0 \ ,  \qquad  \left. n^\alpha \nabla_\beta S_\alpha^{\,\, \beta}\right|^\pm = - K_{ab}^\pm S^{ab} \ .
\end{equation}
Finally, using equations (\ref{gamma2}) and (\ref{Szero2}) we find
\begin{align}
& \left.\left. n^\alpha \nabla_\beta S_\alpha^{\,\, \beta}\right|^+ + n^\alpha \nabla_\beta S_\alpha^{\,\, \beta}\right|^- = 0 \ , \\
& \left.\left. n^\alpha \nabla_\beta S_\alpha^{\,\, \beta}\right|^+ - n^\alpha \nabla_\beta S_\alpha^{\,\, \beta}\right|^- =   8 \pi  \left(S_{ab}S^{ab}-\frac{1}{2} S^2 \right) \ ,
\end{align}
that can be seen as the relativistic analogues of the Newtonian formula for the normal components of mechanical force due to self-attraction of the two faces of a plane layer of surface density $\sigma$:
\begin{equation}
\sigma F^- \cdot n = - \sigma F^+ \cdot n = 2 \sigma^2 \ .
\end{equation}
The tangential force, on the contrary cancels out, in agreement with the first equation of (\ref{Scondizioni}).\\

We can conclude this section by giving a heuristic argument for the definition of the surface energy tensor (\ref{esse}). Let us consider a layer of finite thickness $\varepsilon$ separating the two vacuum regions $V^-$ and $V^+$ and having respective boundaries $\Sigma^-$ and $\Sigma^+$. It is useful to introduce {\it Gaussian coordinates} $x^\alpha$, based on $\Sigma^-$ by setting $x^i = \xi^i$, and letting $x^0$ equal the geodesic distance normal to $\Sigma^-$, taken with sign plus or minus respectively for points in $V^+$ and $V^-$. With these coordinate, the equations for $\Sigma^\pm$ are respectively $x^1=0$ and $x^1=\varepsilon$. Moreover, for a 3-space $x^1= \text{cost}.$, one finds the following expression for the extrinsic curvature
\begin{equation}
K_{ij} = \frac{1}{2}\frac{\partial \, ^{(4)}g_{ij}}{\partial x^1} \ ,
\end{equation}
while 
\begin{equation}
^{(4)}R_{ij} = \frac{\partial K_{ij}}{\partial x^1} + Z_{ij} \ ,
\label{riccilayer}
\end{equation}
where $Z_{ij}$ is defined as
\begin{equation}
Z_{ij} = ^{(3)}R_{ij} - K K_{ij}+ 2 K^h_i K_{hj} \ .
\end{equation}

At this point it is possible integrate Einstein's field equation over the layer
\begin{equation}
R_{\alpha \beta} = -  8 \pi  \left(T_{\alpha \beta} -\frac{1}{2}g_{\alpha \beta} T  \right) \ .
\end{equation}
Remembering expression (\ref{riccilayer}) one gets
\begin{equation}
-  8 \pi \int_0^\varepsilon  \left(T_{\alpha \beta} -\frac{1}{2}g_{\alpha \beta} T  \right) dx^1 = K^+_{ij} -  K^-_{ij} + \int_0^\varepsilon Z_{ij} dx^1 \ .
\end{equation}
In the limit $\varepsilon \to 0$, for given values of $K^\pm_{ij}$, if $K_{ij}$ remains bounded in the layer then the integral of $Z_{ij}$ over the layer will converge to zero. Hence $S_{ij}$ defined in (\ref{esse}) is
\begin{equation}
S_{ij} = \lim_{\varepsilon \to 0} \int_0^\varepsilon T_{ij} dx^1 \ ,
\end{equation}
or in other words, $S_{ij}$ is the integral of the stress tensor $T_{ij}$ over the surface layer.

% Articolo
%_____________________________________________
\chapter{The Surface Layers Dual to Hydrodynamic Boundaries}
%_____________________________________________

The AdS/hydrodynamics correspondence provides a 1-1 map between large wavelength features of AdS black branes and conformal fluid flows, %This AdS/hydrodynamics correspondence provides an explicit black brane solution for every history of a particular conformal fluid
so long as the fluid variables are constant over distances large compared with the inverse temperature.  For example progress towards gravity duals of shock waves and vortices has appeared in Refs.~\cite{shock} and \cite{vortices}.  A case of particular interest is represented by gravity duals to turbulent flows.  Turbulence is generic in fluid flows under a wide range of conditions.  The dual of these fluid conditions then provides some condition on a gravity solution under which it to generically decays into a turbulent configuration.  An example of such a situation was presented in Ref.~\cite{Bhattacharyya:2008ji}.  It would of course be interesting to characterize the gravity duals of turbulent flows, and of the conditions under which turbulence may be expected.  In hydrodynamics, even the most basic scaling laws are altered by turbulence.  If gravitational solutions near, for example, spacelike singularities (where indeed chaotic evolution is expected \cite{BKL}) or certain event horizons do generically decay to turbulent solutions, it would be difficult to overstate the potential consequences for, for example, the horizon problem.

Perhaps the best understood turbulence is steady state turbulence, in which energy is injected into a system at the same rate at which it dissipates.  Richardson's cascade model \cite{Rich} of steady (3+1)-dimensional turbulence is as follows.  Energy is injected into a system at large characteristic distance scales, for example, a lake warms the air.  This creates large vortices, which decay into smaller vortices.  Thus the energy flows to smaller distance scales.  At sufficiently small distance scales, higher order derivative terms in the equations of motion become relevant, such as viscosity terms.  These lead to dissipation of the energy in sufficiently small vortices.  Thus energy cascades from the long length scale in which it is introduced, down to the dissipation scale.

To realize steady state turbulence, one needs to inject energy into a system.  There are two principal ways to do this.  First, one may deform the fluid via external perturbations.  Second, one may apply boundary conditions, for example one may consider fluid flow in a pipe or wind tunnel.  The first approach was applied to the AdS/hydrodynamics correspondence in Ref.~\cite{Bhattacharyya:2008ji}, where it was argued that a laminar fluid flow and the dual gravity solution decay to turbulent configurations.  This approach has the disadvantage that solutions are quite complicated, due to the necessarily inhomogeneous forcing and to the geometric implementation of the forcing on the gravity side. \\ 

In what follows we will take a preliminary step towards a realization of the second approach to creating steady state turbulence, we will investigate boundary conditions in the AdS/hydrodynamic correspondence.  For simplicity, we will consider nonrelativistic, incompressible flows.  Consider the surface which separates a solid object from such a fluid.  The normal velocity of the fluid into the solid must vanish.  If furthermore the fluid is viscous, as fluids in the AdS/hydrodynamics correspondence are \cite{D1979}, then the tangential relative velocity of the fluid must also vanish.

What does this correspond to on the gravity side?  The answer to this question is not necessarily unique, one may define a dual and then attempt to understand its dynamics.  One interesting case, which is already sufficient to generate turbulence, is a solid which is a thin, infinite sheet with a stationary fluid on the left side and a moving fluid on the right.  In this case a natural choice would be to consider the gravity duals of both fluids and then to attempt to glue them together.  Equivalently one may choose to think of the entirety of the left side as a solid wall, filling the left half of spacetime, and a liquid filling the right half.  The wall is stationary and so one chooses the dual to be a stationary black brane in half of AdS.  Whatever one chooses to think, the logic is that one imposes that the left half of the gravity dual be a static black brane in AdS, and that the right side be the gravity dual given by the prescription of Ref.~\cite{Bhattacharyya:2008jc}.

So how does one glue these two vacuum gravity solutions together?  Clearly there are many inequivalent choices.  One possibility is to simply attach them and then use the Israel matching conditions \cite{Israel:1966rt} to determine the stress tensor on the surface layer that separates the two sides.  This is equivalent to letting the gravitational solution continuously interpolate between the two solutions over a finite distance $d$ and then taking the limit as this distance tends to zero.  While there are many ways of performing this interpolation, so long as the extrinsic curvature is kept finite, they all lead to the same stress tensor as the interpolation distance $d\rightarrow 0$.

Another possibility is to let the fluid configuration continuously interpolate between the two solutions, and then take the dual using the prescription of Ref.~\cite{Bhattacharyya:2008jc}.  As the fluid is not a solution of the Navier-Stokes equation in this region, the dual will not be a solution of the vacuum Einstein equations in this region.  Instead it will solve Einstein's equations with a nonvanishing stress tensor supported on a surface layer.  The ultralocality of the duality map implies that the vacuum Einstein equations will however be solved way from the surface layer.  In this case, one cannot take the interpolation distance $d$ to zero, because the dual is not defined when derivatives are large with respect to the inverse of the temperature $T$.  Thus the minimum size of $d$ will be of order $1/T$.  Again there are many inequivalent ways of performing the interpolation.  But we will see that, at least for the quantities at we are able to calculate, when $d$ is large with respect to $1/T$, the difference between these prescriptions is suppressed by powers of $dT$ and so, like Israel's method, there is a single answer.

The perhaps surprising result is that the two methods yield bulk stress tensors which differ by a finite amount.  They did not need to agree, indeed one is derived at small $d$ and the other for large $d$.  The reason that they disagree is as follows.  The construction of the metric from the fluid flow proceeds order by order in the derivatives of the fluid's velocity.  The boundary conditions imply that the velocity of the fluid is the same on both sides of the wall, however the first derivatives differ.  Therefore, whatever regularization scheme one uses on the fluid side, the second derivative of the velocity diverges at small $d$.  This means that the metric corrections derived using the map of \cite{Bhattacharyya:2008jc} will diverge at small $d$, invalidating the finiteness assumption in Israel's derivation.  In fact, we will see that the disagreement between the two calculations of the stress tensor differ only in these higher derivative terms.  Of course the fluid map is not defined at small $d$, as it yields a divergent series, and so no divergences appear within the range of validity of either approach.\\

We will begin in Sec.~\ref{flowsez} by describing the flow of interest.  The velocity will be kept sufficiently arbitrary to allow a general interpolation between the flows on the two sides of the wall, and in Sec.~\ref{gravsez} the na\"ive gravity dual will be calculated using the prescription of \cite{Bhattacharyya:2008jc}.  We will see that those higher order derivative corrections which we calculate are indeed suppressed by factors of $dT$.  Then in Sec.~\ref{tsez} we will calculate the bulk stress tensor of the interpolation between the two gravity solutions.  First it will be calculated for the interpolation dual to a continuously interpolating fluid flow.  It will be seen that contributions from the second derivative of the velocity are $d$-independent, while higher order contributions are suppressed by powers of $dT$.  Thus the result is independent of the interpolation scheme when $d$ is sufficiently large.  The stress tensor will then be calculated directly from the Israel matching conditions on the two solutions of the vacuum Einstein equations.  It will be seen that the two stress tensors agree up to terms corresponding to a divergence in the extrinsic curvature at small $d$, and that only the second stress tensor contains a nonvanishing stress.

\section{The Flow} \label{flowsez}

\subsection{The {\it ansatz}}

We will consider a hydrodynamic flow in 4-dimensional Minkowski space, using a $(-,+,+,+)$ metric. To highlight the essential features of the boundary condition, we will consider the simplest possible flow.  The liquid will only move in the $y$ direction, with a velocity $v=v(x)$ that only depends on the coordinate $x$.  The velocity will be taken to be small, and we will drop all terms which are quadratic in $v$.  In fact, as described in Refs.~\cite{Bhattacharyya:2008kq,vortices} we will work in the nonrelativistic, incompressible limit.  More precisely, we will show that our flow satisfies both the full relativistic equations of motion at order $\mathcal{O}(v)$ and also the incompressible Navier-Stokes equation.

We will set $c=1$.  The conformal fluid which is dual to Einstein gravity with a negative cosmological constant is very particular.  Being conformal, all of its transport coefficients may be expressed in terms of a single dimensionful quantity, such as the temperature $T$, and certain constants which may be calculated from the gravity dual.  In the case at hand for example the shear viscosity $\eta$, pressure $p$ and density $\rho$ have been found in Ref.~\cite{Bhattacharyya:2008jc}
\begin{equation}
\eta=\frac{\pi^2}{16G_N}T^3\hsp
p=\frac{\pi^3}{16G_N}T^4\hsp
\rho=\frac{3\pi^3}{16G_N}T^4 \label{rapporti}
\end{equation}
where $G_N$ is the dual Newton's constant.  %Notice that $G_N$ has dimensions of action, and so dimensional analysis proceeds as in a relativistic quantum field theory and not as in a gravitational theory, although our considerations are purely classical.

\begin{figure}
\begin{center}
\includegraphics[scale=.58]{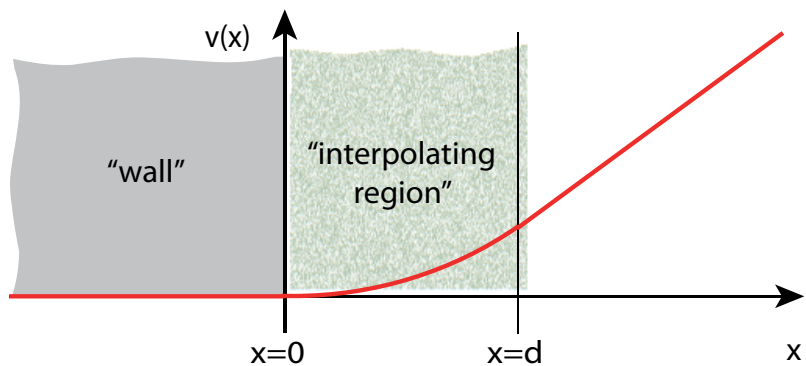}
\caption{The fluid velocity $v$ is in the $y$ direction, and it depends on the $x$ coordinate.  On the left the fluid is stationary, on the right the fluid velocity is linear.  These two regions solve the fluid equations of motion at linear order in $v$.  There is an interpolating region of width $d$, which must be larger than the inverse temperature, in which $v$ does not satisfy the equations of motion.  $v$ and its first derivative $v\p$ are continuous at $x=0$ and $x=d$.}
\label{vy}
\end{center}
\end{figure}

The relativistic velocity 4-vector $u$ is, to linear order in $v$, simply 
\begin{equation}
u_\mu = (\frac{1}{\sqrt{1-v^2}}, 0, \frac{v}{\sqrt{1-v^2}}, 0) \sim (1,0,v,0). \label{u}
\end{equation}
We will be interested in the fluid velocity in three regions, as illustrated in Fig.~\ref{vy}.  First, on the left, where $v=0$.  Second, we will be interested in the velocity on the right, where $v$ will be linear in $x$.  We will show momentarily that this is a solution to the hydrodynamic equations of motion and so will be dual to a vacuum solution of Einstein's equations.  Finally, we will be interested in an interpolating region where $v$ will be arbitrary and we will not impose the equations of motion, therefore the dual metric will not solve the vacuum Einstein equations but, like any metric, will solve Einstein's equations with some stress tensor.

Clearly the left, $v=0$, satisfies the fluid equations of motion.  We will now verify that the region on the right satisfies the relativistic equations of motion, which are simply the conservation of the stress tensor
\begin{equation}
0=\partial_\mu T^{\mu\nu}. \label{edm}
\end{equation}
In accordance with the usual fluid approximation \cite{Landau:1965pi}, we will work at large enough distance scales that only the velocity $v$ and its first derivative $v\p$ need be considered in the stress tensor.  This approximation in general is problematic, leading for example to superluminal propagation \cite{ac}.  However, as we will be interested in velocities well below the speed of light, no problems will arise.  Later, when we will consider the interpolating region, where the second derivative may be large, we will make no such approximation.  We will consider the bulk stress tensor to higher order, calculating all terms up to two derivatives and several terms up to three or four derivatives to check that they are subdominant.  However we do not impose that the interpolating region satisfies the equations of motion, indeed that would lead to a vanishing bulk stress tensor.

\subsection{Relativistic and nonrelativistic equations of motion}

As discussed in Chapter 2 (see Sec. \ref{confluid}), dropping all higher derivatives of the velocity and using the fact that the fluid is conformal to eliminate the bulk viscosity and replace $\rho$ with $3p$, the hydrodynamic stress tensor is
\begin{equation}
T^{\mu\nu}=p(\eta^{\mu\nu}+4u^\mu u^\nu)-2\eta\sigma^{\mu\nu} \label{Tfluido}
\end{equation}
where the shear strain rate $\sigma^{\mu \nu}$ is defined as
\begin{equation}
\sigma^{\mu \nu} = P^{\mu \alpha} P^{\nu \beta} \partial_{(\alpha} u_{\beta)} - \frac{1}{3}\partial_\lambda u^\lambda P^{\mu \nu}. \label{sigdef} 
\end{equation}
Here parenthesis denote, as usually, symmetrization with a factor of one half and $P^{\mu\nu}$ is the projector onto the spacelike directions in the reference frame of the fluid
\begin{equation}
P_{\mu \nu} = \eta_{\mu \nu} + u_\mu u_\nu 
= \left(
\begin{array}{cccc}
 \frac{v^2}{1- v^2} & 0 & \frac{v}{1- v^2} & 0 \\
 0 & 1  & 0 & 0 \\
 \frac{v}{1- v^2} & 0 & \frac{1}{1- v^2} & 0\\
 0 & 0 & 0 & 1
\end{array}
\right). \label{proj}
\end{equation}

Substituting the velocity ansatz (\ref{u}) into the definition (\ref{sigdef}) one easily finds the shear strain at linear order in $v$
\begin{equation}
\sigma^{\mu \nu} \simeq \left(
\begin{array}{cccc}
 0 & 0 & 0 & 0 \\
 0 & 0  & \frac{1}{2} v' & 0 \\
 0 & \frac{1}{2} v' & 0 & 0\\
 0 & 0 & 0 & 0
\end{array}
\right) \ .\label{sigma} 
\end{equation}
The constants of proportionality (\ref{rapporti}) in this particular fluid can then be inserted into the general formula (\ref{Tfluido}) for $T^{\mu\nu}$ to express the stress tensor in terms of the temperature $T$ and the velocity $v$
\begin{equation}
T^{\mu\nu}=\frac{\pi^3T^4}{16G_N}\left(
\begin{array}{cccc}
 3 & 0 & 4v & 0 \\
 0 & 1  & -\frac{v\p}{\pi T} & 0 \\
 4v & -\frac{v\p}{\pi T} & 1 & 0\\
 0 & 0 & 0 & 1
\end{array}
\right) \ .
\end{equation}

The velocity only depends on the coordinate $x$.  Let us choose boundary conditions so that the temperature $T$ also only depends on $x$.  Then the equations of motion (\ref{edm}) are simply
\begin{equation}
0=\partial_x T^{x\nu}.
\end{equation}
However all of the components $T^{x\nu}$ are constants except for $T^{xx}$ and $T^{xy}$.  Therefore the only nontrivial equation of motion at linear order in $v$ is
\begin{equation}
0=\partial_x T^{xx}+\partial_x T^{xy}=\frac{\pi^3T^3}{4G_N}T\p-\frac{\pi^2}{16G_N}\partial_x (T^3 v\p).
\end{equation}
When $v$ is linear in $x$, its second derivative vanishes.  Therefore this equation of motion, the conservation of momentum in the $y$ direction, implies that the temperature is constant.  More precisely it implies that $(\partial T)/T$ is negligible in this approximation.  In light of the relations (\ref{rapporti}) this is consistent with the incompressibility assumption that we have imposed on our fluid.  

Therefore we recover the fact that at sufficiently small velocities, a constant temperature and a $y$-velocity which is linear in $x$ solve the equations of motion (\ref{edm}).  Intuitively this is clear.  Further to the right, the fluid is moving faster.  Therefore the viscous force on a unit of fluid exerted by the faster fluid on its right (in the $+x$ direction) will accelerate it in the $+y$ direction, whereas the slower fluid on its left will exert a viscous force that decelerates it.  A steady flow occurs when these two forces cancel, which implies that the second derivative of the flow vanishes.  Had there been a temperature gradient, then the viscosity to density ratio would also have been stronger on one side by (\ref{rapporti}), and so this balance could only be maintained by introducing a second derivative of the velocity. 

Clearly a linear velocity also satisfies the nonrelativistic Navier-Stokes equation for an incompressible, Newtonian fluid
\begin{equation}
\rho(\partial_t v_k + (v\cdot\partial) v_k)=-\partial_k p + \nu \nabla^2 v_k. 
\label{nonrel}
\end{equation} 
In fact, each term vanishes independently.  Note that the vanishing of the $\partial p$ term is not merely a consequence of incompressibility.  In incompressible flows it may be of the same order as the viscous term.  It vanishes in this case because this provides a solution to (\ref{nonrel}) and it is consistent with the various nonrelativistic, small gradient and incompressible limits taken above.

In conclusion, we have considered fluid flows with an $x$-dependent velocity $v$ in the $y$ direction.  We have verified that, to linear order in $v$, these satisfy both the relativistic and nonrelativistic equations of motion when $v$ is linear in $x$ and the temperature $T$ is constant.  Our flows of interest will have $v=0$ on the left, $v$ linear on the right and an interpolating region inbetween.  Thus the equations of motion will be satisfied on the left and the right but not in the interpolating region, leading to a dual gravity solution which solves Einstein's vacuum equations on the left and right, but inbetween requires material described by a nontrivial stress tensor.  We have also found formula (\ref{proj}) and (\ref{sigma}) for the projector $P^{\mu\nu}$ and the shear tensor $\sigma^{\mu\nu}$ for general functions $v$, and so these results may be applied to the interpolating region.

Clearly if the fluid velocity is linear over a large enough distance, it will eventually approach the speed of light and the nonrelativistic approximation will break down.  Therefore our analysis is only relevant near the boundary.  The solution may be made global by introducing a second boundary, such that the velocity is constant on the other side of the second boundary.  We will see below that the stress tensor on the second boundary, to linear order in $v$, will be minus the stress tensor of the first boundary.  Of course at higher orders one may expect an attraction between the two boundaries, and so this solution will not be stationary.  The underlying assumption in this note is that the walls on the gravity side are built of a solid which remains fixed.  The fact that the null energy condition is violated may be a sign that such a material would be inconsistent, as some null energy violating configurations lead to superluminal propagation or instabilities \cite{Dubovsky:2005xd}, although some do not \cite{tranquilli}, in which case this configuration should only be considered over a sufficiently short timescale.  It may be that this timescale is never sufficient for turbulence to develop.

\section{The Gravity Dual} \label{gravsez}

In Chapter 3 we gave a detailed presentation of the AdS/hydrodynamics correspondence, that yields a black brane metric dual to arbitrary flows in very particular conformal fluids, which for example obey the relations (\ref{rapporti}).  If the flow satisfies the hydrodynamic equations of motion (\ref{edm}), the dual satisfies the vacuum Einstein equations, in this case with a negative cosmological constant.  

\subsection{A note on ultralocality}

As mentioned in the construction of the gravity dual to a generic conformal fluid, the correspondence, at least in the incarnation in Refs.~\cite{Bhattacharyya:2008jc} and \cite{Bhattacharyya:2008mz}, is ultralocal.  It is useful at this point to explain a bit further what this means. Consider the set of ingoing null geodesics which run from the boundary to the black brane horizon.  Clearly each point in the bulk is on precisely one such geodesic.  Also each point on the boundary is on one such geodesic.  Therefore these geodesics can be used to associate a fixed single boundary point to each bulk point.  This association is not one to one, there is an entire geodesic worth of bulk points associated to each boundary point.  

To implement the map, one identifies the boundary with the Minkowski space on which the fluid lives.  The metric and its derivatives at a point in the bulk are determined entirely by the fluid velocity and temperature and their derivatives at the associated boundary point.  One does not need to know the behavior of the fluid elsewhere.  This is the ultralocality of the correspondence.  In particular, the fact that the fluid satisfies the hydrodynamic equations of motion on the left and right (at $x<0$ and $x>d$) implies that the dual metric will satisfy the vacuum Einstein equations on the left and right, so long as the characteristic distance over which these quantities vary is greater than the inverse temperature.

Our fluid does not satisfy the hydrodynamic equations of motion at $0<x<d$.  This means that the dual gravitational configuration will not satisfy the vacuum Einstein equations, instead it will only satisfy Einstein's equations with a nonzero stress tensor, which we will calculate in Sec.~\ref{tsez}.

This ultralocality is somewhat different from the ultralocality that one encounters in classical field theories, or in the BKL limit of gravity theories, in that the bulk geometry is ultralocal in terms of null and not temporal evolution.  This ingoing null identification identifies the temporal evolution of the fluid with outward radial evolution for the gravity theory.  That is to say, the metric at larger radii but the same time is determined by the fluid in the future but at the same location.  In particular, a timeslice of the bulk geometry is determined by the evolution of the boundary fluid during a fixed interval of time.  This interval is of order the inverse temperature, and so no appreciable evolution may occur during this interval if the temperature is large enough for the correspondence to hold.  In this sense any fixed timeslice of the gravity dual contains only as much information as a fixed timeslice of the fluid, despite being one dimension greater.

As each event in the fluid is identified with an inward null geodesic in the bulk, the metric corresponding to this event appears to be falling towards the black hole at the speed of light.  This is not at all to say that there is a Killing vector in the inward null direction, the metric changes in that direction, but in a fashion which is fixed by the map.  Thus a disturbance on the boundary creates gravity waves which fly inward at the speed of light to the horizon.  Similarly, pasting together an infinite sequence of bulk timeslices which are separated by time intervals $1/T$, one obtains a pattern which falls from the boundary in to the horizon at the speed of light.  Although each individual timeslice is too small to see any evolution, the entire pattern is dual to the entire history of the flow.  Like a movie reel, the pattern in turn allows one to reconstruct the gravity dual, as it contains the timeslices.

One may use this identification to speculate on the gravitational dual of decaying turbulence.  For example, the inverse cascade of decaying (2+1)-dimensional turbulence consists of a chaotic period during which the fluid is subjected to random external forces followed by a relaxation period, characterized by the merging of well-separated vortices \cite{Kraichnan, mcwilliams}.  This would then be dual to a kind of forest of gravity waves falling from the boundary to the horizon, beginning when the boundary is subjected to a random perturbation.  First the canopy, representing the chaotic period, falls out of the boundary into the horizon.  When the external perturbation is turned off, it is followed by the branches representing the vortices, and then the branches merge into trunks as the vortices merge.  The usual cascade \cite{Kol1,Kol2} in (3+1)-dimensional turbulence may be similarly described, but when the boundary is randomly perturbed, the trunk falls out first.  This leads to the rather bizarre observation that black brane geometries in AdS$_4$ and AdS$_5$ respond very differently to random perturbations of their boundaries.  Needless to say, it would be interesting to make this picture precise, or to see whether it is inconsistent with the various approximations involved in the duality.

\subsection{The metric}

We will now calculate the metric dual to the flow (\ref{u}) using the map in Ref.~\cite{Bhattacharyya:2008jc} with the simplified notation of Refs.~\cite{Bhattacharyya:2008mz, Rangamani:2009xk}.  This map takes regions in which the flow satisfies the fluid equations to regions in which the metric satisfies the source-free Einstein equations.  Acting on the region in which the fluid does not satisfy the hydrodynamic equation, the map is not known to have any special properties other than continuity, which will produce an interpolation between the vacuum Einstein metrics on the two sides.  Therefore the choice of this map corresponds to a rather arbitrary choice of interpolation.  However we will see that this interpolation has two nice properties.  First, it is reasonably independent of the interpolating velocity function chosen.  In particular, the third and higher derivatives of the velocity will yield contributions to the integrated stress tensor which are suppressed by powers of $dT$, while the leading contribution is independent of $d$.  Second, the resulting stress tensor is simpler than the Israel stress tensor, it will have zero stress, whereas Israel's stress tensor has shear stress.

Ultralocality in the ingoing null direction implies that the simplest coordinates in which to express the metric are Gaussian null coordinates, in which $r$ parametrizes the ingoing null lines.  In these coordinates, the bulk 5-dimensional metric (\ref{metric-3}) corresponding to an $x$-dependent 4-dimensional fluid flow is \cite{Bhattacharyya:2008mz}
\begin{equation}
ds^2 =  G_{MN} dX^M dX^N = - 2 u_\mu(x) \, dx^\mu \, (dr + \mathcal{V}_\nu(r, x) \, dx^\nu) + \mathfrak{G}_{\mu \nu}(r, x) dx^\mu dx^\nu \label{met}
\end{equation}
where, up to second derivatives in $v$, $ \mathcal{V}_\nu$ and $ \mathfrak{G}_{\mu \nu}$ are defined as
\begin{multline}
\mathcal{V}_\nu = r \mathcal{A}_\mu - \mathcal{S}_{\mu \lambda} u^\lambda - {\mathfrak v}_1(br) P_{\mu}^\nu \mathcal{D}_\lambda \sigma^\lambda_\nu \\
+ u_\mu \left[\frac{1}{2} r^2 \left(1-\frac{1}{(br)^4}\right) - \frac{1}{4(br)^4}\omega_{\alpha \beta} \omega^{\alpha \beta} + {\mathfrak v}_2(br) \frac{\sigma^{\alpha \beta} \sigma_{\alpha \beta}}{d-1} \right] \label{vdef}
\end{multline}
and
\begin{multline}
\mathfrak{G}_{\mu \nu} = r^2 P_{\mu \nu} - \omega_\mu^\lambda \omega_{\lambda \nu} + 2 (br)^2 {\mathfrak g}_1(br)\left[\frac{1}{b}\sigma_{\mu \nu} + {\mathfrak g}_1(br) \sigma_{\mu}^\lambda \sigma_{\lambda \nu} \right] - {\mathfrak g}_2(br)  \frac{\sigma^{\alpha \beta} \sigma_{\alpha \beta}}{d-1} P_{\mu \nu} \\
- {\mathfrak g}_3 (br) \left[{\mathfrak T}_{1 \mu \nu} + \frac{1}{2} {\mathfrak T}_{3 \mu \nu}+ 2 {\mathfrak T}_{2 \mu \nu} \right] + {\mathfrak g}_4(br)[{\mathfrak T}_{1 \mu \nu} +{\mathfrak T}_{4 \mu \nu} ] \ . \label{gdef}
\end{multline}
The functions that appear in this definition were introduced in Sec. \ref{defmetric}, (one can refer also to the Appendix).

Eq.~(\ref{met}) is the bulk metric dual to a fluid flow in an arbitrary curved space.  The fluid is conformally invariant, and the conformal invariance has been used to write the metric in a compact form using objects which transform covariantly under the conformal symmetry.  We are interested in a flat boundary, and so many of these objects will vanish.  In fact, since $u$ only depends on $x$, but only has nonvanishing components in the $t$ and $y$ directions, even the gauge field for a Weyl transformation will vanish
\begin{equation}
\mathcal{A}_\mu = u^\lambda \nabla_\lambda u_\mu - \frac{1}{d-1}u_\mu \nabla^\lambda u_\lambda=0 
\end{equation}
as $u^\lambda \nabla_\lambda u_\mu=0$ and $ \nabla_\lambda u^\lambda =0$. This implies that the Weyl-covariant derivative reduces to the ordinary derivative
\begin{equation}
\mathcal{D}_\mu  = \partial_\mu \ .
\end{equation}

The Weyl-covariant Schouten tensor $\mathcal{S}$ (defined in eq. (\ref{Schouten})) is proportional to the Weyl-covariant curvature of the boundary.  As the Weyl-covariant derivative is just the ordinary derivative, this is just the ordinary curvature.  As the boundary is Minkowski space, the curvature vanishes, and so the Weyl-covariant Schouten tensor also vanishes
\begin{equation}
\mathcal{S}_{\mu \nu} = 0 \ .
\end{equation}
Similarly the Weyl-covariant Weyl curvature $\mathcal{C}$ is the sum of the ordinary Weyl curvature and the curvature of the Weyl tensor (see eq. (\ref{Weyltensor})), which both vanish and so
\begin{equation}
\mathcal{C}_{\mu \nu \lambda \sigma} = 0\ .
\end{equation}

The vorticity $\omega$ does not vanish, however like the shear strain $\sigma$ it is of first order in $v$.  Therefore $\omega^2$, $\omega\sigma$ and $\sigma^2$ terms are all of order $\mathcal{O}(v^2)$ and so do not contribute at order $\mathcal{O}(v)$.  Thus only the third and fourth terms of (\ref{vdef}) contribute to $\mathcal{V}_\mu$.  The third term is easily evaluated
\begin{equation}
\mathcal{D}_\lambda \sigma^\lambda_\nu=\partial_\lambda \sigma^\lambda_\nu =  \partial_x \sigma^x_\nu
\simeq \left(
\begin{array}{c}
 0\\
 0\\
 \frac{1}{2} v''\\
 0
\end{array}
\right) \ .
\end{equation}
Adding the third and fourth terms one finds $\mathcal{V}_\mu$ at order $\mathcal{O}(v)$
\begin{equation}
\mathcal{V}_\mu \simeq {\mathfrak v}_1(br) 
 \left(
\begin{array}{c}
 0\\
 0\\
 \frac{1}{2} v''\\
 0
\end{array}
\right) + 
 \left(
\begin{array}{c}
 1\\
 0\\
 v\\
 0
\end{array}
\right) \frac{1}{2}\left(r^2-\frac{1}{b^4r^2}\right). \label{v2}
\end{equation}

The functions ${\mathfrak T}$, reported in Table 2.1, are easily expressed in terms of the shorthand notation $<>$, which symmetrizes and contracts with the projectors $P_{\mu \nu}$.  The first two are identically zero, while the second two are of order $\mathcal{O}(v^2)$
\begin{align}
{\mathfrak T}_{1 \mu \nu} &= 2 u^\alpha \mathcal{D}_\alpha \sigma_{\mu \nu} = 0 \\
{\mathfrak T}_{2 \mu \nu} &= C_{\mu \alpha \nu \beta} u^\alpha u^\beta = 0\\
{\mathfrak T}_{3 \mu \nu} &= 4 \sigma^{\alpha \langle \mu} \sigma^{\nu\rangle}_\alpha \sim 0 \\
{\mathfrak T}_{4 \mu \nu} &= 4 \sigma^{\alpha \langle \mu} \omega^{\nu\rangle}_\alpha \sim 0. %\ , \\
%{\mathfrak T}_{5 \mu \nu} &= 4 \omega^{\alpha \langle \mu} \omega^{\nu\rangle}_\alpha \simeq 0 \ , 
\end{align}
Therefore only the first and third terms of (\ref{gdef}) contribute to $\mathfrak{G}_{\mu \nu}$
\begin{equation}
\mathfrak{G}_{\mu \nu} \simeq  
r^2 \left(
\begin{array}{cccc}
 0 & 0 & v & 0 \\
 0 & 1  & 0 & 0 \\
 v & 0 & 1 & 0\\
 0 & 0 & 0 & 1
\end{array}
\right) 
+  b r^2 {\mathfrak g}_1(br)  
\left(
\begin{array}{cccc}
 0 & 0 & 0 & 0 \\
 0 & 0  &  v' & 0 \\
 0 & v' & 0 & 0\\
 0 & 0 & 0 & 0
\end{array}
\right) \ . \label{g2}
\end{equation}

Finally, inserting Eqs.~(\ref{v2}) and (\ref{g2}) into (\ref{met}), one finds the final form of the metric \cite{Evslin:2010ss}
\begin{multline}
ds^2 = - r^2 \left(1-\frac{1}{b^4 r^4}\right)dt^2 - 2 dt dr  + \left(2 \frac{v}{b^4 r^2}+{\mathfrak v}_1(br) v'' \right) dt dy \\
 +r^2(dx^2 + dy^2+dz^2)+ 2 r^2 b\, {\mathfrak g} _1(br) v' dx dy - 2 v dy dr
\end{multline}
where $b=1/{\pi T}$ and the functions ${\mathfrak v}_1$ and ${\mathfrak g}_1$ are defined in Eqs.~(\ref{v1}) and (\ref{g1}).  Using the basis $(t,x,y,z,r)$, we may write the metric in matrix form as
\[
g_{\mu \nu}=
\begin{pmatrix}
-\left(r^2-\frac{1}{b^4 r^2}\right) & 0 & \frac{v}{b^4 r^2}+\frac{1}{2}{\mathfrak v}_1(br) v''  & 0 & -1\\
0 & r^2 & r^2 b\, {\mathfrak g} _1(br) v'  & 0 & 0\\
 \frac{v}{b^4 r^2}+\frac{1}{2}{\mathfrak v}_1(br) v''  &  r^2 b\, {\mathfrak g} _1(br) v'  & r^2 & 0 & - v\\ 
 0 & 0 & 0 & r^2 & 0\\
 -1 & 0 & -v & 0 & 0
\end{pmatrix} \   \label{metfin}
\]
while the inverse metric is
\[
g^{\mu \nu}=
\begin{pmatrix}
 0 & 0 & -\frac{v}{r^2} & 0 & -1\\
 0 & \frac{1}{r^2} & - \frac{b\, {\mathfrak g} _1(br) v'}{r^2} & 0 & 0\\
 -\frac{v}{r^2} & - \frac{b\, {\mathfrak g} _1(br) v'}{r^2} & \frac{1}{r^2} & 0 & \frac{{\mathfrak v}_1(br) v''}{2 r^2} + v\\
 0 & 0 & 0 & \frac{1}{r^2} & 0\\
 -1 & 0 & \frac{{\mathfrak v}_1(br) v''}{2 r^2} + v & 0 & r^2 - \frac{1}{b^4 r^2}
\end{pmatrix} \ .
\]
 
\subsection{Christoffel symbols}
  
In Sec.~\ref{tsez} we will see that the leading contribution to the stress tensor comes from the second derivative of the velocity.  Contributions at that order come from the second derivative of the velocity in the curvature, which in turn contains contributions from first and second derivatives of the velocity in the Christoffel symbols, as well as from Christoffel symbols which are velocity independent as these are multiplied by velocity-dependent terms when calculating the curvature.  We will now calculate all of the Christoffel symbols up to first order in $v$, $v\p$ and $v\p\p$, although the $v$ terms will not contribute to the stress tensor.

We will begin with the terms at order $\mathcal{O}(v^0)$, these are just the Christoffel symbols of the static black brane 
\begin{align}
\Gamma_{tt}^t &= -\left(r+\frac{1}{b^4r^3}\right), & \Gamma_{tt}^r &=r^3 - \frac{1}{b^8 r^5}, & \nonumber\\
\Gamma_{xx}^t &= \Gamma_{yy}^t = \Gamma_{zz}^t = r, & \Gamma_{tr}^r &=  \Gamma_{rt}^r= r + \frac{1}{b^4 r^3}, &\nonumber\\
\Gamma_{xr}^x &= \Gamma_{rx}^x = \frac{1}{r}, & \Gamma_{yr}^y &=\Gamma_{ry}^y= \frac{1}{r}, & \nonumber\\
\Gamma_{zr}^z &= \Gamma_{rz}^z = \frac{1}{r}, & \Gamma_{xx}^r &= \Gamma_{yy}^r = \Gamma_{zz}^r = -\left(r^3-\frac{1}{b^4 r}\right). \label{Chris0}
\end{align}

The new terms, at order $\mathcal{O}(v)$, are
\begin{align}
\Gamma_{ty}^t &= \Gamma_{yt}^t = \frac{1}{4}b\,{\mathfrak v}'_1(br) v'' - \frac{v}{b^4 r^3}, \nonumber \\
\Gamma_{yr}^x &= \Gamma_{ry}^x = \frac{v'}{2 r^2}+\frac{b^2}{2} {\mathfrak g}_1(br) v'\nonumber\\
\Gamma_{tt}^y &= \left(r+ \frac{1}{b^4 r^3}\right)\left(\frac{{\mathfrak v}_1(br) v''}{2 r^2} + v\right) ,\nonumber \\
\Gamma_{tx}^y &= \Gamma_{xt}^y = \frac{v'}{2 b^4 r^4}\nonumber\\
\Gamma_{xx}^y &= b {\mathfrak g}_1(br) v'' - vr- \frac{{\mathfrak v}_1(br) v''}{2 r},\nonumber \\
\Gamma_{tr}^y &= \Gamma_{rt}^y = \frac{v}{r} + \frac{1}{4}\frac{{\mathfrak v}'_1(br) b v''}{r^2} \nonumber\\
\Gamma_{xr}^y &= \Gamma_{rx}^y = \frac{1}{2} b^2 {\mathfrak g}'_1(br) v' - \frac{v'}{2 r^2}, \nonumber 
\end{align}
\begin{align}
\Gamma_{yy}^y &= \Gamma_{zz}^y = - r v - \frac{{\mathfrak v}_1(br) v''}{2r}\nonumber\\
\Gamma_{ty}^r &= \Gamma_{yt}^r = \left(r^2-\frac{1}{b^4 r^2}\right)\left(\frac{v}{b^4 r^3}-\frac{1}{4} b {\mathfrak v}'_1(br) v'' \right),\nonumber \\
\Gamma_{ty}^x &= \Gamma_{yt}^x  =-\frac{v'}{2 b^4 r^4} \nonumber\\
\Gamma_{yr}^r &= \Gamma_{ry}^r = \frac{v}{b^4 r^3} - \frac{1}{4} b {\mathfrak v}'_1(br) v'' + \frac{{\mathfrak v}_1(br) v''}{2 r}+ v r, \nonumber \\
\Gamma_{yr}^t &= \Gamma_{ry}^t  =-\frac{v}{r}\nonumber \\
\Gamma_{xy}^r &= \Gamma_{yx}^r = -\frac{1}{2} r^2 v' - \left(r^2-\frac{1}{b^4 r^2}\right)\left(r b \, {\mathfrak g}_1(br) v' + \frac{1}{2} r^2 b^2 \, {\mathfrak g}'_1(br) v' \right)  \nonumber\\
\Gamma_{xy}^t &= \Gamma_{yx}^t = \frac{1}{2}\left(v'+2 r b\,{\mathfrak g}_1(br)v'+r^2 b^2 \,{\mathfrak g}'_1(br) v' \right)  . \label{Chris1}
\end{align}

\subsection{The Riemann tensor and the Ricci tensor and scalar}

Using the Christoffel symbols one can now easily compute the Riemann tensor.  The order $\mathcal{O}(v^0)$ terms again are just those of the static AdS black brane
\begin{align}
R_{trtr} &=  1 - \frac{3}{b^4 r^4} \\
R_{txtx} &= R_{tyty} = R_{tztz} = r^4 - \frac{1}{b^8 r^4} \\
R_{txrx} &= R_{tyry} = R_{tzrz} = r^2 + \frac{1}{b^4 r^2} \\
R_{xyxy} &= R_{xzxz} = R_{yzyz} = - r^4 + \frac{1}{b^4}. 
\end{align} 

The bulk stress tensor is entirely determined by the contributions to the Riemann tensor which do not solve the fluid equations of motion, as it is these that do not solve the vacuum Einstein equations.  If $v$ is a constant, this yields the boosted black brane, which satisfies the vacuum Einstein equations.  If $v$ is linear, then again this is a solution of the linear order fluid equations as we have checked above, and therefore as we will check below $v\p$ will not contribute to the gravitational stress tensor at order $\mathcal{O}(v)$.  Therefore the first nontrivial contributions to the stress tensor arise from the Riemann tensor at linear order in $v''$ ($v= v' = 0$) 
\begin{align}
R_{trty} &= - \frac{1}{2}\left(1+ \frac{1}{b^4 r ^4}\right) {\mathfrak v}_1(br) v'' \\
R_{xrxy} &= \frac{v''}{4}\left(2+ 2{\mathfrak v}_1(br)+2b^2 r^2\, {\mathfrak g}'_1(br)- b r {\mathfrak v}'_1(br)\right)\  \\
R_{zrzy} &=  \frac{v''}{4}\left(2{\mathfrak v}_1(br)- b r {\mathfrak v}'_1(br)\right)\  \\
R_{yxtx} &= - \frac{v''}{4 b^4 r^2}(2+(b^5 r^5- br){\mathfrak v}'_1(br)) \	  \\
R_{yrtr} &=  \frac{v''}{4}\left({\mathfrak v}'_1\,\frac{b}{r} - b^2 {\mathfrak v}''_1 \right) .  
\end{align}

As a check on our calculation, we will also calculate the contributions to the various tensors at linear order in the nondifferentiated velocity $v$
\begin{align}
R_{txyx} &= R_{tzyz} = \frac{1}{b^4}\left(1- \frac{1}{b^4 r^4}\right) v \  \\
R_{tytr} &= - \left(r^2+ \frac{1}{b^4 r^2}\right) v \  \\
R_{tryr} &= \left(1- \frac{3}{b^4 r^4}\right) v \   \\
R_{xyxr} &= R_{zyzr} = \left(r^2 + \frac{1}{b^4 r^2}\right) v 
\end{align}
and in $v\p$
\begin{align}
R_{txty} &= (b^3 r^3 + b^7 r^7 + (b^8 r^8 -1)(2 {\mathfrak g}_1(br)+ br {\mathfrak g}'_1(br))) \frac{v'}{2 b^7 r^4} \  \\
R_{txyr} &=  (-2 + b^4 r^4 + (b^5 r^5 + b r)(2 {\mathfrak g}_1(br)+ br {\mathfrak g}'_1(br))) \frac{v'}{2 b^4 r^3} \  \\
R_{tyxr} &=  - (2 + b^4 r^4 + (b^5 r^5 + b r)(2 {\mathfrak g}_1(br)+ br {\mathfrak g}'_1(br))) \frac{v'}{2 b^4 r^3} \  \\
R_{trxy} &= - \frac{2 v'}{b^4 r^3} \  \\
R_{xzyz} &= - (b^3 r^3 + (b^4 r^4 -1)(2 {\mathfrak g}_1(br)+ br {\mathfrak g}'_1(br))) \frac{v'}{2 b^3} \  \\
R_{xryr} &= - \frac{1}{2} b^2 r v' (2 {\mathfrak g}'_1(br)+ br {\mathfrak g}''_1(br)) \ .
\end{align}

The Ricci tensor is now easily calculated.  Again the order $\mathcal{O}(v^0)$ terms are those of the static black brane solution
\begin{align}
R_{tt} &= 4 r^2 - \frac{4}{b^4 r^2}  \\
R_{tr} &= R_{rt} = 4 \\
R_{xx} &= R_{yy} = R_{zz} = - 4r^2 \ .
\end{align}

Contributions to the stress tensor will arise from the $v''$ terms, ($v= v' = 0$) 
\begin{align}
R_{ty} &= R_{yt} = -\frac{v''}{4 b^4 r^4}(2 + 4(1+ b^4 r^4) {\mathfrak v}_1(br)+ (b^5r^5 -br)({\mathfrak v}'_1(br)+b r{\mathfrak v}''_1(br))) \\
R_{ry} &= R_{yr} = \frac{v''}{4 r^2}(2+ 4 {\mathfrak v}_1  + br(2 br \,{\mathfrak g}'_1(br)- {\mathfrak v}'_1(br)+ b r {\mathfrak v}''_1(br) ) \ .
\end{align}

The terms in the Ricci tensor proportional to $v$ are those of a rigidly boosted black brane
\begin{equation}
R^{(v)}_{ty} = - \frac{4 v}{b^4 r^2} \hsp
R^{(v)}_{yr} = 4 v \ 
\end{equation}
which provides an exact solution both to the hydrodynamic equations and to Einstein's equations with a negative cosmological constant.  Again the terms linear in $v'$ yield a solution to the fluid equations and so Einstein's equations, although only to linear order $\mathcal{O}(v)$
\begin{equation}
R_{xy} = - (8 b^3 r^3 {\mathfrak g}_1(br)+ (5 b^4 r^4 -1){\mathfrak g}'_1(br))\frac{v'}{2 b^2 r} \ .
\end{equation}

Using the large $r$ asymptotic expansions of Ref.~\cite{Bhattacharyya:2008mz} 
\begin{align}
{\mathfrak g}_1 &\sim \frac{1}{b r} - \frac{1}{4 b^4 r^4} + \dots \  \\
{\mathfrak v}_1 &\sim - \frac{1}{12 b^4 r^4}+ \frac{2}{5 b^3 r^3} + \dots \ 
\end{align}
we find that the asymptotic behaviors of the $v\p\p$ terms in the Ricci tensor are
\begin{align}
R_{ty} &\sim \frac{13}{10} \frac{v''}{b^3 r^3} \  \\
R_{ry} &\sim \frac{1}{5} \frac{v''}{b^2 r^4} \ . 
\end{align}

The Ricci scalar is
\begin{equation}
R = -20 \ .
\end{equation}
There is no contribution at order $\mathcal{O}(v)$ to the Ricci scalar.  This is guaranteed for any solution of the vacuum Einstein equations with cosmological constant $\Lambda=-6$, and so there could not have been any corrections from the $v$ and $v\p$ terms.  There are no corrections from the $v\p\p$ terms at linear order because the corresponding components of the inverse metric are themselves of order $\mathcal{O}(v)$, and so the contributions to the Ricci tensor are of order $\mathcal{O}(v^2)$.

\subsection{Contributions to the Ricci tensor at $\mathcal{O}(v^{(3)})$ and $\mathcal{O}(v^{(4)})$}

Before continuing with the calculation of the bulk stress tensor, we will pause to discuss some of the approximations that we have made.  We have made two truncations.  First, we have calculated everything at order $\mathcal{O}(v)$.  As we are working in units in which $c=1$, $v$ is small for nonrelativistic speeds and so this is a valid approximation in a region in which the flow is sufficiently slow.

A more dangerous truncation is that of higher derivatives of the velocity.  The gravity/hydrodynamics correspondence is a one to one map between gravitational and fluid solutions in a derivative expansion.  More precisely, the $k$th order map relates the truncation of the fluid equations to $k$ derivatives and that of the gravity equations to $(k+1)$ derivatives.  The iterative procedure described in Ref.~\cite{Bhattacharyya:2008jc} in principle determines this map for all $k$, however in practice this map has only been determined to order $k=2$.  In other words, it provides a metric as a function of $v$, $v\p$ and $v\p\p$, however a perfect matching with Einstein's equations would require also corrections involving the higher derivatives $v^{(k)}$ which are not known.

General arguments based on dimensional analysis suggest that these corrections become smaller at higher $k$.  In general one expects that each derivative leads to a contribution which is subdominant by a factor of $Tl$ with respect the previous derivative, where $l$ is the distance scale of the derivative.  Ideally one would like to check this claim for all terms with, say, three or four derivatives.  However this would require a knowledge of the map at orders $k=3$ and $k=4$.

The map at order $k=2$, which we have used, does produce some terms in the curvature which depend on the third and fourth derivatives of $v$.  In this subsection we will verify that two of these have the expected convergence scaling, and determine the corresponding condition on our fluid flow.  In other words, we determine a necessary condition for the derivative expansion to apply to our flow.

The Ricci tensor components $R_{xy}$ and $R_{ty}$ have corrections from the third and fourth derivatives of the velocity respectively 
\begin{align}
R_{xy}^{(3)} &= -\frac{v^{(3)}(x) \left(b r {\mathfrak v}'_1(b r)+{\mathfrak v}_1(b r)\right)}{4 r} \  \\
R_{ty}^{(4)} &= -\frac{v^{(4)}(x) {\mathfrak v}_1(b r)}{4 r^2} \ .
\end{align}
We want to determine the condition under which $R_{ty}^{(4)}$ is subdominant to $R_{xy}^{(3)}$.  As the higher derivatives of $v$ define an interpolating function between two solutions over an interval of length $d$, each derivative is larger than the previous one by about $1/d$.  In other words, $\partial_x\sim 1/d$.  

To test the subdominance of $R_{ty}^{(4)}$, it is sufficient to compare it to the similar term in $R_{xy}^{(3)}$, which contains ${\mathfrak v}_1$.  The ratio of these terms is
\begin{equation}
\frac{R_{ty}^{(4)}}{R_{xy}^{(3)}} \sim \frac{v^{(4)}(x)}{rv^{(3)}(x)} \sim \frac{1}{r d}
\end{equation}
therefore the fourth order term is subdominant if $d\gg1/r$ in the entire bulk.  The bulk extends from the horizon at $r=1/b=\pi T$ to the boundary at $r=\infty$.  Therefore convergence requires
\begin{equation}
d\gg\frac{1}{\pi T}. \label{spesso}
\end{equation}
This fourth order term is suppressed by $\pi d T$ with respect to the third order term, in line with the above expectations from dimensional analysis.  This means that the gravity duality procedure is only convergent when $d$ is sufficiently large.  Of course, the duality never yields a solution of the vacuum Einstein equations, and so one may argue that its convergence is immaterial.  Nonetheless, it is only well-defined as a series when $d$ satisfies (\ref{spesso}).

\subsection{The static black brane solution}

As a check on our calculation and conventions, we recover that the static $(v=0)$ black brane satisfies the vacuum Einstein equations with cosmological constant $\Lambda=-6$
\begin{equation}
R_{\mu \nu} - \frac{1}{2} R g_{\mu \nu} = \left(
\begin{array}{ccccc}
 \frac{6}{b^4 r^2}-6 r^2 & 0 & 0 & 0 & -6 \\
 0 & 6 r^2 & 0 & 0 & 0 \\
 0 & 0 & 6 r^2 & 0 & 0 \\
 0 & 0 & 0 & 6 r^2 & 0 \\
 -6 & 0 & 0 & 0 & 0
\end{array}
\right) \ . \label{bn}
\end{equation}

\section{Two Calculations of the Stress Tensor} \label{tsez}

In this section we will calculate the bulk stress tensor of the surface layer interpolating between the vacuum gravity solutions using two different methods, corresponding to two different metrics.  First, we will apply the duality map of Ref.~\cite{Bhattacharyya:2008jc} to a fluid flow which interpolates between the two solutions, the stationary solution on the left and the linear velocity solution on the right.  In this case, as we have seen, the interpolating region is necessarily larger than the inverse temperature.  Next, we will directly interpolate between the gravitational solutions using the Israel matching conditions \cite{Israel:1966rt}.  This method requires the interpolating region to be very thin, and uses the assumption that in this limit the extrinsic curvature remains bounded.

\subsection{Interpolating between the hydrodynamic flows} \label{T1sez}

The duality map of Ref.~\cite{Bhattacharyya:2008jc} takes a fluid flow and yields a dual metric.  This dual metric solves the vacuum Einstein equations when the fluid flow satisfies the hydrodynamic equations of motion (\ref{edm}).  If the flow does not satisfy the equations of motion, the dual metric does not satisfy the vacuum Einstein equations.  Thus apparently there is no benefit in using this map over any other map.  However we will use the map, and observe the consequences.  The resulting dual metric will necessarily solve Einstein's equations with some value of the stress tensor
\begin{equation}
8\pi G_N T_{\mu \nu} = R_{\mu \nu} - \frac{1}{2} R g_{\mu \nu} + \Lambda g_{\mu \nu} \ . \label{ein}
\end{equation}
We will determine this value. 

We saw in Eq.~(\ref{bn}) that there is no contribution to the stress tensor at order $\mathcal{O}(v^0)$.  We have argued that, at order $\mathcal{O}(v)$, the dominant contributions to the stress tensor are proportional to $v\p\p$.  These are easily found from (\ref{ein}) to be 
\begin{align}
T_{ty} &= \frac{v''(x) \left(4 \left(b^4 r^4-1\right) {\mathfrak v}_1(b r)-b r \left(b^4 r^4-1\right) \left({\mathfrak v}'_1(b r)+b r {\mathfrak v}''_1(b
   r)\right)-2\right)}{32\pi G_N b^4 r^4} \, \\
T_{ry} &= \frac{v''(x) \left(4 {\mathfrak v}_1(b r)+b r \left(2 b r {\mathfrak g}'_1(b r)-{\mathfrak v}'_1(b r)-b r
   {\mathfrak v}''_1(b r)\right)+2\right)}{32\pi G_N r^2}\ .
\end{align}
There appears to also be a contribution proportional to $v\p$
\begin{equation}
T_{xy} = -\frac{v'(x) \left(b r \left(\left(b^4 r^4-1\right) {\mathfrak g}''_1(b r)+3 b r\right)+\left(5 b^4 r^4-1\right) {\mathfrak g}'_1(b r)\right)}{16\pi G_N b^2 r}. \label{txy}
\end{equation}
At order $v\p$ one expects no contributions to the stress tensor, as a solution with a linear velocity satisfies the fluid equations at order $\mathcal{O}(v)$.  Therefore a nontrivial contribution would be in contradiction with the gravity/hydrodynamics correspondence.  We will see shorty that this contribution is in fact equal to zero.

The functions ${\mathfrak v}_1(r)$ and ${\mathfrak g}_1(r)$ are defined as
\begin{align}
{\mathfrak v}_1(r) &= \frac{2}{r^2} \int_{r}^{\infty} dx \, x^3 \int_x^{\infty} dy \frac{y-1}{y^3(y^4-1)} \label{v1}\\
{\mathfrak g}_1(r) &= \int_{r}^{\infty} dx \, \frac{x^3-1}{x(x^4-1)} \ . \label{g1}
\end{align}
Integrating we \cite{Bhattacharyya:2008mz} obtain analytical expressions for ${\mathfrak v}_1(r)$ and ${\mathfrak g}_1(r)$ 
\begin{align}
{\mathfrak v}_1 &= - \frac{1}{4} + \frac{r}{2} + \frac{1}{8 r^2}(r^4-1)\left(\log{\frac{(r^2+1)}{(r+1)^2}}+2 \tan ^{-1}(r)-\pi\right) \\
{\mathfrak g}_1 &= \frac{1}{4} \left(\log{\left(\frac{(1+r)^2(1+r^2)}{r^4}\right)}-2 \tan ^{-1}(r)+\pi \right) \  .
\end{align}
The derivatives of these expressions are
\begin{align}
{\mathfrak g}'_1 &= \frac{1}{2} \left(\frac{r}{r^2+1}-\frac{1}{r^2+1}+\frac{1}{r+1}-\frac{2}{r}\right)\  \label{gp} \\
{\mathfrak g}''_1 &= -\frac{r^2}{\left(r^2+1\right)^2}+\frac{r}{\left(r^2+1\right)^2}+\frac{1}{2(r^2+1)}+\frac{1}{r^2}-\frac{1}{2(r+1)^2} \  \label{gpp}
\end{align}
and
\begin{multline}
{\mathfrak v}'_1 = -\frac{1}{4 r^3} \left( \pi  r^4+2 \left(r^4+1\right) \log (r+1)-2 \left(r^4+1\right) \tan ^{-1}(r) \right. \\ 
\left. -4 r^3+2 r^2-\left(r^4+1\right) \log \left(r^2+1\right)+\pi\right) 
\end{multline}

\begin{multline}
{\mathfrak v}''_1 = -\frac{1}{4 r^4 (r+1) \left(r^2+1\right)}   \left(3 \log \left(r^2+1\right)+\pi  (r+1) \left(r^2+1\right) \left(r^4-3\right)  \right.\\
+\left(r (r+1) \left(r^4+r^2-3\right)-3\right) r \left(2 \log
   (r+1) \right. \\
     -\log\left.  \left(r^2+1\right)\right)-2 (r+1) \left(r^2+1\right) \left(r^4-3\right) \tan ^{-1}(r)\\
    \left. -2 \left(2 r^4+r^3+r^2+r+3\right) r^2-6 \log
   (r+1)\right) .
\end{multline}

The explicit formula Eqs.~(\ref{gp}) and (\ref{gpp}) for the derivatives of ${\mathfrak g}_1$ can be combined to show that
\begin{equation}
r \left(\left(r^4-1\right) {\mathfrak g}''_1(r)+3 r\right)+\left(5 r^4-1\right){\mathfrak g}'_1(r) = 0 \ .
\end{equation}
This combination is proportional to formula (\ref{txy}) for $T_{xy}$, therefore
\begin{equation}
T_{xy}=0
\end{equation}
and there are no contributions proportional to $v\p$.  

Similarly one may evaluate the combination of functions that appears in $T_{ry}$ 
\begin{align}
r \left(2 r {\mathfrak g}'_1(r)-r {\mathfrak v}''_1(r)-{\mathfrak v}'_1(r)\right)+4 {\mathfrak v}_1(r)+2 = 0 \ .
\end{align}
This implies that
\begin{equation}
T_{ry}=0
\end{equation}
leaving only $T_{ty}$, the momentum in the $y$ direction.  Thus the bulk stress tensor contains no stress, only momentum.

We may use the exact expressions for the functions ${\mathfrak g}_1$ and ${\mathfrak v}_1$ to simplify the only nonvanishing component of the stress tensor \cite{Evslin:2010ss}
\begin{equation}
T_{ty}= -\frac{v\p\p (x)}{16\pi G_N b^3 r^3} \ . \label{tty}
\end{equation}
Using the fundamental theorem of calculus, this may be integrated over the interpolating region to obtain
\begin{equation}
\int_0^d dx\ T_{ty}= -\frac{v\p}{16\pi G_N b^3 r^3} 
\end{equation}
where $v\p$ is the derivative of the velocity in the region $x>d$.  In particular, at this leading order the integrated stress tensor of the surface layer is independent of the interpolation and independent of $d$.  Of course it still depends on the map that we used to generate the dual metric.

Had the $v\p$ term been the dominant contribution, the stress tensor would have been constant, and so the integral would be have proportional to $d$.  Similarly a $v^{(3)}$ term would have led to a stress tensor proportional to $1/d$, and higher powers of $v$ to other scalings.  Therefore it is somewhat nontrivial that the leading contribution to the integrated stress tensor is in fact $d$-independent.  Clearly this $d$-independence is desirable, as $d$ is not a physical quantity but merely an artifact of the scheme that we used to regularize the divergent second derivative of the fluid velocity.

The bulk stress tensor does not satisfy any of energy conditions we discussed in Sec.~\ref{Energysec}, not even the null energy condition.  As the only nonvanishing component is $T_{ty}$, the only nonvanishing product of a null vector $w$ and the stress tensor is
\begin{equation}
w^\perp Tw=2w^t T_{ty} w^y. \label{prod}
\end{equation}
As $T_{ty}$ is already of order $\mathcal{O}(v)$, at order $\mathcal{O}(v)$ one need only consider the terms in $w$ of order $\mathcal{O}(v^0)$.  That is to say, $w$ only needs to be null with respect to the static black brane metric.  Consider for example the null vectors $w_\pm$
\begin{equation}
w^t_{\pm}=r\hsp
w^y_{\pm}=\pm r\sqrt{1-\frac{1}{b^4r^4}}. \label{w}
\end{equation}
The product (\ref{prod}) is 
\begin{equation}
w_\pm^\perp Tw_\pm=\mp\frac{v''(x)}{8\pi G_N b^3 r}\sqrt{1-\frac{1}{b^4r^4}} \label{prod2}
\end{equation}
which is nonzero.  However $w_+$ and $w_-$ yield opposite signs, as incidentally do the two choices of signs of $v$.  Therefore at least one of these will yield a negative product, and so the bulk stress tensor does not satisfy the null energy condition.  This may or may not mean that no external matter may be consistently added which produces such a surface layer.

%\subsection{Asymptotic behavior}
 
% The component $G_{ty}$ of Einstein Tensor is
% \begin{equation}
% G_{ty} = \frac{v''(x) \left(-b r \left(b^4 r^4-1\right) \left(b r {\mathfrak v}''_1(b r)+{\mathfrak v}'_1(b r)\right)+4 \left(4 b^4 r^4-1\right) {\mathfrak v}_1(b
%   r)-2\right)}{4 b^4 r^4} \ ,
% \end{equation}
% and using exact solution for ${\mathfrak v}_1$ and its derivatives we get
% \begin{multline}
% G_{ty}  =  \frac{1}{8 b^3 r^3}  \left[v''(x) \left(3 b r \left(\pi  \left(-b^4\right) r^4+2 b^2 r^2 (2 b r-1)+\pi \right) \right. \right. \\
%\left. \left. +3 b r \left(b^4 r^4-1\right) \left(\log \left(b^2
%   r^2+1\right)-2 \log (b r+1)+2 \tan ^{-1}(b r)\right)-4\right) \right] ,
% \end{multline}
% that is non zero and has asymptotic behaviour 
%\begin{equation}
% G_{ty}  \sim \frac{7}{10} \frac{v''(x)}{b^3 r^3} \ .
%\end{equation}
 
% The component $T_{ty}$ of stress tensor (Einstein Tensor minus the cosmological constant term) is
% \begin{equation}
% T_{ty} = \frac{v''(x) \left(-b r \left(b^4 r^4-1\right) \left(b r {\mathfrak v}''_1(b r)+{\mathfrak v}'_1(b r)\right)+4 \left(b^4 r^4-1\right) {\mathfrak v}_1(b
%   r)-2\right)}{32\pi b^4 r^4} \ ,
% \end{equation}
% and using exact solution for ${\mathfrak v}_1$ and its derivatives we get
%\begin{equation}
%T_{ty}= -\frac{v''(x)}{16\pi b^3 r^3} \ .
%\end{equation}
%Using instead the asymptotic behaviour of $ {\mathfrak v}_1$ we get
%\begin{equation}
%T_{ty} \sim -\frac{v''(x)}{16\pi b^3 r^3} \ .
%\end{equation}

\subsection{Israel's matching conditions on the gravity duals}

We will now calculate the bulk stress tensor in a different geometry.  Following Ref.~\cite{Israel:1966rt} and discussion in Sec.~\ref{Israelsec}, we will consider the vacuum Einstein solution corresponding to a static fluid on the left and that corresponding to a linear velocity flow on the right.  These solutions will be glued together by interpolating continuously between the two metrics over a distance $d$ and taking the limit $d\rightarrow 0$ such that the extrinsic curvature remains bounded.  In Ref.~\cite{Israel:1966rt}, Israel has shown that the resulting configuration contains two solutions separated by a surface layer whose bulk stress tensor is independent of the interpolation used.

Following Ref.~\cite{Israel:1966rt}, the first step in the calculation of the stress tensor is the definition of the unit normal vector to the hyperplane
\begin{equation}
n_\mu = \{0, r, 0, 0, 0\} \ ,
\end{equation}
which satisfies the normalization condition
\begin{equation}
n_\mu g^{\mu \nu}n_\nu = \frac{1}{r^2}(n_x)^2 = 1 \ .
\end{equation}
The surface layer $\Sigma$ extends along all of the directions except for the $x$ direction.  A basis of tangent vectors to $\Sigma$ is
\begin{equation}
ds = e_{(i)} d x^i 
\end{equation}
where
\begin{align}
e_{(t)} &= \{1, 0, 0, 0, 0\} \  \\ 
e_{(y)} &= \{0, 0, 1, 0, 0\} \  \\
e_{(z)} &= \{0, 0, 0, 1, 0\} \   \\
e_{(r)} &= \{0, 0, 0, 0, 1\} \ .
\end{align}
In terms of these tangent vectors the extrinsic curvature may be calculated as
\begin{equation}
K_{ij} = e_{(j)}\cdot \nabla_j n = \frac{\partial n_j}{\partial x^i} - n^m \Gamma_{m,ji} = \frac{\partial n_j}{\partial x^i} - n_m \Gamma^m_{ji} \ . \label{K}
\end{equation}

On the left, where the fluid is static ($v = 0$), substituting (\ref{Chris0}) into (\ref{K}) one finds no extrinsic curvature $K^{(-)}$
\begin{align}
K_{ty}^{(-)} &= - r \Gamma^x_{ty} = 0 \  \\
K_{yr}^{(-)} &= - r \Gamma^x_{yr} =  0 \  .
\end{align}
On the right, where the fluid velocity is linear, the Christoffel symbols of Eq.~(\ref{Chris1}) yield a nontrivial extrinsic curvature $K^{(+)}$.
\begin{align}
K^{(+)}_{ty} &= - r \Gamma^x_{ty} = \frac{v'}{2 b^4 r^3} \  \\
K^{(+)}_{yr} &= - r \Gamma^x_{yr} =  \frac{v'}{2 r}\left(1 + b^2 r^2 {\mathfrak g}'_1(br) \right) \ . 
\end{align}

The tensor $\gamma_{ij}$ (\ref{gammaij}) is defined to be the difference between the extrinsic curvatures on the two sides of the surface layer
\begin{equation}
\gamma_{ij} = K^{(+)}_{ij} - K^{(-)}_{ij} \ .
\end{equation}
The bulk stress tensor integrated over $x$ is equal to the tensor $S_{ij}$, defined by 
\begin{equation}
-8 \pi G_N  S_{ij} = \gamma_{ij} - g_{ij} \gamma_{m}^m \ . \label{Sdef}
\end{equation}
%where $\chi$ is Einstein's constant of gravitation
%\begin{equation}
%\chi = 8 \pi G_N \ .
%\end{equation} 

The expression (\ref{Sdef}) for the integrated bulk stress tensor was derived in \cite{Israel:1966rt} for a 4-dimensional space with no cosmological constant.  While several factors in the derivation change in our current 5-dimensional situation, Eq.~(\ref{Sdef}) remains unchanged.  The cosmological constant term yields a contribution proportional to the integral of $\Lambda$ times the metric integrated over the thickness $d$ of the surface layer.  As the metric is taken to be finite, this term vanishes in the $d\rightarrow 0$ limit.

The trace of $\gamma$ is $O(v^2)$, therefore (\ref{Sdef}) yields the integrated bulk stress tensor
\begin{align}
S_{ty} &= - \frac{v'}{16\pi G_N b^4 r^3} \  \\
S_{yr} &=  - \frac{v'}{16\pi G_N r}\left(1 + b^2 r^2 {\mathfrak g}'_1(br) \right) \ .
\end{align}
These are equal to the integrals over the $x$ direction\footnote{Note that, following Ref.~\cite{Israel:1966rt}, the measure of this integral must be that of $x$ rescaled to normal coordinates.  Therefore the integral contains an additional factor of $r=\sqrt{g_{xx}}$.} of the stress tensors $T^{(1)}$ of Subsec.~\ref{T1sez} at order $k=1$, in other words, without the ${\mathfrak v}_1$ term that entered into the metric (\ref{metfin}) multiplied by $v\p\p$ \cite{Evslin:2010ss}
\begin{align}
T^{(1)}_{ty} &= - \frac{v''}{16\pi G_N b^4 r^4} \  \\
T^{(1)}_{yr} &=  - \frac{v''}{16\pi G_N r^2}\left(1 + b^2 r^2 {\mathfrak g}'_1(br) \right) \ .
\end{align}
The ${\mathfrak v}_1$ terms arose from the dualization of the interpolating region, which did not satisfy the equations of motion.  It therefore cannot enter into the Israel calculation, which uses only the solutions of the vacuum Einstein equations.  Indeed, the ${\mathfrak v}_1$ terms in (\ref{metfin}) are singular in the limit $d\rightarrow 0$ as $v\p\p$ diverges as $1/d$, and therefore the boundedness of the extrinsic curvature assumed in Israel's derivation fails for the metric interpolation (\ref{metfin}).

Like the stress tensor (\ref{tty}) calculated by interpolating the hydrodynamic flow, the Israel stress tensor does not satisfy the null energy conditions.  Again, to linear order in $v$, one may consider vectors which are null with respect to the static black brane metric.  Therefore, again one may consider the null vectors $w_\pm$ of Eq.~(\ref{w}).  As $T_{ty}$ is, at least for any finite $d$, equal to that of Subsec.~\ref{T1sez} divided by the positive combination $br$, the sign of the inner product (\ref{prod2}) is unchanged.  Therefore the null energy condition is also violated by this stress tensor.  

The main difference between the two stress tensors is then that $T_{yr}$ does not vanish for the Israel tensor.  Remembering that in our Gaussian null coordinates the $r$ direction is the sum of a spatial and temporal piece, the spatial component implies that there is a nonzero stress.  More precisely, while both Israel's thin surface layer and the thick fluid surface layer have a nonvanishing $y$ momentum, the Israel surface layer also has a flux of this $y$ momentum in the radial direction, from the boundary into the horizon of the black brane.  As the black brane is infinite in the $x$ direction, this is not problematic for the time-independence of the solution.

\chapter*{Conclusions \& future directions}
\addcontentsline{toc}{chapter}{Conclusions \& future directions}

Turbulence often arises as a result of the boundary conditions placed on a fluid.  As a preliminary step towards an understanding of turbulence in gravity, we have proposed two gravitational duals of such boundaries.  Both of these duals involve the addition of a surface layer of matter, with a certain stress tensor.  These proposals are in a sense trivial, as the dynamics of the duals is defined not by any known equations of motion, but by the duality map itself.  It remains to be shown whether such matter can exist.  For example, even if the equations of motion which it obeys can be found, the existence of a UV completion of the matter theory may be fundamentally obstructed as in Ref.~\cite{nima}.  Or the failure of the null energy condition may imply that, whatever the ultraviolet theory may be, the wall simply disintegrates before it has any significant effect on the fluid.  

Of course, an ultraviolet completion is not necessarily a prerequisite for learning something interesting about whatever the gravitational dual to turbulence may be.  After all, no ultraviolet completion of Einstein gravity is used in this correspondence.  The surface layer implies the existence of equations of motion which are distinct from the Einstein vacuum equations and perhaps pathological.  However the interesting part of the fluid, the turbulent part, is not at the wall.  For example, if we consider the motion of a fluid in a pipe, the flow may be turbulent throughout the interior of the pipe.  The ultralocality of the duality map implies that, at a distance greater than $1/T$ from the pipe, the vacuum Einstein equations are still satisfied by the gravity dual.   Thus in a sense the ultralocality decouples the problem of understanding turbulence in gravity from the problem of defining a gravity dual of a boundary.

Besides trying to characterize the gravitational dual of turbulent flow, the other interesting question is to find the gravitational dual of the conditions under which turbulence can occur.  In nonrelativistic, incompressible flows, turbulence is expected when the product of a system's characteristic scale $L$ times the characteristic velocity $v$ of a fluid is much greater than the kinematic viscosity.  In Ref.~\cite{Bhattacharyya:2008ji}, the authors claim that for the conformal fluids dual to AdS black branes, turbulence is expected when $LTv\gg1$, where $T$ is the temperature of the fluid.  The AdS/hydrodynamics correspondence is expected to be reliable at scales $L$ such that $LT\gg1$.  Therefore since $v<1$, it appears that whenever turbulence is expected, $LT>LTv\gg1$ and so the correspondence can be trusted at least for quantities that vary over a distance $L$.  $(3+1)$-dimensional turbulence is characterized by vortices of various sizes from $L$ down to the dissipation scale \cite{Rich}.  Thus the duality appears to be reliable at least for the largest vortices in a turbulent flow.  The dissipation scale is a function of $L$, $T$ and $v$, and so in principle one may determine whether or not the duality is reliable for vortices all of the way down to this scale and so for the entire flow.

Understanding the gravity duals of turbulent flows, as described above, may yield new insights into the dynamics of black branes in AdS space, perhaps revealing a surprising difference between branes in AdS$_4$ and AdS$_5$, or indicating that generically they come with funnels attached as in Refs.~\cite{MarolfRang}.  The main weakness of this program is the dependence on asymptotically AdS geometry in the duality map of Ref.~\cite{Bhattacharyya:2008jc}.  There was no such restriction in the original correspondence of Ref.~\cite{D1979}, nor in other identifications of black holes and viscous fluids such as the blackfold program of Refs.~\cite{bf1,bf2} and the Wilsonian identification of Ref.~\cite{Strominger}.  An extension of turbulence to asymptotically Minkowski space could relate (3+1)-dimensional fluid dynamics to wealth of studies of asymptotically Minkowski 5d black objects, such as Refs.~\cite{5d}.  More importantly, relaxing the asymptotically AdS condition may mean that fluid mechanics, perhaps in only 2+1 dimensions, has something to teach us about real world gravity.  
 
 \appendix
 
 \chapter{Conventions \& Notation}
 
We work in the $(-++\ldots)$ signature. $\mu,\nu$ denote space-time indices, $i,j=1 \ldots k$ label the $k$ different conserved charges. The dimension of the spacetime in which the conformal fluid lives is denoted by $d$. In the context of AdS/CFT, the dual AdS$_{d+1}$ space has $d+1$ spacetime dimensions.\\
 
 We adopt standard symmetrization and anti-symmetrization conventions. For any tensor $F_{ab}$ we define the symmetric part as
 \begin{equation}
 F_{(ab)} = \frac{1}{2}\, \left(F_{ab} + F_{ba}\right) 
 \end{equation}
 and the anti-symmetric part as
 \begin{equation}
 F_{[ab]} = \frac{1}{2}\left( F_{ab} - F_{ba}\right) \ .
 \end{equation}
 We also use ${\mathscr D}$ to indicate the velocity projected covariant derivative: 
 \begin{equation}
 {\mathscr D}\equiv u^\mu \, \nabla_\mu \ .
 \end{equation}
 For any two tensor ${\mathcal T}^{\mu\nu}$ we denote the symmetric traceless projections transverse to the velocity field as:
\begin{equation}
{\mathcal T}^{\langle\mu\nu\rangle} = P^{\mu\alpha} \, P^{\nu\beta}\, {\mathcal T}_{(\alpha \beta)} - \frac{1}{d-1}\, P^{\mu\nu} \, P^{\alpha \beta}\, {\mathcal T}_{\alpha\beta} \ .
\label{angleb}
\end{equation}	

\section{Integral functions}

We list in the following the two different conventions on integral functions entering in the second order metric dual to fluid solutions on the boundary. The definitions adopted in many works ({\it e.g. \cite{Bhattacharyya:2008mz}}) are

 \begin{align}
F(br) &\equiv \int_{br}^{\infty}\frac{y^{d-1}-1}{y(y^{d}-1)}dy ,\\
H_1(br)&\equiv \int_{br}^{\infty}\frac{y^{d-2}-1}{y(y^{d}-1)}dy ,\\
H_2(br)&\equiv \int_{br}^{\infty}\frac{d\xi}{\xi(\xi^d-1)}
\int_{1}^{\xi}y^{d-3}dy \left[1+(d-1)y F(y) +2 y^{2} F'(y) \right] \\
&=\frac{1}{2} F(br)^2-\int_{br}^{\infty}\frac{d\xi}{\xi(\xi^d-1)} \int_{1}^{\xi}\frac{y^{d-2}-1}{y(y^{d}-1)}dy , \\
K_1(br) &\equiv \int_{br}^{\infty}\frac{d\xi}{\xi^2}\int_{\xi}^{\infty}dy\ y^2 F'(y)^2 , \\
K_2(br) &\equiv \int_{br}^{\infty}\frac{d\xi}{\xi^2}\left[1-\xi(\xi-1)F'(\xi) -2(d-1)\xi^{d-1} \right. \nonumber \\
&\left. \quad +\left(2(d-1)\xi^d-(d-2)\right)\int_{\xi}^{\infty}dy\ y^2 F'(y)^2 \right] ,\\
L(br) &\equiv \int_{br}^\infty\xi^{d-1}d\xi\int_{\xi}^\infty dy\ \frac{y-1}{y^3(y^d-1)} \ .
\end{align}

Definitions adopted in \cite{Rangamani:2009xk} are

 \begin{align}
{\mathfrak g}_1(br)  &\equiv \int_{br}^{\infty}\frac{y^{d-1}-1}{y(y^{d}-1)}dy ,\\
{\mathfrak g}_2(br)  &\equiv 2\,(rb)^2 \int_{br}^{\infty}\frac{d\xi}{\xi^2}\int_{\xi}^{\infty}dy\ y^2 F'(y)^2 , \\
{\mathfrak g}_3(br) &\equiv  (br)^2 \int_{br}^{\infty}\frac{y^{d-2}-1}{y(y^{d}-1)}dy ,\\
{\mathfrak g}_4(br) &\equiv (br)^2 \int_{br}^{\infty}\frac{d\xi}{\xi(\xi^d-1)}
\int_{1}^{\xi}y^{d-3}dy \left[1+(d-1)y F(y) +2 y^{2} F'(y) \right] \\
{\mathfrak v}_1(br) &\equiv \frac{2}{(br)^{d-2}} \int_{br}^\infty\xi^{d-1}d\xi\int_{\xi}^\infty dy\ \frac{y-1}{y^3(y^d-1)} \ . \\
{\mathfrak v}_2(br)  &\equiv \frac{1}{2\, (br)^{d-2}}  \int_{br}^{\infty}\frac{d\xi}{\xi^2}\left[1-\xi(\xi-1)F'(\xi) -2(d-1)\xi^{d-1} \right. \nonumber \\
&\left. \quad +\left(2(d-1)\xi^d-(d-2)\right)\int_{\xi}^{\infty}dy\ y^2 F'(y)^2 \right] ,
\end{align}

and therefore one can easily move across definitions noticing that 
 
 \begin{align}
{\mathfrak g}_1(br) &= F(br) , \qquad \qquad \, \, \, \, \, \,{\mathfrak g}_2(br) = 2\,(rb)^2 K_1(br), \nonumber \\
{\mathfrak g}_3(br) & = (br)^2 H_1(br) , \qquad \, \, \, {\mathfrak g}_4(br) = (br)^2 H_2(br) , \nonumber \\
{\mathfrak v}_1(br) &=\frac{2}{(br)^{d-2}} L(br) , \qquad{\mathfrak v}_2(br) = \frac{1}{2\, (br)^{d-2}} K_2(br) .
\end{align}

\section{List of symbols adopted}

We summarise in table A.1 some of the symbols we adopted. The relevant equations defining symbols are denoted by their respective equation numbers appearing inside parenthesis.\\

\begin{table}[h!] \footnotesize
  \centering
   \begin{tabular}{||r|l||r|l||}
    \hline

    Symbol & Definition & Symbol & Definition \\
    \hline
    $d$ & dimensions of spacetime & $p$, ${\rm p}$ & Pressure \\
    $s$ & Proper entropy density & $\rho$ & Energy density  \\
    $T$ & Fluid temperature & $\mu_I$ & Chemical potentials of the fluid \\
    $\zeta$ & Bulk viscosity & $\eta$ & Shear viscosity measured at \\
    $q_I$ & charge density &  \, & zero shear and vorticity \\
    $\widetilde{\gamma}_I$ & Contr. en. density to charge curr. & $\mho_I$  & Pseudo-vector transport coeff. \\
    $\widetilde{\varkappa}_{IJ}$ & Matrix of charge diffusion coeff. & \, & \\
    \hline

    $T^{\mu\nu}$ & Energy-momentum tensor & $J^\mu_S$ & Entropy current \\
    $J^\mu_i$ & Charge currents & $u^\mu$ & Fluid velocity ($u^\mu u_\mu =-1$) \\
    $g_{\mu\nu}$ & Spacetime metric & $P^{\mu\nu}$ & Projection tensor, $g^{\mu\nu}+u^\mu u^\nu$ \\
    $a^\mu$ & Fluid acceleration (\ref{a}), (\ref{transdyquant}) &$\vartheta$&  Fluid expansion (\ref{theta}), (\ref{transdyquant}) \\
    $\sigma_{\mu\nu}$ & Shear strain rate (\ref{sigma}) and (\ref{transdyquant}) & $\omega_{\mu\nu}$ & Fluid vorticity (\ref{omega}), (\ref{transdyquant})\\
    $\ell^\mu$ & See (\ref{elle})  and (\ref{transdyquant}) &  &\\
    \hline

     $\mathcal{D}_\mu$ & Weyl-cov. deriv. (\ref{wderiv}) and \ref{wderiv2}) & $\mathcal{A}_\mu$ & See (\ref{wconn})\\
     $\nabla_{\mu}$ & Lorentz-covariant derivative & $\Gamma_{\mu\nu}{}^\lambda$ & Christoffel connection  \\
     $R_{\mu\nu\lambda}{}^{\sigma}$ & Riemann Curvature & $\mathcal{R}_{\mu\nu\lambda}{}^{\sigma}$ & See(\ref{Reimannconf})\\
     $\mathcal{F}_{\mu\nu}$ & $\nabla_\mu\mathcal{A}_\nu-\nabla_\nu\mathcal{A}_\mu$  & $\mathcal{R}$& See (\ref{ricciscalconf}) \\
     $R_{\mu\nu},R$ & Ricci tensor/scalar & $\mathcal{R_{\mu\nu}}$ & See (\ref{ricciconf}) \\
      $G_{\mu\nu}$ & Einstein tensor & $C_{\mu\nu\lambda\sigma},$ & Weyl Curvature (\ref{Weyltensor})\\
      $S_{\mu\nu},$ & Schouten tensor (\ref{Schouten}) &  $\mathcal{C}_{\mu\nu\lambda\sigma}$&  See (\ref{Weyltensor}) \\
      $\mathcal{S}_{\mu\nu}$ &  See (\ref{Schouten}) &  $ {\mathfrak T}_i^{\mu\nu}$ & See Table 2.1 \\
     \hline
\end{tabular}
\caption{List of principal symbols \& definitions adopted.}
\end{table}

 \backmatter

% \chapter{On the theory of strings\\ as a research programme}

% \epigraph{The scientific status of a theory is its falsifiability, or refutability, or testability.}{K. Popper\\ Conjectures and Refutations}

%\epigraph{One may rationally stick to a degenerating research programme until it is overtaken by a rival and even after. What one must not do is to deny its poor public record [...]. It is perfectly rational to play a risky game: what is irrational is to deceive oneself about the risk.}{I. Lakatos\\ History of science and its rational reconstruction} 

%_____________________________________________
% Bibliography
%_____________________________________________


\newpage
\begin{thebibliography}{99}

  \bibitem{EIH}
  A.~Einstein, L.~Infeld and B.~Hoffmann,
  ``The Gravitational equations and the problem of motion,''
  Annals Math.\  {\bf 39} (1938) 65.

\bibitem{D1979}
T. Damour,
in: ``Quelques propri\'et\'es m\'ecaniques, \'electromagn\'etiques, thermodynamiques
et quantiques des trous noirs''; Th\`ese de Doctorat d'Etat, Universit\'e Pierre et Marie Curie, Paris VI, 1979.
available (see files these1.pdf to these6.pdf) on http://www.ihes.fr/\~{}damour/Articles/

\bibitem{HH1972}
  S.~W.~Hawking and J.~B.~Hartle,
  ``Energy And Angular Momentum Flow Into A Black Hole,''
  Commun.\ Math.\ Phys.\  {\bf 27} (1972) 283.
  %%CITATION = CMPHA,27,283;%%

  %\cite{Bhattacharyya:2008jc}
\bibitem{Bhattacharyya:2008jc}
  S.~Bhattacharyya, V.~E.~Hubeny, S.~Minwalla and M.~Rangamani,
  ``Nonlinear Fluid Dynamics from Gravity,''
  JHEP {\bf 0802} (2008) 045
  [arXiv:0712.2456 [hep-th]].
  %%CITATION = JHEPA,0802,045;%%

  %\cite{Bhattacharyya:2008mz}
\bibitem{Bhattacharyya:2008mz}
  S.~Bhattacharyya, R.~Loganayagam, I.~Mandal, S.~Minwalla and A.~Sharma,
  ``Conformal Nonlinear Fluid Dynamics from Gravity in Arbitrary Dimensions,''
  JHEP {\bf 0812} (2008) 116
  [arXiv:0809.4272 [hep-th]].
  %%CITATION = JHEPA,0812,116;%%    
  


%\cite{Polyakov:1992yw}
\bibitem{Polyakov:1992yw}
  A.~M.~Polyakov,
  ``Conformal turbulence,''
  arXiv:hep-th/9209046.
  %%CITATION = HEP-TH/9209046;%%

\bibitem{BKL}
  V.~A.~Belinsky, I.~M.~Khalatnikov and E.~M.~Lifshitz,
  ``Oscillatory approach to a singular point in the relativistic cosmology,''
  Adv.\ Phys.\  {\bf 19} (1970) 525.


\bibitem{Rich}
L.F. Richardson, ``Weather Prediction by Numerical Process.'' Cambridge: Cambridge University Press, 1922.

  
      %\cite{Bhattacharyya:2008ji}
\bibitem{Bhattacharyya:2008ji}
  S.~Bhattacharyya, R.~Loganayagam, S.~Minwalla, S.~Nampuri, S.~P.~Trivedi and S.~R.~Wadia,
  ``Forced Fluid Dynamics from Gravity,''
  JHEP {\bf 0902} (2009) 018
  [arXiv:0806.0006 [hep-th]].
  %%CITATION = JHEPA,0902,018;%%
  
  %\cite{Evslin:2010ss}
\bibitem{Evslin:2010ss}
  J.~Evslin and G.~Ricco,
  %``The Surface Layers Dual to Hydrodynamic Boundaries,''
  arXiv:1009.0175 [hep-th].
  %%CITATION = ARXIV:1009.0175;%%
  
  
  %\cite{Israel:1966rt}
\bibitem{Israel:1966rt}
  W.~Israel,
  ``Singular hypersurfaces and thin shells in general relativity,''
  Nuovo Cim.\  B {\bf 44S10} (1966) 1
  [Erratum-ibid.\  B {\bf 48} (1967) 463]
  [Nuovo Cim.\  B {\bf 44} (1966) 1].
  %%CITATION = NUCIA,B44,1;%%

  
 %__________________________________________________________

%\cite{Penrose:1969pc}
\bibitem{P1969}
  R.~Penrose,
  ``Gravitational collapse: The role of general relativity,''
  Riv.\ Nuovo Cim.\  {\bf 1} (1969) 252
  [Gen.\ Rel.\ Grav.\  {\bf 34} (2002) 1141].
  %%CITATION = GRGVA,34,1141;%%

%\cite{Christodoulou:1970wf}
\bibitem{C1970}
  D.~Christodoulou,
  ``Reversible and irreversible transforations in black hole physics,''
  Phys.\ Rev.\ Lett.\  {\bf 25} (1970) 1596.
  %%CITATION = PRLTA,25,1596;%%

%\cite{Christodoulou:1972kt}
\bibitem{CR1971}  D.~Christodoulou and R.~Ruffini,
  ``Reversible transformations of a charged black hole,''
  Phys.\ Rev.\  D {\bf 4} (1971) 3552.
  %%CITATION = PHRVA,D4,3552;%%

%\cite{Hawking:1971tu}
\bibitem{H1971}
  S.~W.~Hawking,
  ``Gravitational radiation from colliding black holes,''
  Phys.\ Rev.\ Lett.\  {\bf 26} (1971) 1344.
  %%CITATION = PRLTA,26,1344;%%

%\cite{Bardeen:1973gs}
\bibitem{BCH1973}  J.~M.~Bardeen, B.~Carter and S.~W.~Hawking,
  ``The Four laws of black hole mechanics,''
  Commun.\ Math.\ Phys.\  {\bf 31} (1973) 161.
  %%CITATION = CMPHA,31,161;%%

%\cite{Hanni:1973fn}
\bibitem{HR1973}
  R.~S.~Hanni and R.~Ruffini,
  ``Lines of Force of a Point Charge near a Schwarzschild Black Hole,''
  Phys.\ Rev.\  D {\bf 8} (1973) 3259.
  %%CITATION = PHRVA,D8,3259;%%

%\cite{Damour:1978cg}
\bibitem{D1978}
  T.~Damour,
  ``Black Hole Eddy Currents,''
  Phys.\ Rev.\  D {\bf 18} (1978) 3598.
  %%CITATION = PHRVA,D18,3598;%%

\bibitem{D1982}
T. Damour,
in: ``Surface Effects in Black Hole Physics'';
Proceedings of the Second Marcel Grossmann Meeting on General Relativity,
(edited by R. Ruffini, North Holland, 1982) pp 587-608;
available (see file surfaceeffects.pdf) on 
 http://www.ihes.fr/\~{}damour/Articles/

%\cite{Blandford:1977ds}
\bibitem{Z1978}
  R.~D.~Blandford and R.~L.~Znajek,
  ``Electromagnetic Extractions Of Energy From Kerr Black Holes,''
  Mon.\ Not.\ Roy.\ Astron.\ Soc.\  {\bf 179} (1977) 433.
  %%CITATION = MNRAA,179,433;%%
  
  %\cite{Thorne:1986iy}
\bibitem{Thorne:1986iy}
  K.~S.~.~Thorne, R.~H.~.~Price and D.~A.~.~Macdonald,
  ``Black holes: the membrane paradigm,''
%\href{http://www.slac.stanford.edu/spires/find/hep/www?irn=2064235}{SPIRES entry}
{\it  New Haven, USA: Yale Univ. Pr. (1986) 367p}

%\cite{Bekenstein:1973ur}
\bibitem{Bekenstein:1973ur}
  J.~D.~Bekenstein,
  ``Black holes and entropy,''
  Phys.\ Rev.\  D {\bf 7} (1973) 2333.
  %%CITATION = PHRVA,D7,2333;%%

%\cite{Hawking:1974sw}
\bibitem{Hawking:1974sw}
  S.~W.~Hawking,
  ``Particle Creation By Black Holes,''
  Commun.\ Math.\ Phys.\  {\bf 43} (1975) 199
  [Erratum-ibid.\  {\bf 46} (1976) 206].
  %%CITATION = CMPHA,43,199;%%
  
  %\cite{Damour:1976jd}
\bibitem{Damour:1976jd}
  T.~Damour and R.~Ruffini,
  ``Black Hole Evaporation In The Klein-Sauter-Heisenberg-Euler Formalism,''
  Phys.\ Rev.\  D {\bf 14} (1976) 332.
  %%CITATION = PHRVA,D14,332;%%
  
  %\cite{Damour:2004kwa}
\bibitem{Damour:2004kwa}
  T.~Damour,
  ``The entropy of black holes: A primer,''
  arXiv:hep-th/0401160.
  %%CITATION = HEP-TH/0401160;%%
  
  %\cite{Gourgoulhon:2005ch}
\bibitem{Gourgoulhon:2005ch}
  E.~Gourgoulhon,
  ``A generalized Damour-Navier-Stokes equation applied to trapping horizons,''
  Phys.\ Rev.\  D {\bf 72} (2005) 104007
  [arXiv:gr-qc/0508003].
  %%CITATION = PHRVA,D72,104007;%%

%\cite{Gourgoulhon:2005ng}
\bibitem{Gourgoulhon:2005ng}
  E.~Gourgoulhon and J.~L.~Jaramillo,
  ``A 3+1 perspective on null hypersurfaces and isolated horizons,''
  Phys.\ Rept.\  {\bf 423} (2006) 159
  [arXiv:gr-qc/0503113].
  %%CITATION = PRPLC,423,159;%%
  
  %\cite{Kovtun:2004de}
\bibitem{Kovtun:2004de}
  P.~Kovtun, D.~T.~Son and A.~O.~Starinets,
  ``Viscosity in strongly interacting quantum field theories from black hole physics,''
  Phys.\ Rev.\ Lett.\  {\bf 94} (2005) 111601
  [arXiv:hep-th/0405231].
  %%CITATION = PRLTA,94,111601;%%
  
  %\cite{Son:2007vk}
\bibitem{Son:2007vk}
  D.~T.~Son and A.~O.~Starinets,
  ``Viscosity, Black Holes, and Quantum Field Theory,''
  Ann.\ Rev.\ Nucl.\ Part.\ Sci.\  {\bf 57} (2007) 95
  [arXiv:0704.0240 [hep-th]].
  %%CITATION = ARNUA,57,95;%%
   
   
%________________________________________________________________ 
   
   
   \bibitem{Fefferman:2000wo}
C.~Fefferman, { {``Existence and smoothness of the Navier-Stokes equation''}},
\newblock Clay Millenium Problems (2000).
   
   \bibitem{Landau:1965pi}
   L.~Landau and E.~Lifshitz, { {``Fluid Mechanics: Course of Theoretical
  Physics'', Vol. 6}},
\newblock Butterworth-Heinemann, 1965.
      
   %\cite{Andersson:2006nr}
\bibitem{Andersson:2006nr}
  N.~Andersson and G.~L.~Comer,
  ``Relativistic fluid dynamics: Physics for many different scales,''
  Living Rev.\ Rel.\  {\bf 10} (2005) 1
  [arXiv:gr-qc/0605010].
  %%CITATION = 00222,10,1;%%
  
   \bibitem{Weinberg}
S.~Weinberg, { {``Gravitation and Cosmology, Principle and applications of the General Theory of Relativity''}},
\newblock John Wiley \& Sons, 1972.
  
       %\cite{Rangamani:2009xk}
\bibitem{Rangamani:2009xk}{rang}
  M.~Rangamani,
  ``Gravity and Hydrodynamics: Lectures on the fluid-gravity correspondence,''
  Class.\ Quant.\ Grav.\  {\bf 26} (2009) 224003
  [arXiv:0905.4352 [hep-th]].
  %%CITATION = CQGRD,26,224003;%%  
   
   %\cite{Erdmenger:2008rm}
\bibitem{Erdmenger:2008rm}
  J.~Erdmenger, M.~Haack, M.~Kaminski and A.~Yarom,
  ``Fluid dynamics of R-charged black holes,''
  JHEP {\bf 0901} (2009) 055
  [arXiv:0809.2488 [hep-th]].
  %%CITATION = JHEPA,0901,055;%%
   
   %\cite{Banerjee:2008th}
\bibitem{Banerjee:2008th}
  N.~Banerjee, J.~Bhattacharya, S.~Bhattacharyya, S.~Dutta, R.~Loganayagam and P.~Surowka,
  ``Hydrodynamics from charged black branes,''
  arXiv:0809.2596 [hep-th].
  %%CITATION = ARXIV:0809.2596;%%
   
     %\cite{Bhattacharyya:2008kq}
\bibitem{Bhattacharyya:2008kq}
  S.~Bhattacharyya, S.~Minwalla and S.~R.~Wadia,
  ``The Incompressible Non-Relativistic Navier-Stokes Equation from Gravity,''
  JHEP {\bf 0908} (2009) 059
  [arXiv:0810.1545 [hep-th]].
  %%CITATION = JHEPA,0908,059;%%
   
   
   
   

   
   
   
   
   
   
   
  
%________________________________________________________________  
    
 %\cite{Maldacena:1997re}
\bibitem{Maldacena:1997re}
  J.~M.~Maldacena,
  ``The large N limit of superconformal field theories and supergravity,''
  Adv.\ Theor.\ Math.\ Phys.\  {\bf 2}, 231 (1998)
  [Int.\ J.\ Theor.\ Phys.\  {\bf 38}, 1113 (1999)]
  [arXiv:hep-th/9711200].
  %%CITATION = IJTPB,38,1113;%%

%\cite{Gubser:1998bc}
\bibitem{Gubser:1998bc}
  S.~S.~Gubser, I.~R.~Klebanov and A.~M.~Polyakov,
  ``Gauge theory correlators from non-critical string theory,''
  Phys.\ Lett.\  B {\bf 428}, 105 (1998)
  [arXiv:hep-th/9802109].
  %%CITATION = PHLTA,B428,105;%%

%\cite{Witten:1998qj}
\bibitem{Witten:1998qj}
  E.~Witten,
  ``Anti-de Sitter space and holography,''
  Adv.\ Theor.\ Math.\ Phys.\  {\bf 2}, 253 (1998)
  [arXiv:hep-th/9802150].
  %%CITATION = 00203,2,253;%%
  
  
  %\cite{VanRaamsdonk:2008fp}
\bibitem{VanRaamsdonk:2008fp}
  M.~Van Raamsdonk,
  ``Black Hole Dynamics From Atmospheric Science,''
  JHEP {\bf 0805}, 106 (2008)
  [arXiv:0802.3224 [hep-th]].
  %%CITATION = JHEPA,0805,106;%%
  
  %\cite{Bhattacharyya:2008xc}
\bibitem{Bhattacharyya:2008xc}
  S.~Bhattacharyya {\it et al.},
  ``Local Fluid Dynamical Entropy from Gravity,''
  JHEP {\bf 0806}, 055 (2008)
  [arXiv:0803.2526 [hep-th]].
  %%CITATION = JHEPA,0806,055;%%
  
  %\cite{Dutta:2008gf}
\bibitem{Dutta:2008gf}
  S.~Dutta,
  ``Higher Derivative Corrections to Locally Black Brane Metrics,''
  JHEP {\bf 0805}, 082 (2008)
  [arXiv:0804.2453 [hep-th]].
  %%CITATION = JHEPA,0805,082;%%
  
  %\cite{Loganayagam:2008is}
\bibitem{Loganayagam:2008is}
  R.~Loganayagam,
  ``Entropy Current in Conformal Hydrodynamics,''
  JHEP {\bf 0805}, 087 (2008)
  [arXiv:0801.3701 [hep-th]].
  %%CITATION = JHEPA,0805,087;%%
    
    %\cite{Haack:2008cp}
\bibitem{Haack:2008cp}
  M.~Haack and A.~Yarom,
  ``Nonlinear viscous hydrodynamics in various dimensions using AdS/CFT,''
  JHEP {\bf 0810} (2008) 063
  [arXiv:0806.4602 [hep-th]].
  %%CITATION = JHEPA,0810,063;%  

%\cite{Chamblin:1999by}
\bibitem{Chamblin:1999by}
  A.~Chamblin, S.~W.~Hawking and H.~S.~Reall,
  ``Brane-World Black Holes,''
  Phys.\ Rev.\  D {\bf 61}, 065007 (2000)
  [arXiv:hep-th/9909205].
  %%CITATION = PHRVA,D61,065007;%%
  
  %\cite{Benincasa:2007tp}
\bibitem{Benincasa:2007tp}
  P.~Benincasa, A.~Buchel, M.~P.~Heller and R.~A.~Janik,
  ``On the supergravity description of boost invariant conformal plasma at strong coupling,''
  Phys.\ Rev.\  D {\bf 77}, 046006 (2008)
  [arXiv:0712.2025 [hep-th]].
  %%CITATION = PHRVA,D77,046006;%%
  
  %\cite{Buchel:2008kd}
\bibitem{Buchel:2008kd}
  A.~Buchel and M.~Paulos,
  ``Second order hydrodynamics of a CFT plasma from boost invariant expansion,''
  Nucl.\ Phys.\  B {\bf 810}, 40 (2009)
  [arXiv:0808.1601 [hep-th]].
  %%CITATION = NUPHA,B810,40;%%
  
  %\cite{Henningson:1998gx}
\bibitem{Henningson:1998gx}
  M.~Henningson and K.~Skenderis,
  ``The holographic Weyl anomaly,''
  JHEP {\bf 9807} (1998) 023
  [arXiv:hep-th/9806087].
  %%CITATION = JHEPA,9807,023;%%
  
  %\cite{Balasubramanian:1999re}
\bibitem{Balasubramanian:1999re}
  V.~Balasubramanian and P.~Kraus,
  ``A stress tensor for anti-de Sitter gravity,''
  Commun.\ Math.\ Phys.\  {\bf 208} (1999) 413
  [arXiv:hep-th/9902121].
  %%CITATION = CMPHA,208,413;%%
  
  %\cite{Chesler:2008hg}
\bibitem{Chesler:2008hg}
  P.~M.~Chesler and L.~G.~Yaffe,
  ``Horizon formation and far-from-equilibrium isotropization in supersymmetric Yang-Mills plasma,''
  Phys.\ Rev.\ Lett.\  {\bf 102} (2009) 211601
  [arXiv:0812.2053 [hep-th]].
  %%CITATION = PRLTA,102,211601;%%
  
  %\cite{Friess:2006kw}
\bibitem{Friess:2006kw}
  J.~J.~Friess, S.~S.~Gubser, G.~Michalogiorgakis and S.~S.~Pufu,
  ``Expanding plasmas and quasinormal modes of anti-de Sitter black holes,''
  JHEP {\bf 0704}, 080 (2007)
  [arXiv:hep-th/0611005].
  %%CITATION = JHEPA,0704,080;%%
  
  %\cite{Figueras:2009iu}
\bibitem{Figueras:2009iu}
  P.~Figueras, V.~E.~Hubeny, M.~Rangamani and S.~F.~Ross,
  ``Dynamical black holes and expanding plasmas,''
  JHEP {\bf 0904}, 137 (2009)
  [arXiv:0902.4696 [hep-th]].
  %%CITATION = JHEPA,0904,137;%%
  
  
  %_________________________________________________________________________
  
    %\cite{Hawking:1973uf}
\bibitem{Hawking:1973uf}
  S.~W.~Hawking and G.~F.~R.~Ellis,
  ``The Large scale structure of space-time,''
%\href{http://www.slac.stanford.edu/spires/find/hep/www?irn=6991262}{SPIRES entry}
{\it  Cambridge University Press, Cambridge, 1973}
  
  %\cite{Wald:1984rg}
\bibitem{Wald:1984rg}
  R.~M.~Wald,
  ``General Relativity,''
%\href{http://www.slac.stanford.edu/spires/find/hep/www?irn=1334239}{SPIRES entry}
{\it  Chicago, Usa: Univ. Pr. ( 1984) 491p}

%\cite{Hsu:2004vr}
\bibitem{Hsu:2004vr}
  S.~D.~H.~Hsu, A.~Jenkins and M.~B.~Wise,
  ``Gradient instability for w<-1,''
  Phys.\ Lett.\  B {\bf 597} (2004) 270
  [arXiv:astro-ph/0406043].
  %%CITATION = PHLTA,B597,270;%%
  
  %\cite{Dubovsky:2005xd}
\bibitem{Dubovsky:2005xd}
  S.~Dubovsky, T.~Gregoire, A.~Nicolis and R.~Rattazzi,
  ``Null energy condition and superluminal propagation,''
  JHEP {\bf 0603} (2006) 025
  [arXiv:hep-th/0512260].
  %%CITATION = JHEPA,0603,025;%%
  
  %\cite{Buniy:2005vh}
\bibitem{Buniy:2005vh}
  R.~V.~Buniy and S.~D.~H.~Hsu,
  ``Instabilities and the null energy condition,''
  Phys.\ Lett.\  B {\bf 632} (2006) 543
  [arXiv:hep-th/0502203].
  %%CITATION = PHLTA,B632,543;%%
  
%\cite{Misner:1974qy}
\bibitem{Misner:1974qy}
  C.~W.~Misner, K.~S.~Thorne and J.~A.~Wheeler,
  ``Gravitation,''
%\href{http://www.slac.stanford.edu/spires/find/hep/www?irn=6627595}{SPIRES entry}
{\it  San Francisco 1973, 1279p}
  
  
%___________________________________________________________________
  
\bibitem{shock}
  S.~Khlebnikov, M.~Kruczenski and G.~Michalogiorgakis,
  ``Shock waves in strongly coupled plasmas,''
  arXiv:1004.3803 [hep-th].


\bibitem{vortices}
  J.~Evslin and C.~Krishnan,
  ``Vortices in (2+1)d Conformal Fluids,''
  arXiv:1007.4452 [hep-th].

\bibitem{ac}
  N.~Andersson and G.~L.~Comer,
  ``Relativistic fluid dynamics: Physics for many different scales,''
  Living Rev.\ Rel.\  {\bf 10} (2005) 1
  [arXiv:gr-qc/0605010].


\bibitem{tranquilli}
  P.~Creminelli, M.~A.~Luty, A.~Nicolis and L.~Senatore,
  ``Starting the universe: Stable violation of the null energy condition and
  non-standard cosmologies,''
  JHEP {\bf 0612} (2006) 080
  [arXiv:hep-th/0606090].


\bibitem{Kraichnan}
R. H. Kraichnan, ``Inertial ranges in two dimensional turbulence'', 
Phys. Fluids {\bf 10} (1967) 1417-1423.

\bibitem{mcwilliams} 
J.~C.~McWilliams, ``The vortices of two-dimensional turbulence,'' 
J. of Fluid Mech., {\bf 219} (1990) 361-385.

\bibitem{Kol1}
A. N. Kolmogorov, ``The local structure of turbulence in incompressible viscous fluid for very large Reynolds numbers''. Proceedings of the USSR Academy of Sciences {\bf{30}} (1941) 299-303. 

\bibitem{Kol2}
A. N. Kolmogorov, ``Dissipation of energy in locally isotropic turbulence"''. Proceedings of the USSR Academy of Sciences {\bf{32}} (1941) 16-18.

\bibitem{nima}
  A.~Adams, N.~Arkani-Hamed, S.~Dubovsky, A.~Nicolis and R.~Rattazzi,
  ``Causality, analyticity and an IR obstruction to UV completion,''
  JHEP {\bf 0610} (2006) 014
  [arXiv:hep-th/0602178].

\bibitem{MarolfRang}
  V.~E.~Hubeny, D.~Marolf and M.~Rangamani,
  ``Black funnels and droplets from the AdS C-metrics,''
  Class.\ Quant.\ Grav.\  {\bf 27} (2010) 025001
  [arXiv:0909.0005 [hep-th]].

\bibitem{bf1}
  R.~Emparan, T.~Harmark, V.~Niarchos and N.~A.~Obers,
  ``Essentials of Blackfold Dynamics,''
  JHEP {\bf 1003} (2010) 063
  [arXiv:0910.1601 [hep-th]].


\bibitem{bf2}
  R.~Emparan, T.~Harmark, V.~Niarchos and N.~A.~Obers,
  ``New Horizons for Black Holes and Branes,''
  JHEP {\bf 1004} (2010) 046
  [arXiv:0912.2352 [hep-th]].


\bibitem{Strominger}
  I.~Bredberg, C.~Keeler, V.~Lysov and A.~Strominger,
  ``Wilsonian Approach to Fluid/Gravity Duality,''
  arXiv:1006.1902 [hep-th].
 
\bibitem{5d}
  R.~Emparan and H.~S.~Reall,
  ``A rotating black ring in five dimensions,''
  Phys.\ Rev.\ Lett.\  {\bf 88}, 101101 (2002)
  [arXiv:hep-th/0110260].
  %%CITATION = PRLTA,88,101101;%
  A.~A.~Pomeransky and R.~A.~Sen'kov,
  ``Black ring with two angular momenta,''
  arXiv:hep-th/0612005.
  H.~Elvang and P.~Figueras,
  ``Black Saturn,''
  JHEP {\bf 0705}, 050 (2007)
  [arXiv:hep-th/0701035].
  J.~Evslin and C.~Krishnan,
  ``Metastable Black Saturns,''
  JHEP {\bf 0809}, 003 (2008)
  [arXiv:0804.4575 [hep-th]].
  H.~Iguchi and T.~Mishima,
  ``Black di-ring and infinite nonuniqueness,''
  Phys.\ Rev.\  D {\bf 75}, 064018 (2007)
  [Erratum-ibid.\  D {\bf 78}, 069903 (2008)]
  [arXiv:hep-th/0701043].
  J.~Evslin and C.~Krishnan,
  ``The Black Di-Ring: An Inverse Scattering Construction,''
  Class.\ Quant.\ Grav.\  {\bf 26}, 125018 (2009)
  [arXiv:0706.1231 [hep-th]].
  K.~Izumi,
  ``Orthogonal black di-ring solution,''
  Prog.\ Theor.\ Phys.\  {\bf 119}, 757 (2008)
  [arXiv:0712.0902 [hep-th]].
  H.~Elvang and M.~J.~Rodriguez,
  ``Bicycling Black Rings,''
  JHEP {\bf 0804}, 045 (2008)
  [arXiv:0712.2425 [hep-th]].
  %%CITATION = JHEPA,0804,045;%%
  R.~Emparan and H.~S.~Reall,
  ``Black Holes in Higher Dimensions,''
  Living Rev.\ Rel.\  {\bf 11}, 6 (2008)
  [arXiv:0801.3471 [hep-th]].
  %%CITATION = 00222,11,6;%%
  J.~Evslin,
  ``Geometric Engineering 5d Black Holes with Rod Diagrams,''
  JHEP {\bf 0809}, 004 (2008)
  [arXiv:0806.3389 [hep-th]].
  %%CITATION = JHEPA,0809,004;%%


  
  
  
 \end{thebibliography}
\end{document}